\documentclass{aa}
\usepackage[varg]{txfonts}
\usepackage{graphicx}
\usepackage{natbib,twoopt}
\usepackage[breaklinks=true]{hyperref} 
\bibpunct{(}{)}{;}{a}{}{,} 
\makeatletter
 \newcommandtwoopt{\citeads}[3][][]{\href{https://ui.adsabs.harvard.edu/abs/#3/abstract}%
 {\def\hyper@linkstart##1##2{}%
 \let\hyper@linkend\@empty\citealp[#1][#2]{#3}}}
 \newcommandtwoopt{\citepads}[3][][]{\href{https://ui.adsabs.harvard.edu/abs/#3/abstract}%
 {\def\hyper@linkstart##1##2{}%
 \let\hyper@linkend\@empty\citep[#1][#2]{#3}}}
 \newcommandtwoopt{\citetads}[3][][]{\href{https://ui.adsabs.harvard.edu/abs/#3/abstract}%
 {\def\hyper@linkstart##1##2{}%
 \let\hyper@linkend\@empty\citet[#1][#2]{#3}}}
 \newcommandtwoopt{\citeyearads}[3][][]%
 {\href{https://ui.adsabs.harvard.edu/abs/#3/abstract}
 {\def\hyper@linkstart##1##2{}%
 \let\hyper@linkend\@empty\citeyear[#1][#2]{#3}}}
\makeatother

\begin{document}
   \title{Constraining the orientation of the spin axes
          of extrasolar minor bodies 1I/2017~U1~(`Oumuamua) and 2I/Borisov}
   \author{C. de la Fuente Marcos$^{1}$
           \and
           R. de la Fuente Marcos$^{2}$}
   \authorrunning{C. de la Fuente Marcos \and R. de la Fuente Marcos}
   \titlerunning{Spin axes of 1I/2017~U1~(`Oumuamua) and 2I/Borisov}
   \offprints{C. de la Fuente Marcos, \email{nbplanet@ucm.es}}
   \institute{$^{1}$ Universidad Complutense de Madrid,
              Ciudad Universitaria, E-28040 Madrid, Spain \\
              $^{2}$AEGORA Research Group,
              Facultad de Ciencias Matem\'aticas,
              Universidad Complutense de Madrid,
              Ciudad Universitaria, E-28040 Madrid, Spain}
   \date{Received 5 January 2020 / Accepted 16 September 2020}

   \abstract
      {The orientation of the spin axis of a comet is defined by the values of 
       its equatorial obliquity and its cometocentric longitude of the Sun at
       perihelion. These parameters can be computed from the components of the
       nongravitational force caused by outgassing if the cometary activity is 
       well characterized. The trajectories of known interstellar bodies 
       passing through the Solar System show nongravitational accelerations. 
       }
      {The spin-axis orientation of 1I/2017~U1~(`Oumuamua) remains to be 
       determined; for 2I/Borisov, the already released results are mutually 
       exclusive. In both cases, the values of the components of the 
       nongravitational force are relatively well constrained. Here, we 
       investigate ---within the framework of the forced precession model of 
       a nonspherical cometary nucleus--- the orientation of the spin axes 
       of `Oumuamua and 2I/Borisov using public orbit determinations that 
       consider outgassing.
       }
      {We applied a Monte Carlo simulation using the covariance matrix method 
       together with Monte Carlo random search techniques to compute the 
       distributions of equatorial obliquities and cometocentric longitudes of 
       the Sun at perihelion of `Oumuamua and 2I/Borisov from the values of 
       the nongravitational parameters.
       }
      {We find that the equatorial obliquity of `Oumuamua could be about 
       93{\degr}, if it has a very prolate (fusiform) shape, or close to 
       16{\degr}, if it is very oblate (disk-like). Different orbit 
       determinations of 2I/Borisov gave obliquity values of 59{\degr} and 
       90{\degr}. The distributions of cometocentric longitudes were in 
       general multimodal. 
       }
      {Our calculations suggest that the most probable spin-axis direction of 
       `Oumuamua in equatorial coordinates is $(280{\degr},~+46{\degr})$ 
       if very prolate or $(312{\degr},~-50{\degr})$ if very oblate. Our 
       analysis favors a prolate shape. For the orbit determinations of 
       2I/Borisov used here, we find most probable poles pointing near 
       $(275{\degr},~+65{\degr})$ and $(231{\degr},~+30{\degr})$, 
       respectively. Although our analysis favors an oblate shape for 
       2I/Borisov, a prolate one cannot be ruled out.
       }

   \keywords{
      methods: data analysis -- methods: numerical -- celestial mechanics -- 
      comets: general -- comets: individual: 1I/2017~U1~(`Oumuamua) -- 
      comets: individual: 2I/Borisov  
            }

   \maketitle

   \section{Introduction\label{Intro}}
      Outgassing causes nongravitational accelerations that affect both the trajectories and rotational properties of minor bodies (see for 
      example \citealt{1981AREPS...9..113S}). The spin-axis orientation of a comet is defined by two angles (see for example fig.~1 in 
      \citealt{2004come.book..137Y}), the equatorial obliquity ($\epsilon_{\rm c}$) and the cometocentric longitude or argument of the 
      subsolar meridian at perihelion ($\phi_{\rm c}$). The subsolar meridian is the great circle intersecting the object's rotation axis 
      and the subsolar point, where the Sun is directly overhead as seen from the minor body. 

      The value of $\epsilon_{\rm c}$ gives the angle between the orbital and equatorial planes of the comet. Retrograde comets have 
      $\epsilon_{\rm c}\in(90\degr,~180\degr)$; if the direction of rotation is consistent with that of its orbital motion, 
      $\epsilon_{\rm c}<90\degr$ and the comet is prograde \citep{1981AREPS...9..113S}. The value of $\phi_{\rm c}$ is measured from the 
      vernal equinox of the comet (projection of the ascending node of the orbital plane of the comet on its equator) to the subsolar 
      meridian at perihelion in the direction of increasing true anomaly, $f$. Therefore, it gives the hour angle of the Sun as seen from 
      the comet at perihelion. When $\phi_{\rm c}\in(180\degr,~360\degr)$, the sunlit pole is the southern one; if 
      $\phi_{\rm c}\in(0\degr,~180\degr)$, the northern pole faces the Sun at perihelion \citep{1981AREPS...9..113S}. 

      In addition to the spin-axis orientation defined by $(\epsilon_{\rm c},~\phi_{\rm c})$, the thermal lag angle ($\eta_{\rm c}$) 
      measures how the region that dominates outgassing is shifted behind the subsolar meridian \citep{2004come.book..137Y}, and its value 
      depends on the thermophysical properties of the sublimation phenomena produced in the comet and its rotational period 
      \citep{1981AREPS...9..113S}. Maximum insolation takes place at comet's noon. If $\eta_{\rm c}\in(-90\degr, 0\degr)$, sublimation takes 
      place mainly when the Sun is rising as seen from the surface of the comet; when the peak of activity is shifted toward the afternoon 
      and evening hours, as seen from the comet, $\eta_{\rm c}\in(0\degr, 90\degr)$. The interval $\eta_{\rm c}\in(-90\degr, 90\degr)$ 
      corresponds to the comet's dayside; outside this range, we have the comet's nightside. Sublimation of volatile materials, such as 
      H$_{2}$O or CO, generates nongravitational accelerations that can be detected by analyzing astrometric data. 

      The nongravitational force is the result of directional mass loss and, therefore, its detectability decreases as the rotation rate 
      increases; the orbital evolution of fast cometary nuclei with rotation periods of a few hours is nearly unaffected by outgassing
      (see for example \citealt{2004come.book..281S}). \citet{1973AJ.....78..211M} discussed a mathematical formalism that developed the 
      model originally proposed by \citet{1950ApJ...111..375W} to quantify the nongravitational effects on the orbital evolution of comets. 
      Vaporization of volatiles produces a force with radial, transverse, and normal components of the form $A\ g(r)$ with $g(r) = \alpha\ 
      (r/r_0)^{-m} (1 + (r/r_0)^{n})^{-k}$, where $r$ is the heliocentric distance of the comet \citep{1973AJ.....78..211M}. For subsurface 
      water ice, $r_0=2.808$~AU, $k=4.6142$, $m=2.15$, $n=5.093$, and $\alpha=0.1112620426$ \citep{1973AJ.....78..211M}; for carbon monoxide 
      ice, $r_0=5.0$~AU, $k=2.6$, $m=2.0$, $n=3.0$, and $\alpha=0.0408373333128795$ \citep{1996MNRAS.283..347Y,2018Natur.559..223M}. The 
      values of the nongravitational parameters ---$A_1$ (radial), $A_2$ (transverse), and $A_3$ (normal)--- depend on those of the angles 
      $\epsilon_{\rm c}$, $\phi_{\rm c}$, and $\eta_{\rm c}$.

      Assuming a spherically symmetric nucleus and applying a linear precession model, the nongravitational parameters can be computed from
      astrometric observations of the comet, and $\epsilon_{\rm c}$, $\phi_{\rm c}$, and $\eta_{\rm c}$ can be obtained from the components
      of the nongravitational force by using an iterative least squares algorithm, as shown by \citet{1990AcA....40..405S}. However, many
      observed comets and active asteroids are known to be nonspherical; in particular, interstellar object 1I/2017~U1~(`Oumuamua) could be 
      a very elongated (see for example \citealt{2017ApJ...851L..31K,2017Natur.552..378M,2018ApJ...852L...2B,2018NatAs...2..407D,
      2018NatAs...2..383F}) or flattened \citep{2019MNRAS.489.3003M} body. 

      \citet{1984AJ.....89.1573S} developed the forced precession model of a nonspherical cometary nucleus to reproduce the effects of the 
      torque that anisotropic outgassing exerts. This model includes additional parameters, and the direction of the spin pole 
      ($\epsilon_{\rm c}$, $\phi_{\rm c}$) now depends on the oblateness of the nucleus: $p=1-R_{\rm b}/R_{\rm a}$, where $R_{\rm a}$ is the 
      equatorial radius of the object and $R_{\rm b}$ is its polar radius. When the nonspherical body rotates on its shorter axis, $p>0$, 
      we speak of an oblate nucleus; if the rotation of the nucleus is along its longer axis, $p<0$, we have a prolate one. 

      In this work, we use the formulae of the forced precession model of a nonspherical nucleus described by \citet{1998A&A...335..757K}. 
      This model assumes that the nongravitational effects due to outgassing come from one dominant active area located on the surface of
      an ellipsoidal object of oblateness $p$ that is spinning about a precessing axis at a given epoch with direction defined by 
      $(\epsilon_{\rm c},~\phi_{\rm c})$; the outgassing area reaches its maximum activity at an angle $\eta_{\rm c}$ from the subsolar 
      meridian as seen from the surface of the object. 

      The relevant part of the set of equations (3) in \citet{1998A&A...335..757K} is:
      \begin{eqnarray}
         C_1 & = & \frac{\cos{\eta_{\rm c}} + (1 - S - \cos{\eta_{\rm c}}) \ \sin^{2}{\epsilon_{\rm c}} \ \sin^{2}{\Lambda}} 
                        {\sqrt{1 - S_1 \ \sin^{2}{\Psi}}} \nonumber \,, \\
         C_2 & = & \frac{\sin{\eta_{\rm c}} \ \cos{\epsilon_{\rm c}} + (1 - S - \cos{\eta_{\rm c}}) \ \sin^{2}{\epsilon_{\rm c}} 
                                             \ \sin{\Lambda} \ \cos{\Lambda}} 
                        {\sqrt{1 - S_1 \ \sin^{2}{\Psi}}} \label{equationsFPM} \,, \\ 
         C_3 & = & \frac{(\sin{\eta_{\rm c}} \ \cos{\Lambda} + (1 - S - \cos{\eta_{\rm c}}) \ \cos{\epsilon_{\rm c}} \ \sin{\Lambda}) 
                                              \ \sin{\epsilon_{\rm c}}} 
                        {\sqrt{1 - S_1 \ \sin^{2}{\Psi}}} \nonumber \,, 
      \end{eqnarray}
      where $S = p \ (2 - p)$, $S_1 = S \ (2 - S)$, $\Lambda = \phi_{\rm c}+f$, and $\sin{\Psi} = \sin{\epsilon_{\rm c}} \ \sin{\Lambda}$. 
      For a spherical nucleus, $p=0$, the set of equations (\ref{equationsFPM}) collapses into the one shown on page 140 of 
      \citet{2004come.book..137Y} and originally derived by \citet{1981AREPS...9..113S}. In both cases, the direction cosines are also 
      given by the expressions $C_{1}$=$A_{1}/A$, $C_{2}$=$A_{2}/A$, and $C_{3}$=$A_{3}/A$, where $A$=$\sqrt{A_1^2+A_2^2+A_3^2}$. The 
      nongravitational parameters ($A_1$, $A_2$, and $A_3$) can be computed from astrometric observations of the comet, and 
      $\epsilon_{\rm c}$, $\phi_{\rm c}$, $\eta_{\rm c}$, and $p$ can be obtained from the components of the nongravitational force by 
      using an iterative least squares algorithm, as shown in \citet{1998A&A...335..757K}. 
%
%
     \begin{table*}
        \fontsize{8}{12pt}\selectfont
        \tabcolsep 0.15truecm
        \caption{\label{elements1I}Heliocentric and barycentric orbital elements, and 1$\sigma$ uncertainties of interstellar object 
                                   1I/2017~U1~(`Oumuamua).
                }
        \centering
        \begin{tabular}{lccc}
           \hline\hline
            Orbital parameter                                                        &   & Heliocentric                           & Barycentric  \\
           \hline
            Eccentricity, $e$                                                        & = &    1.20113$\pm$0.00002                 &    1.20315   \\
            Perihelion distance, $q$ (AU)                                            & = &    0.255912$\pm$0.000007               &    0.258847  \\
            Inclination, $i$ (\degr)                                                 & = &  122.7417$\pm$0.0003                   &  123.0322    \\
            Longitude of the ascending node, $\Omega$ (\degr)                        & = &   24.5969$\pm$0.0003                   &   24.7778    \\
            Argument of perihelion, $\omega$ (\degr)                                 & = &  241.8105$\pm$0.0012                   &  242.0634    \\
            Time of perihelion passage, $\tau$ (TDB)                                 & = & 2458006.0073$\pm$0.0003                & 2458005.7829 \\
           \hline
            Nongravitational radial acceleration parameter, $A_1$ (AU d$^{-2}$)      & = &  $2.8\times10^{-7}\pm3.6\times10^{-8}$ &              \\
            Nongravitational transverse acceleration parameter, $A_2$ (AU d$^{-2}$)  & = &  $1.4\times10^{-8}\pm2.4\times10^{-8}$ &              \\
            Nongravitational normal acceleration parameter, $A_3$ (AU d$^{-2}$)      & = &  $1.6\times10^{-8}\pm2.2\times10^{-8}$ &              \\
           \hline
        \end{tabular}
        \tablefoot{This solution is hyperbolic with a statistical significance of 9644-$\sigma$ (barycentric, using all the decimal figures provided 
                   by the data source) and it is based on 207 observations that span a data-arc of 80~d. Although `Oumuamua reached perihelion on 
                   September 9, 2017, the first observation used in the calculations was acquired on October 14, 2017. The orbit determination has 
                   been computed by D.~Farnocchia at epoch JD 2458080.5 that corresponds to 00:00:00.000 TDB, Barycentric Dynamical Time, on 2017 
                   November 23, J2000.0 ecliptic and equinox. Source: JPL's SSDG SBDB (solution date, 2018-Jun-26 12:17:57 PDT).
                  }
     \end{table*}
%
%
%
%
     \begin{table*}
        \fontsize{8}{12pt}\selectfont
        \tabcolsep 0.15truecm
        \caption{\label{elements2Ia}Heliocentric and barycentric orbital elements, and 1$\sigma$ uncertainties of interstellar comet 2I/Borisov (I).
                }
        \centering
        \begin{tabular}{lccc}
           \hline\hline
            Orbital parameter                                                        &   & Heliocentric                            & Barycentric  \\
           \hline
            Eccentricity, $e$                                                        & = &    3.3571$\pm$0.0002                    &    3.3641    \\
            Perihelion distance, $q$ (AU)                                            & = &    2.00662$\pm$0.00002                  &    2.01315   \\
            Inclination, $i$ (\degr)                                                 & = &   44.05349$\pm$0.00014                  &   44.06149   \\
            Longitude of the ascending node, $\Omega$ (\degr)                        & = &  308.1500$\pm$0.0003                    &  308.0993    \\
            Argument of perihelion, $\omega$ (\degr)                                 & = &  209.1244$\pm$0.0004                    &  209.2067    \\
            Time of perihelion passage, $\tau$ (TDB)                                 & = & 2458826.0543$\pm$0.0006                 & 2458826.3146 \\
           \hline
            Nongravitational radial acceleration parameter, $A_1$ (AU d$^{-2}$)      & = & $-4.1\times10^{-8}\pm3.5\times10^{-8}$  &              \\
            Nongravitational transverse acceleration parameter, $A_2$ (AU d$^{-2}$)  & = & $-2.9\times10^{-8}\pm5.9\times10^{-8}$  &              \\
            Nongravitational normal acceleration parameter, $A_3$ (AU d$^{-2}$)      & = & $-1.15\times10^{-7}\pm1.7\times10^{-8}$ &              \\
           \hline
        \end{tabular}
        \tablefoot{This solution is hyperbolic with a statistical significance of 10950-$\sigma$ (barycentric, using all the decimal figures provided 
                   by the data source) and it is based on 1191 observations that span a data-arc of 389~d. Interstellar comet 2I/Borisov reached 
                   perihelion on December 8, 2019; the last observation used in these calculations was acquired on January 6, 2020 (the first one 
                   corresponds to December 13, 2018). The orbit determination has been computed by D. Farnocchia at epoch JD 2458853.5 that 
                   corresponds to 00:00:00.000 TDB on 2020 January 5, J2000.0 ecliptic and equinox. Source: JPL's SSDG SBDB (solution date, 
                   2020-Jan-09 10:42:19 PST).
                  }
     \end{table*}
%
%
%
%
     \begin{table*}
        \fontsize{8}{12pt}\selectfont
        \tabcolsep 0.15truecm
        \caption{\label{elements2Ib}Heliocentric and barycentric orbital elements, and 1$\sigma$ uncertainties of interstellar comet 2I/Borisov (II).
                }
        \centering
        \begin{tabular}{lccc}
           \hline\hline
            Orbital parameter                                                        &   & Heliocentric                           & Barycentric   \\
           \hline
            Eccentricity, $e$                                                        & = &    3.356191$\pm$0.000015               &    3.358810   \\
            Perihelion distance, $q$ (AU)                                            & = &    2.006624$\pm$0.000002               &    2.011869   \\
            Inclination, $i$ (\degr)                                                 & = &   44.052626$\pm$0.000011               &   44.062226   \\
            Longitude of the ascending node, $\Omega$ (\degr)                        & = &  308.14892$\pm$0.00003                 &  308.10039    \\
            Argument of perihelion, $\omega$ (\degr)                                 & = &  209.12461$\pm$0.00005                 &  209.16747    \\
            Time of perihelion passage, $\tau$ (TDB)                                 & = & 2458826.04489$\pm$0.00005              & 2458826.26179 \\
           \hline
            Nongravitational radial acceleration parameter, $A_1$ (AU d$^{-2}$)      & = & $7.31\times10^{-8}\pm4.2\times10^{-9}$ &               \\
            Nongravitational transverse acceleration parameter, $A_2$ (AU d$^{-2}$)  & = & $-3.3\times10^{-8}\pm1.1\times10^{-8}$ &               \\
            Nongravitational normal acceleration parameter, $A_3$ (AU d$^{-2}$)      & = & 0                                      &               \\
           \hline
        \end{tabular}
        \tablefoot{This solution is hyperbolic with a statistical significance of 155164-$\sigma$ (barycentric, using all the decimal figures provided
                   by the data source) and it is based on 1310 observations that span a data-arc of 444~d. The last observation used in these 
                   calculations was acquired on March 1, 2020. The orbit determination has been computed by D. Farnocchia at epoch JD 2459061.5 that 
                   corresponds to 00:00:00.000 TDB on 2020 July 31, J2000.0 ecliptic and equinox. Source: JPL's SSDG SBDB (solution date, 2020-Mar-19 
                   08:23:45 PST).
                  }
     \end{table*}

%
%

      The orbit determinations of extrasolar minor bodies `Oumuamua and 2I/Borisov (see Tables~\ref{elements1I}, \ref{elements2Ia}, 
      \ref{elements2Ib}, and \ref{elements2Ic}) include the usual orbital elements ---eccentricity, $e$, perihelion distance, $q$, 
      inclination, $i$, longitude of the ascending node, $\Omega$, argument of perihelion, $\omega$, and time of perihelion passage, 
      $\tau$--- and the nongravitational parameters (see above). If an orbit determination is meant to reproduce actual observations 
      accurately and to make reliable ephemeris predictions into the past and the future, one has to consider how the orbital parameters 
      (nongravitational ones included) affect one another using the covariance matrix whose elements indicate the level to which two given 
      parameters vary together. In this context, the mean values of the parameters and the covariance matrix define a hyperellipsoid in 
      multidimensional space. Such data can be used to derive distributions of the angles ($\epsilon_{\rm c}$, $\phi_{\rm c}$, and 
      $\eta_{\rm c}$), and the oblateness parameter, $p$, compatible with the observations via Monte Carlo random search starting with input 
      data generated as described by \citet{2015MNRAS.453.1288D}. 

      Here, we outline a procedure to derive the spin-axis orientation and thermal lag angle of a comet from its nongravitational orbit 
      determination that includes the mutual uncertainties between the various parameters. This approach is applied to study the orientation 
      of the spin axes of extrasolar minor bodies `Oumuamua and 2I/Borisov. This paper is organized as follows. In Sect.~\ref{Data}, we 
      present the data used in our analyses, discussing and validating our techniques. In Sect.~\ref{1I}, we apply our methodology to 
      `Oumuamua's data. The spin-axis orientation of 2I/Borisov is studied in Sect.~\ref{2I}. Our results are placed within the context of 
      previous work in Sect.~\ref{Discussion} and our conclusions are summarized in Sect.~\ref{Summary}.

   \section{Data description and methods\label{Data}}
      In this work, we find two types of assumptions that affect the results obtained. On the one hand, we have those linked to the model 
      used to constrain the three angles ($\epsilon_{\rm c}$, $\phi_{\rm c}$, and $\eta_{\rm c}$), and the value of the oblateness; on the 
      other, we find those associated with the public orbit determination, that has not been computed by us, such as the one species
      dominating the outgassing. 

      The assumptions behind the forced precession model of a nonspherical cometary nucleus are described in detail in 
      \citet{1984AJ.....89.1573S} and \citet{1998A&A...335..757K}. The ones affecting the orbit determination are validated numerically by 
      their ability to reproduce the available observations and predict the future positions of the object. If the orbit determination is 
      capable of providing the actual position of the object within a reasonable level of uncertainty (often less than a few arcseconds), 
      then one has to assume that the starting hypotheses (used to compute the orbit determination) are sufficiently justified. However, it 
      is certainly possible that some starting hypotheses could be formally inconsistent with evidence from other sources and still produce 
      numerically correct results, in the sense of being consistent with the observations.

      \subsection{Data sources}
         In this paper, we use publicly available data from Jet Propulsion Laboratory's (JPL) Small-Body Database (SBDB, 
         \citealt{2015IAUGA..2256293G})\footnote{\href{https://ssd.jpl.nasa.gov/sbdb.cgi}{https://ssd.jpl.nasa.gov/sbdb.cgi}} to 
         investigate the spin-axis orientations of 1I/2017~U1~(`Oumuamua) and 2I/Borisov. The data include the orbital elements and
         nongravitational parameters shown in Tables~\ref{elements1I}, \ref{elements2Ia}, and \ref{elements2Ib}, and their associated 
         covariance matrices. The data were used to construct the distributions of the angles $\epsilon_{\rm c}$, $\phi_{\rm c}$, and 
         $\eta_{\rm c}$, and in some cases that of the oblateness. Additional data have been retrieved from JPL's SBDB or the Minor Planet
         Center (MPC, \citealt{2016IAUS..318..265R}) using the tools provided by the Python package Astroquery \citep{2019AJ....157...98G}
         and the Python module sbpy \citep{2019JOSS....4.1426M}. 

         JPL's SBDB explicitly states that the orbit determination of `Oumuamua in Table~\ref{elements1I} is identical to solution 7c in 
         \citet{2018Natur.559..223M} that assumes a nongravitational acceleration that uses normalization factors appropriate for CO driven 
         sublimation (see above). However, we have to acknowledge that outgassing of carbon based molecules, specifically CO and CO$_2$ has 
         been definitively ruled out as accelerants by NASA's Spitzer Space Telescope non-detection of `Oumuamua \citep{2018AJ....156..261T}.
         The authors of this analysis suggest that outgassing of another gas species, perhaps H$_2$O, may be responsible for the observed
         nongravitational acceleration; however, the producer of the orbit determination in Table~\ref{elements1I}, D.~Farnocchia, is 
         listed as a coauthor of both \citet{2018Natur.559..223M} and \citet{2018AJ....156..261T}. In their table~1, 
         \citet{2019NatAs...3..594O} show that the upper limits for CO from different sources are clearly orders of magnitude lower than 
         those of other species. In any case, this inconsistency may not have any major impact on our results. 

         The primary objective of this work is not in contesting the orbit determinations computed by others, but in exploring their 
         associated effects on the possible rotational properties of the objects. As we use the numerical mean values and their associated 
         covariance matrices, and they are able to reproduce observational results (the astrometry) within reasonable accuracy levels, we 
         consider their numerical values as essentially valid even if the original hypotheses used to compute them could be invalid. The 
         subject of the actual driver of the detected outgassing remains controversial in the case of `Oumuamua (see for example 
         \citealt{2019ApJ...885L..41F,2019AJ....158..256H,2020ApJ...896L...8S,2020ApJ...899L..23H}).

         As for the orbit determinations of 2I/Borisov in Tables~\ref{elements2Ia} and \ref{elements2Ib}, JPL's SBDB states that the 
         standard model ---in other words, assuming a nongravitational acceleration that uses normalization factors appropriate for 
         H$_{2}$O driven sublimation (see above)--- has been applied to compute the nongravitational parameters. A water ice driver is 
         supported by \citet{2019arXiv191106271S}. However, it has already been confirmed that this comet exhibits an unusually high CO 
         abundance \citep{2020NatAs...4..861C}. Although the orbital elements in Tables~\ref{elements2Ia} and \ref{elements2Ib} are, in 
         statistical terms, mutually consistent, we cannot say the same about the nongravitational parameters. The orbit determination in 
         Table~\ref{elements2Ia} ---for which the last observation used was acquired on January 6, 2020--- shows a statistically significant 
         (above the 6.8-$\sigma$ level) value of the nongravitational normal acceleration parameter (but the others are statistically 
         compatible with zero). This is often the case when cometary outgassing comes from a single active region located at a rotation pole 
         and when the obliquity of the orbit plane to the equatorial plane is near 50{\degr} or 130{{\degr} (see sect. 3.1 in 
         \citealt{2004come.book..137Y}). 

         In sharp contrast, the orbit determination in Table~\ref{elements2Ib} has no value of the normal nongravitational parameter; 
         therefore, we have to assume that it is compatible with zero. In addition, the value of $A_2$ is consistent with that in 
         Table~\ref{elements2Ia} but now the dominant nongravitational acceleration is radial (statistically significant above the 
         17-$\sigma$ level), as it is in the case of `Oumuamua (see Table~\ref{elements1I}). The orbit determination in 
         Table~\ref{elements2Ib} includes observations acquired in January and February. Assuming that both orbit determinations are 
         correct, they may represent two snapshots in the dynamical life of 2I/Borisov that may reflect the rapidly evolving nature of the 
         outgassing processes taken place at the surface of the comet. In fact, there is robust observational evidence to support this 
         interpretation: A sequence of outbursts was observed by multiple observers in February-March \citep{2020ATel13613....1B,
         2020ATel13549....1D,2020ATel13611....1J,2020ApJ...896L..39J,2020ATel13618....1Z}.

      \subsection{Methodology}
         The first step of the approach used in this paper is an implementation of the Monte Carlo using the Covariance Matrix (MCCM, 
         \citealt{2001CeMDA..80..227B,2007SoSyR..41..413A}) methodology in which a Monte Carlo process generates control or clone orbits 
         based on the nominal orbit, but adding random noise on each orbital parameter by making use of the covariance matrix. Considering a 
         covariance matrix as computed by JPL's Solar System Dynamics Group, Horizons On-Line Ephemeris System, the vector including the 
         mean values of the orbital parameters at a given epoch $t_{0}$ is of the form $\vec{v} = (e, q, \tau, \Omega, \omega, i, A_1, A_2, 
         A_3)$. If \textbf{\textsf{C}} is the covariance matrix at the same epoch associated with the nominal orbital solution that is 
         symmetric and positive-semidefinite, then \textbf{\textsf{C}} = \textbf{\textsf{A}} \textbf{\textsf{A}}$^{\textbf{\textsf{T}}}$ 
         where \textbf{\textsf{A}} is a lower triangular matrix with real and positive diagonal elements, 
         \textbf{\textsf{A}}$^{\textbf{\textsf{T}}}$ is the transpose of \textbf{\textsf{A}}. In the case studied here, these matrices were 
         9$\times$9 and the expressions of the individual elements $a_{ij}$ of matrix \textbf{\textsf{A}} are shown in 
         Appendix~\ref{Aelements}.

         If $\vec{r}$ is a vector made of univariate Gaussian random numbers ---components $r_{i}$ with $i=1,9$ generated using the 
         Box-Muller method \citep{BM58}--- the sets of initial orbital elements of the control orbits can be computed using the expressions 
         (assuming the structure provided by JPL's Horizons On-Line Ephemeris System pointed out above), $\vec{v}_{\rm c} = \vec{v} + 
         \textbf{\textsf{A}}\,\vec{r}$:
         \begin{equation}
            \begin{aligned}
               e_{\rm c} & = e + a_{11}\,r_{1} \\
               q_{\rm c} & = q + a_{22}\,r_{2} + a_{21}\,r_{1} \\
               \tau_{\rm c} & = \tau + a_{33}\,r_{3} + a_{32}\,r_{2} + a_{31}\,r_{1} \\
               \Omega_{\rm c} & = \Omega + a_{44}\,r_{4} + a_{43}\,r_{3} + a_{42}\,r_{2} + a_{41}\,r_{1} \\
               \omega_{\rm c} & = \omega + a_{55}\,r_{5} + a_{54}\,r_{4} + a_{53}\,r_{3} + a_{52}\,r_{2} + a_{51}\,r_{1} \\
               i_{\rm c} & = i + a_{66}\,r_{6} + a_{65}\,r_{5} + a_{64}\,r_{4} + a_{63}\,r_{3} + a_{62}\,r_{2} + a_{61}\,r_{1} \\
               A_{1 {\rm c}} & = A_{1} + a_{77}\,r_{7} + a_{76}\,r_{6} + a_{75}\,r_{5} + a_{74}\,r_{4} + a_{73}\,r_{3} + a_{72}\,r_{2} \\
                             & + a_{71}\,r_{1} \\
               A_{2 {\rm c}} & = A_{2} + a_{88}\,r_{8} + a_{87}\,r_{7} + a_{86}\,r_{6} + a_{85}\,r_{5} + a_{84}\,r_{4} + a_{83}\,r_{3} \\
                             & + a_{82}\,r_{2} + a_{81}\,r_{1} \\
               A_{3 {\rm c}} & = A_{3} + a_{99}\,r_{9} + a_{98}\,r_{8} + a_{97}\,r_{7} + a_{96}\,r_{6} + a_{95}\,r_{5} + a_{94}\,r_{4} \\
                             & + a_{93}\,r_{3} + a_{92}\,r_{2} + a_{91}\,r_{1} \,.
               \label{good}
            \end{aligned}
         \end{equation}
         This approach and set of equations have already been used to study the dynamics of 2I/Borisov as presented in 
         \citet{2020MNRAS.495.2053D}. 

         In order to reconstruct the distributions of $\epsilon_{\rm c}$, $\phi_{\rm c}$, and $\eta_{\rm c}$, and the oblateness, we 
         generated $10^{5}$ instances of a given orbit (unless stated otherwise) using the set of equations~(\ref{good}). For each one of 
         them, we estimated the associated values of the angles (and $p$, if not imposed by additional evidence) by applying a Monte Carlo 
         random search algorithm. The random search uses sets of values of the three angles (and $p$ in relevant cases) that are generated 
         randomly (uniformly) within the relevant intervals (see Sect.~\ref{Intro}) until a stopping criterion is met. For example, if we 
         consider the direction cosine for the radial component of the force (see equations in Sect.~\ref{Intro}):
         \begin{equation}
            C_1 =  \frac{\cos{\eta_{\rm c}} + (1 - S - \cos{\eta_{\rm c}}) \ \sin^{2}{\epsilon_{\rm c}} \ \sin^{2}{\Lambda}}
                        {\sqrt{1 - S_1 \ \sin^{2}{\Psi}}} \,,
            \label{radial}
         \end{equation}
         that is also given by the expression $C_{1}$=$A_{1}/A$, where $A$=$\sqrt{A_1^2+A_2^2+A_3^2}$. We can construct $C_{1 {\rm c}}$ 
         using data from MCCM and then try values of $\epsilon_{\rm c}$, $\phi_{\rm c}$, $\eta_{\rm c}$, and $p$ (to compute $S$ and $S_1$) 
         on Eq.~(\ref{radial}) until $\lvert{C_{1 {\rm c}} - C_1}\rvert<\Delta$, where $\Delta$ is a certain tolerance value that in our 
         calculations and for practical reasons was set to 0.0001 (our results do not depend significantly on the tolerance value). A set of 
         $\epsilon_{\rm c}$, $\phi_{\rm c}$, $\eta_{\rm c}$, and $p$ is regarded as a valid solution if 
         $\lvert{C_{i {\rm c}} - C_i}\rvert<\Delta$, with $i=1,3$ (the radial, transverse, and normal direction cosines). We applied this 
         technique and stopping criterion to obtain $10^{5}$ sets of angles and $p$ (one per orbital instance) that we used to generate the 
         distributions of angles and oblateness. This approach is robust when working with multimodal distributions that may signal the 
         existence of multiple solutions. An iterative procedure like the one discussed by \citet{1990AcA....40..405S} may not be able to 
         obtain all the statistically viable solutions.

         Our technique delivers the distribution of spin-axis orientations defined by $(\epsilon_{\rm c}, \phi_{\rm c})$, but other 
         alternative approaches discussed in the literature provide the ecliptic coordinates, $(\lambda, \beta)$, of the spin-axis 
         orientation. We can explore the associated distribution of pole directions by taking into account the dot product of the unit 
         vector in the direction of the rotation pole ---$\vec{u}_{\rm r}=(\cos{\beta_{\rm p}} \cos{\lambda_{\rm p}}, \cos{\beta_{\rm p}} 
         \sin{\lambda_{\rm p}}, \sin{\beta_{\rm p}})$--- with that of the orbital pole ---$\vec{u}_{\rm o}=(\sin{i_{\rm c}} 
         \sin{\Omega_{\rm c}}, -\sin{i_{\rm c}} \cos{\Omega_{\rm c}}, \cos{i_{\rm c}})$--- that gives $\vec{u}_{\rm r} \cdot 
         \vec{u}_{\rm o}=\cos{\epsilon_{\rm c}}$. Therefore, the expression that can be used to apply a random search technique with a 
         stopping criterion to find $(\lambda_{\rm p},~\beta_{\rm p})$ is: $\cos{\beta_{\rm p}} \cos{\lambda_{\rm p}} \sin{i_{\rm c}} 
         \sin{\Omega_{\rm c}} - \cos{\beta_{\rm p}} \sin{\lambda_{\rm p}} \sin{i_{\rm c}} \cos{\Omega_{\rm c}} + \sin{\beta_{\rm p}} 
         \cos{i_{\rm c}}=\cos{\epsilon_{\rm c}}$. From here, we can compute the equivalent distribution in equatorial coordinates, 
         $(\alpha,~\delta)$, using the conversion tools provided by Astropy \citep{2013A&A...558A..33A,2018AJ....156..123A}. 

         In order to analyze the results, we produced histograms using the Matplotlib library \citep{2007CSE.....9...90H} with two different 
         sets of bins computed using Astropy \citep{2013A&A...558A..33A,2018AJ....156..123A}: One by applying the Freedman and Diaconis rule 
         \citep{FD81} and a second one using the Bayesian Blocks technique \citep{2013ApJ...764..167S}. While the Freedman and Diaconis rule
         produces a constant bin size, the Bayesian Blocks technique allows varying bin widths that work well for complicated distributions
         like the ones studied here. We consider two different statistically motivated histograms to show that our conclusions do not depend
         on the choice of bin width. On the other hand and instead of using frequency-based histograms, we considered counts to form a 
         probability density so the area under the histogram will sum to one. In addition, we construct diagrams to visualize better the 
         statistical significance of the spin-axis orientations ---as $(\epsilon_{\rm c}, \phi_{\rm c})$, $(\lambda_{\rm p}, \beta_{\rm p})$, 
         or $(\alpha_{\rm p}, \delta_{\rm p})$ maps--- representing kernel-density estimates using Gaussian kernels (see for example 
         \citealt{Scott1992}) implemented by SciPy \citep{2020NatMe..17..261V}. 

         The spin-axis orientations obtained using the methodology described above are referred to the same epoch as the associated orbit 
         determination. We have not studied the time evolution of the rotational properties of minor bodies as described by, for example, 
         \citet{1998A&A...335..757K}. However, for an object following a hyperbolic path and in absence of outbursts, fragmentation 
         events, or close encounters with massive planets, the overall spin-axis orientation is expected to remain relatively stable as it 
         is often the case for each apparition (not for multiple consecutive apparitions) of well-studied comets, see sections 3.5 to 3.8 in 
         \citet{2004come.book..137Y}. In this context, we may interpret the results obtained for a given epoch as the most probable values 
         of the rotational properties around that epoch.
%
%
     \begin{table*}
        \fontsize{8}{12pt}\selectfont
        \tabcolsep 0.15truecm
        \caption{\label{elements51P}Heliocentric orbital elements and 1$\sigma$ uncertainties of comet 51P/Harrington.
                }
        \centering
        \begin{tabular}{lcc}
           \hline\hline
            Orbital parameter                                                        &   & Value$\pm$1$\sigma$                        \\
           \hline
            Eccentricity, $e$                                                        & = &    0.5424370$\pm$0.0000006                 \\
            Perihelion distance, $q$ (AU)                                            & = &    1.699602$\pm$0.000002                   \\
            Inclination, $i$ (\degr)                                                 & = &    5.42439$\pm$0.00002                     \\
            Longitude of the ascending node, $\Omega$ (\degr)                        & = &   83.7008$\pm$0.0011                       \\
            Argument of perihelion, $\omega$ (\degr)                                 & = &  269.2756$\pm$0.0013                       \\
            Time of perihelion passage, $\tau$ (TDB)                                 & = & 2457247.1857$\pm$0.0005                    \\
           \hline
            Nongravitational radial acceleration parameter, $A_1$ (AU d$^{-2}$)      & = &  $3.867\times10^{-8}\pm6.9\times10^{-10}$  \\
            Nongravitational transverse acceleration parameter, $A_2$ (AU d$^{-2}$)  & = &  $3.9969\times10^{-9}\pm9.5\times10^{-12}$ \\
            Nongravitational normal acceleration parameter, $A_3$ (AU d$^{-2}$)      & = & $-7.129\times10^{-9}\pm8.4\times10^{-10}$  \\
           \hline
        \end{tabular}
        \tablefoot{This solution is based on 653 observations that span a data-arc of 5085~d. The orbit determination has been computed by 
                   A.~B. Chamberlin at epoch JD 2457300.5 that corresponds to 00:00:00.000 TDB on 2015 October 5, J2000.0 ecliptic and 
                   equinox. Source: JPL's SSDG SBDB (solution date, 2015-Oct-07 10:04:30 PST).
                  }
     \end{table*}
%
%

      \subsection{Validation: Comet 51P/Harrington}
         The nongravitational motion of comet 51P/Harrington has been studied by \citet{1996AcA....46...29S}. This comet has experienced 
         multiple fragmentation events since its discovery on August 14, 1953 \citep{1997A&A...318L...5S,2002EM&P...90..153K}. The work by
         \citet{1996AcA....46...29S} is referred to the epoch February 12, 1995 and the data-arc spans from discovery time to January 28,
         1995. The values of the relevant properties are $\epsilon_{\rm 51P/Harrington}$=77\fdg6$\pm$1\fdg1, 
         $\phi_{\rm 51P/Harrington}$=92\degr$\pm$4\degr, $\eta_{\rm 51P/Harrington}$=18\fdg4$\pm$1\fdg4, and 
         $p_{\rm 51P/Harrington}$=$-0.32\pm0.08$ \citep{1996AcA....46...29S}, and they have been computed using Sekanina's forced precession 
         model of a nonspherical cometary nucleus. \citet{1996AcA....46...29S} considered asymmetric cometary activity with respect to 
         perihelion in his calculations and $A$=$7.9192\times10^{-9}\pm1.987\times10^{-10}$~AU~d$^{-2}$, but the values of the individual
         nongravitational acceleration parameters were not provided. 

         Table~\ref{elements51P} shows the latest orbit determination available for comet 51P/Harrington. JPL's SBDB states that the orbit 
         determination has been computed using the standard model (see above). When comparing the data in Table~\ref{elements51P} with those 
         in Tables~\ref{elements1I}, \ref{elements2Ia}, and \ref{elements2Ib}, we realize that the statistical relevance of the 
         nongravitational parameters is substantially higher in the case of comet 51P/Harrington. It is also important to emphasize that 
         51P/Harrington is a periodic comet that has been studied for multiple decades (its orbital period is 7.16~yr); interstellar objects 
         can only be studied once and they have comparatively short visibility windows.
%
%
      \begin{figure}
        \centering
         \includegraphics[width=0.99\linewidth]{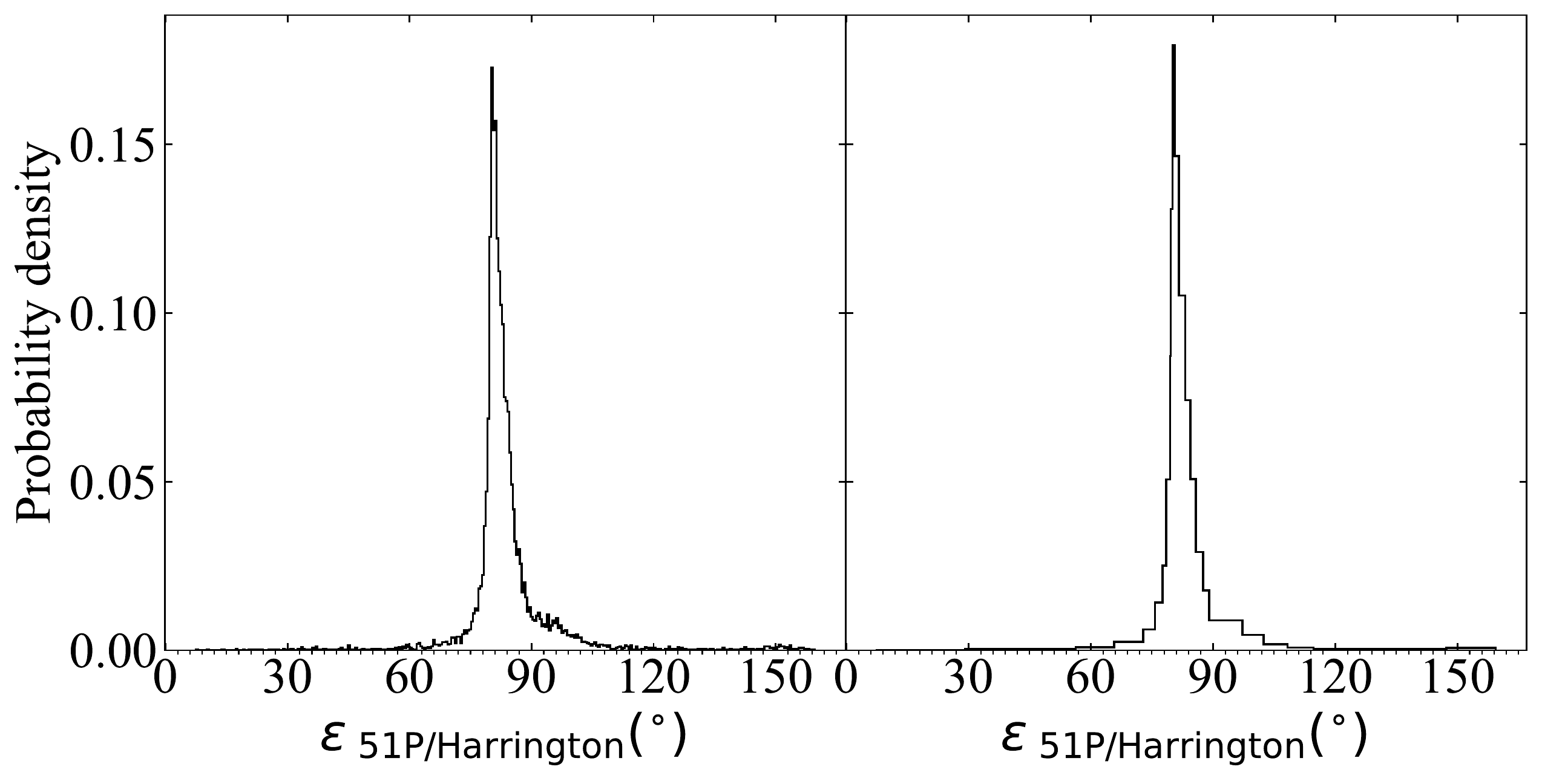}
         \includegraphics[width=0.99\linewidth]{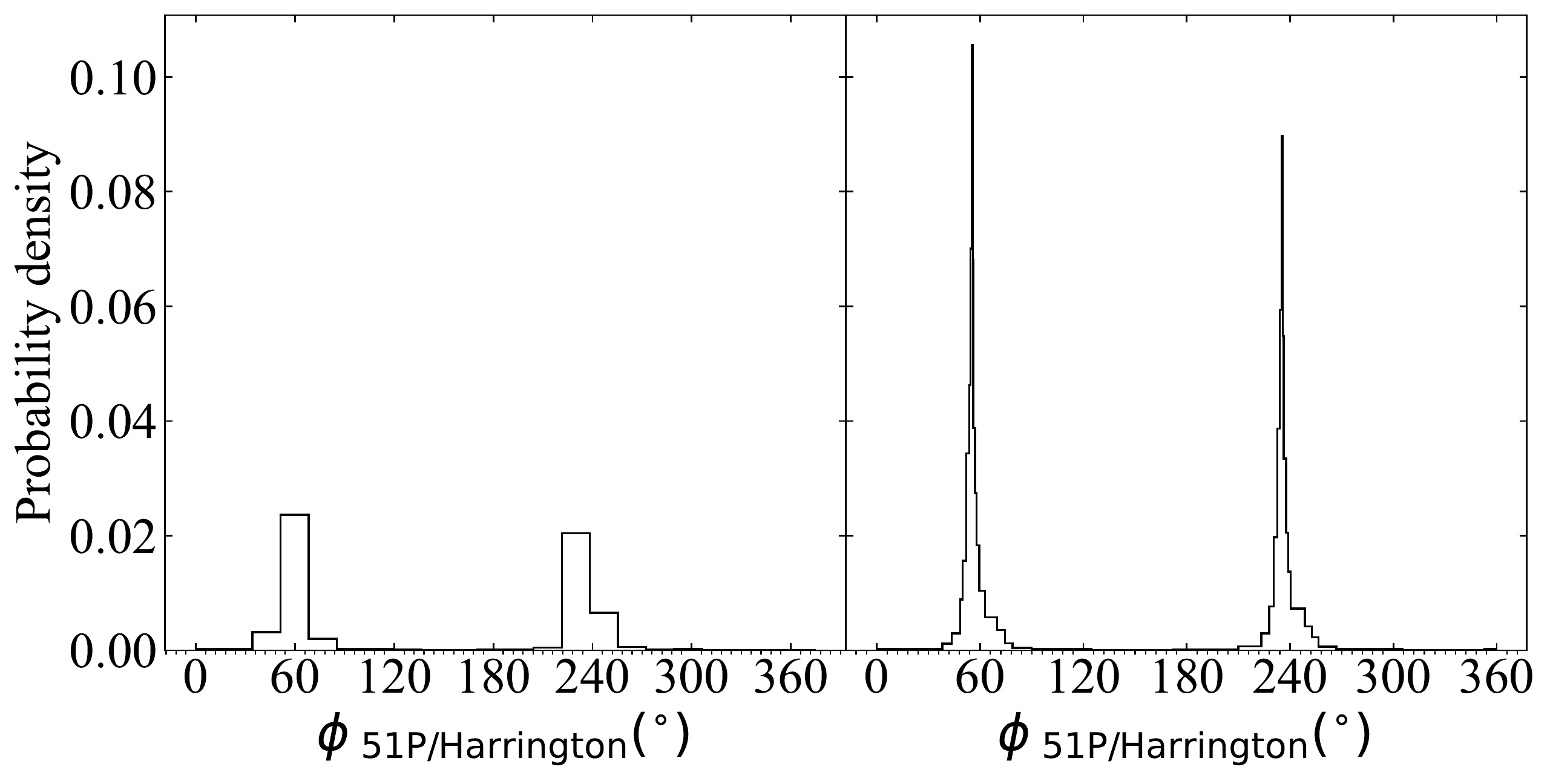}
         \caption{Distributions of equatorial obliquity and cometocentric longitude of the Sun at perihelion for 51P/Harrington. The top 
                  left panel shows the histogram of equatorial obliquity, $\epsilon_{\rm \ 51P/Harrington}$, with bins computed using the 
                  Freedman and Diaconis rule, while the top right panel uses the Bayesian Blocks technique. Similarly, the cometocentric 
                  longitude, $\phi_{\rm \ 51P/Harrington}$, histograms displayed in the lower panels also use the Freedman and Diaconis rule 
                  (left) and Bayesian Blocks (right). Histograms are based on data from Table~\ref{elements51P} (see the text for details). 
                 }
         \label{51Pspin}
      \end{figure}
%
%

         Applying the procedure outlined in Sect.~\ref{Data} using as input data those in Table~\ref{elements51P} and the associated 
         covariance matrix, we generated $10^{4}$ instances of a given orbit using the set of equations~(\ref{good}). The results for the 
         spin-axis orientation $(\epsilon_{\rm \ 51P/Harrington}, \phi_{\rm \ 51P/Harrington})$ are shown in Fig.~\ref{51Pspin}. The top 
         panels display the distribution of the equatorial obliquity that is unimodal with median and 16th and 84th percentiles of 
         82\degr$_{-2\degr}^{+6\degr}$; the bottom panels show a bimodal distribution for the cometocentric longitude with average maxima
         at about 55{\degr} and 235{\degr} (therefore, separated by 180{\degr}). 

         These results indicate that the effective equatorial plane of 51P/Harrington during its most recent perihelion was nearly 
         perpendicular to its orbital plane; on the other hand, if $\phi_{\rm \ 51P/Harrington}$$\sim$55{\degr} the northern pole of the 
         comet was facing the Sun at perihelion, or the southern one if $\phi_{\rm \ 51P/Harrington}$$\sim$235{\degr}. Our result for the 
         value of the equatorial obliquity is fully consistent with that in \citet{1996AcA....46...29S}, in particular with the trend shown 
         in his fig.~1. The value of $\phi_{\rm \ 51P/Harrington}$ in Fig.~\ref{51Pspin}, bottom panels, is however inconsistent with the 
         one in \citet{1996AcA....46...29S} and the trend in his fig.~2; the median and 16th and 84th percentiles of 
         $\phi_{\rm \ 51P/Harrington}$ are 93{\degr}$_{-38\degr}^{+145\degr}$. It can be argued that the differences in 
         $\phi_{\rm \ 51P/Harrington}$ could be the result of the fragmentation events pointed out above. 
%
%
      \begin{figure}
        \centering
         \includegraphics[width=0.99\linewidth]{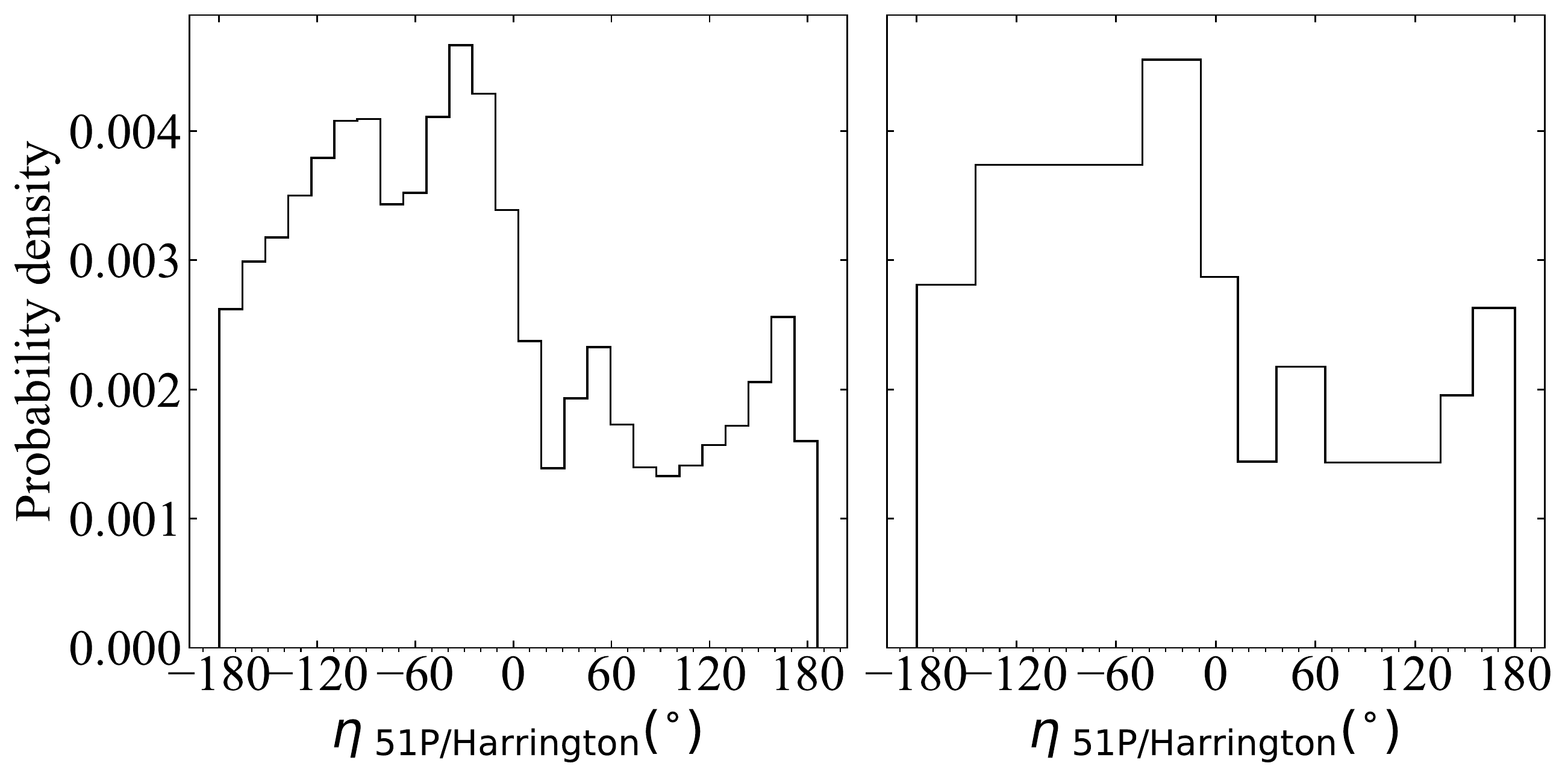}
         \caption{Distribution of thermal lag angle for 51P/Harrington. The left panel shows the histogram with bins computed using the 
                  Freedman and Diaconis rule, while the right panel uses the Bayesian Blocks technique. Histograms are based on data from 
                  Table~\ref{elements51P}.
                 }
         \label{51Plag}
      \end{figure}
%
%

         Figure~\ref{51Plag} shows an asymmetric distribution for $\eta_{\rm \ 51P/Harrington}$, with no clear maximum. The median and 16th 
         and 84th percentiles of the distribution are $-38{\degr}_{-89{\degr}}^{+135{\degr}}$. This implies that the maximum outgassing was 
         taking place when the Sun was rising as seen from the surface of the comet. Again, this result is inconsistent with the one in
         \citet{1996AcA....46...29S}, but it might be a by-product of the fragmentation events pointed out above if the fresh surface has a
         different value of the thermal inertia. 
%
%
      \begin{figure}
        \centering
         \includegraphics[width=0.99\linewidth]{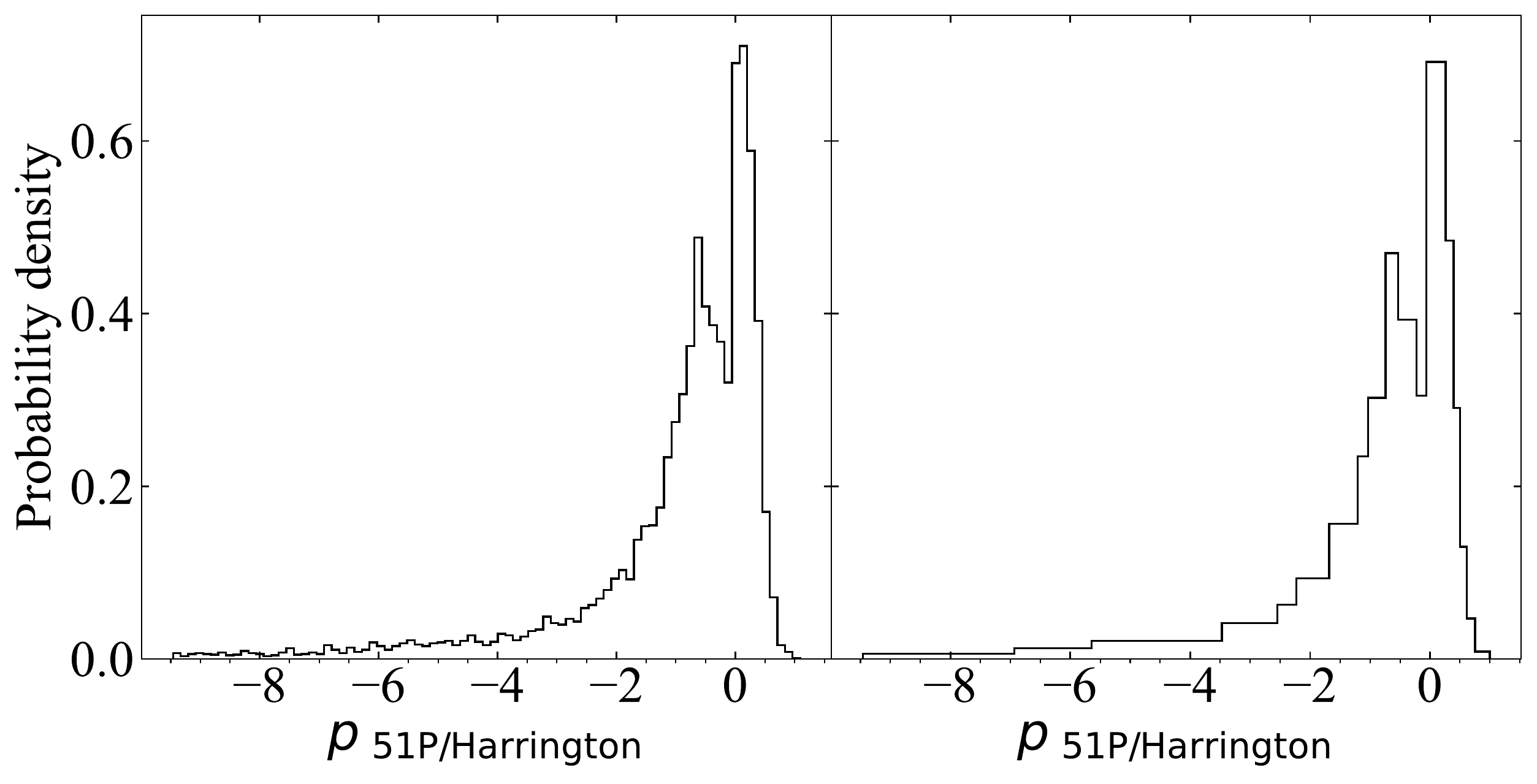}
         \caption{Distribution of values of the oblateness of the nucleus for 51P/Harrington. The left panel shows the histogram with bins 
                  computed using the Freedman and Diaconis rule, while the right panel uses the Bayesian Blocks technique. Histograms are 
                  based on data from Table~\ref{elements51P}.
                 }
         \label{51Pobla}
      \end{figure}
%
%

         Figure~\ref{51Pobla} shows a bimodal distribution for the oblateness of the nucleus of 51P/Harrington. The median and 16th and 84th 
         percentiles of the distribution are $-0.50_{-1.5}^{+0.7}$, with maxima at about 0.06 and $-$0.6. Our results favor an oblate 
         nucleus although a prolate one cannot be discarded. Considering the large dispersion, our median value is statistically consistent
         with the results in \citet{1996AcA....46...29S} and our secondary maximum is marginally consistent with his result. Again, the 
         fact that the surface of the comet has experienced multiple alterations since the data used by \citet{1996AcA....46...29S} were
         acquired and that some of these episodes may have even altered the shape of 51P/Harrington make it difficult to validate our 
         results against those in \citet{1996AcA....46...29S}. This is particularly obvious when we consider that the value of $A$ in 
         Table~\ref{elements51P} is nearly five times larger than the one computed by \citet{1996AcA....46...29S}. In addition, the model
         used to derive the values of the parameters in Table~\ref{elements51P} considered symmetric ---not asymmetric as in 
         \citet{1996AcA....46...29S}--- cometary activity with respect to perihelion.

         Taking into account the dot product of the unit vector in the direction of the rotation pole with that of the orbital pole and its 
         relationship with $\epsilon_{\rm \ 51P/Harrington}$ as pointed out above, we have computed the distributions of $(\lambda_{\rm p}, 
         \beta_{\rm p})$ and $(\alpha_{\rm p}, \delta_{\rm p})$. Figure~\ref{51Ppole} shows the resulting distributions in ecliptic 
         coordinates and Fig.~\ref{51Ppolecoor} shows the maps of statistical significance computed as described in the second to last 
         paragraph of sect.~2.2. The most probable spin-axis direction of 51P/Harrington in equatorial coordinates could be 
         $(260{\degr},~-13{\degr})$ ---or $(120{\degr},~+20{\degr})$. On the other hand, the equatorial coordinates of the Sun when the comet 
         reached perihelion on August 12, 2015 were approximately $(141\fdg8,~+15\fdg0)$, somewhat consistent with the orientation of one of 
         the poles.   
%
%
      \begin{figure}
        \centering
         \includegraphics[width=0.99\linewidth]{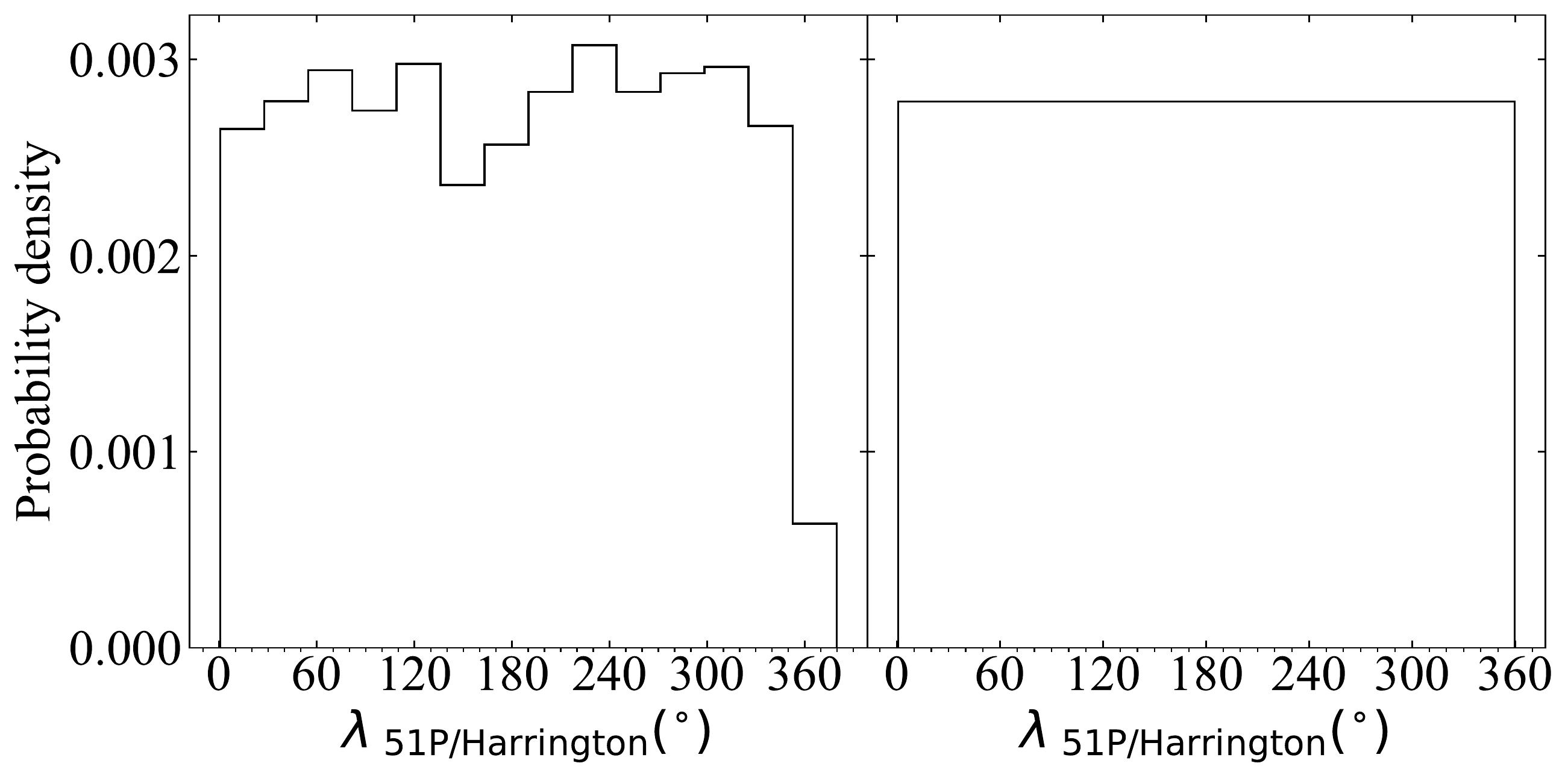}
         \includegraphics[width=0.99\linewidth]{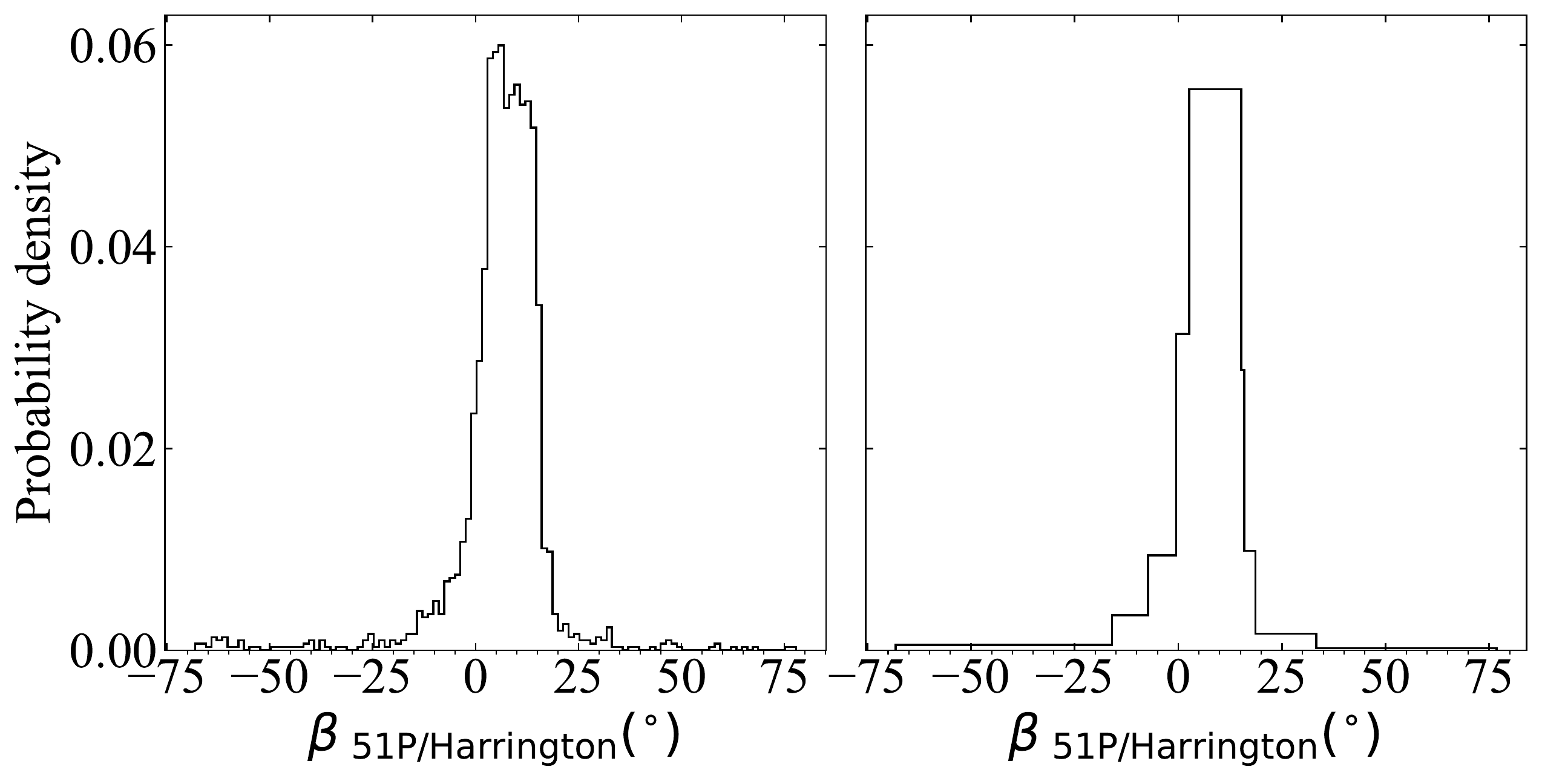}
         \caption{Distributions of spin-axis orientations, $(\lambda_{\rm p}, \beta_{\rm p})$, for 51P/Harrington. The left panel shows the 
                  histogram with bins computed using the Freedman and Diaconis rule, while the right panel uses the Bayesian Blocks 
                  technique. Histograms are based on data from Table~\ref{elements51P}.
                 }
         \label{51Ppole}
      \end{figure}
%
%
%
%
      \begin{figure}
        \centering
         \includegraphics[width=\linewidth]{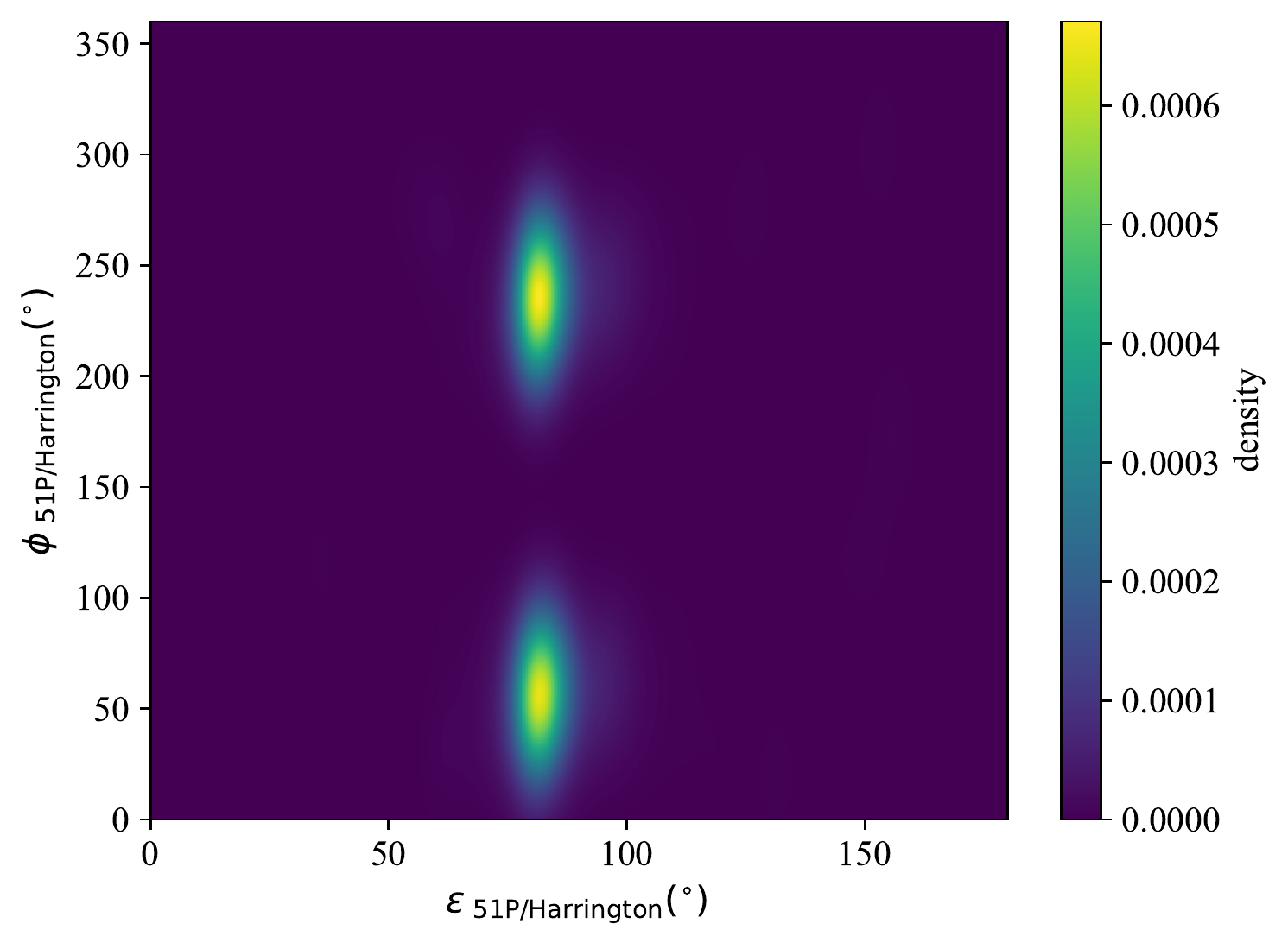}
         \includegraphics[width=\linewidth]{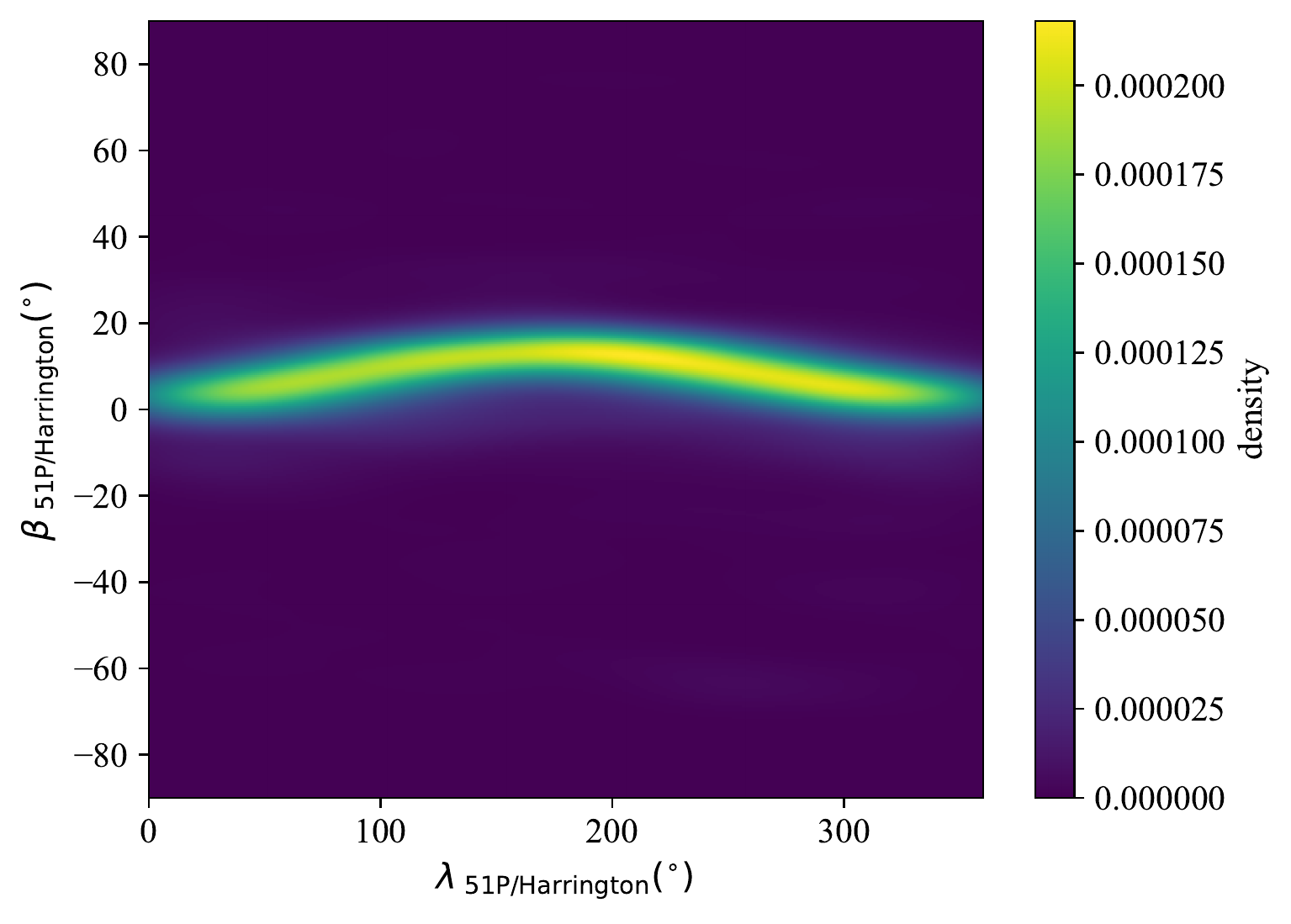}
         \includegraphics[width=\linewidth]{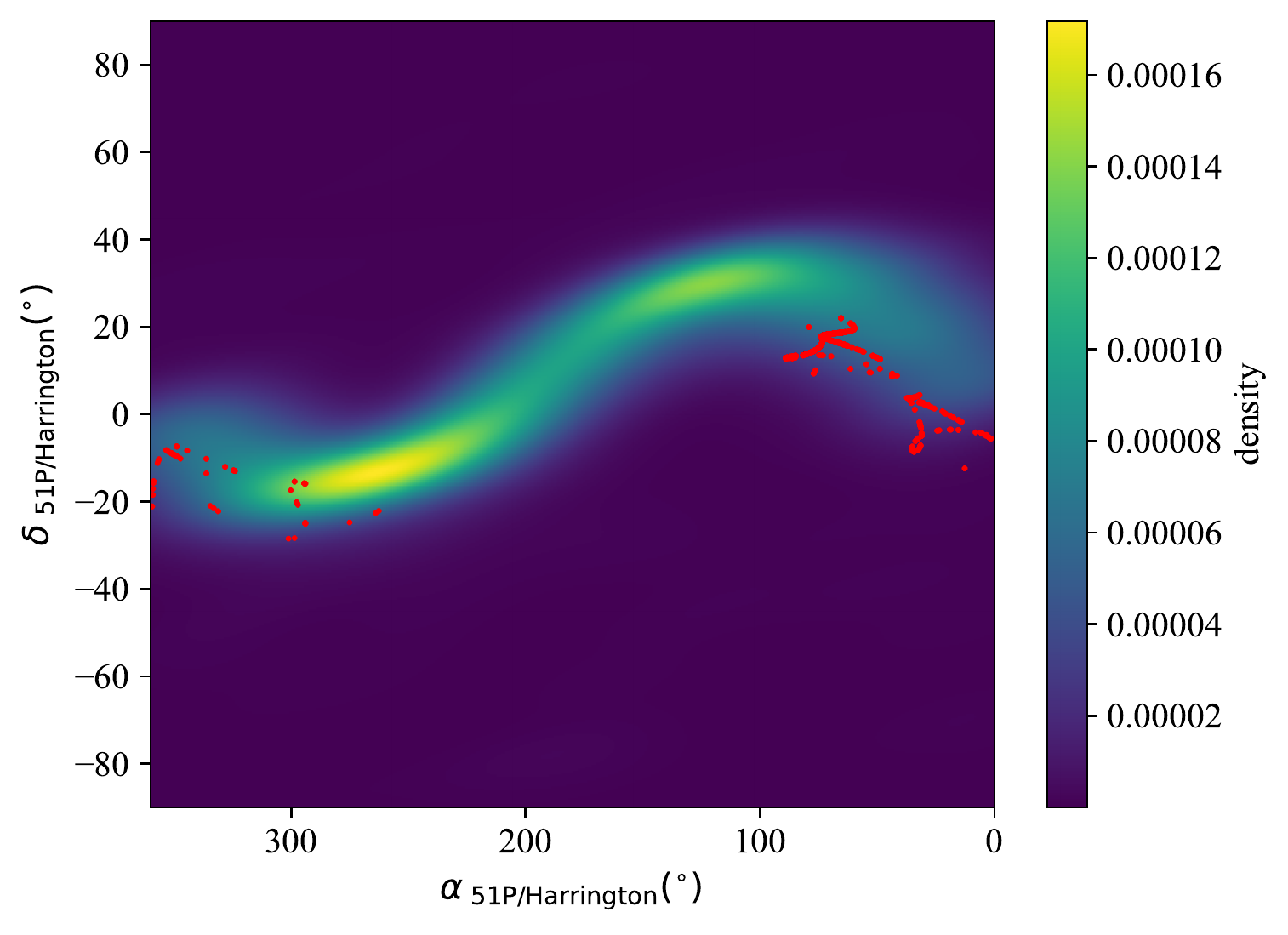}
         \caption{Gaussian kernel density estimation of the spin-axis orientations computed using the SciPy library 
                  \citep{2020NatMe..17..261V}. The top panel shows the $(\epsilon_{\rm c}, \phi_{\rm c})$ map. The results in ecliptic 
                  coordinates, $(\lambda_{\rm p}, \beta_{\rm p})$, are shown in the middle panel and those in equatorial coordinates,
                  $(\alpha_{\rm p}, \delta_{\rm p})$, are displayed in the bottom panel that includes the actual observations (in red) of
                  comet 51P/Harrington available from the MPC.
                 }
         \label{51Ppolecoor}
      \end{figure}
%
%

   \section{Interstellar object 1I/2017~U1~(`Oumuamua)\label{1I}}
      We applied the procedure outlined in Sect.~\ref{Data} to produce the distributions of $\epsilon_{\rm c}$, $\phi_{\rm c}$, 
      $\eta_{\rm c}$, and $p_{\rm c}$ for interstellar object 1I/2017~U1~(`Oumuamua) considering the orbit determination in 
      Table~\ref{elements1I} that is discussed by \citet{2018Natur.559..223M}. The model favored by these authors assumes that `Oumuamua's 
      comet-like outgassing is driven by CO, not H$_{2}$O (see Sect.~\ref{Intro} and the discussion in Sect.~\ref{Data}). The results for 
      the spin-axis orientation $(\epsilon_{\rm \ `Oumuamua}, \phi_{\rm \ `Oumuamua})$ are displayed in Fig.~\ref{oumuamuaSPIN}, which 
      summarizes the results of $10^{5}$ Monte Carlo experiments. 

      Figure~\ref{oumuamuaSPIN}, top panels, shows the distribution of the equatorial obliquity. The median and 16th and 84th percentiles of 
      the distribution are 93\degr$_{-13\degr}^{+8\degr}$. This result indicates that the effective equatorial plane of `Oumuamua was nearly 
      perpendicular to its orbital plane. Figure~\ref{oumuamuaSPIN}, bottom panels, shows a bimodal distribution with average maxima at 
      about 140{\degr} and 320{\degr} (therefore, separated by 180{\degr}). If $\phi_{\rm \ `Oumuamua}$$\sim$140{\degr} the northern pole of 
      `Oumuamua was facing the Sun at perihelion, if $\phi_{\rm \ `Oumuamua}$$\sim$320{\degr} was the southern one. In general, our 
      methodology cannot always distinguish which pole was facing the Sun when `Oumuamua reached perihelion.   
%
%
      \begin{figure}
        \centering
         \includegraphics[width=0.99\linewidth]{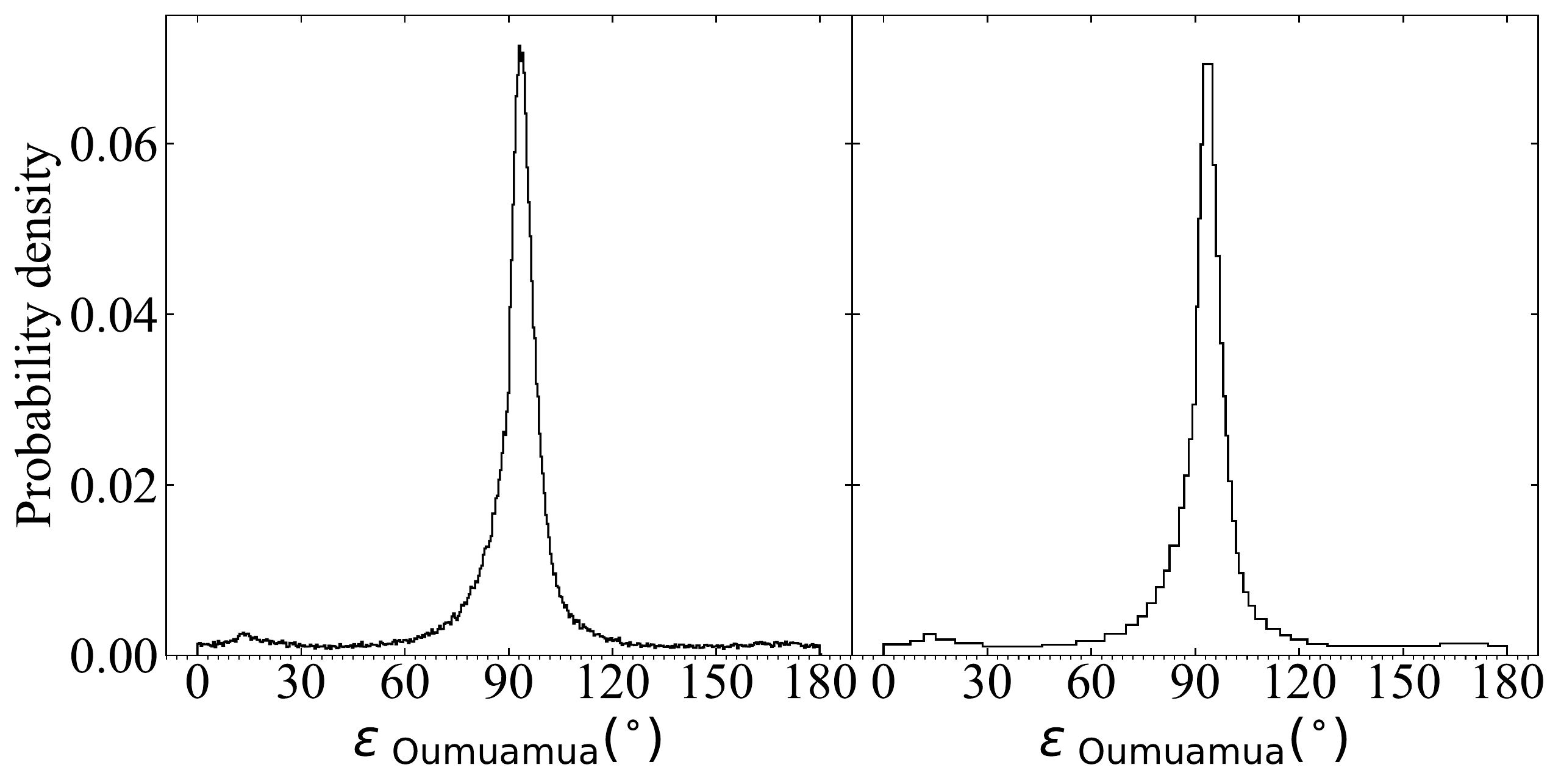}
         \includegraphics[width=0.99\linewidth]{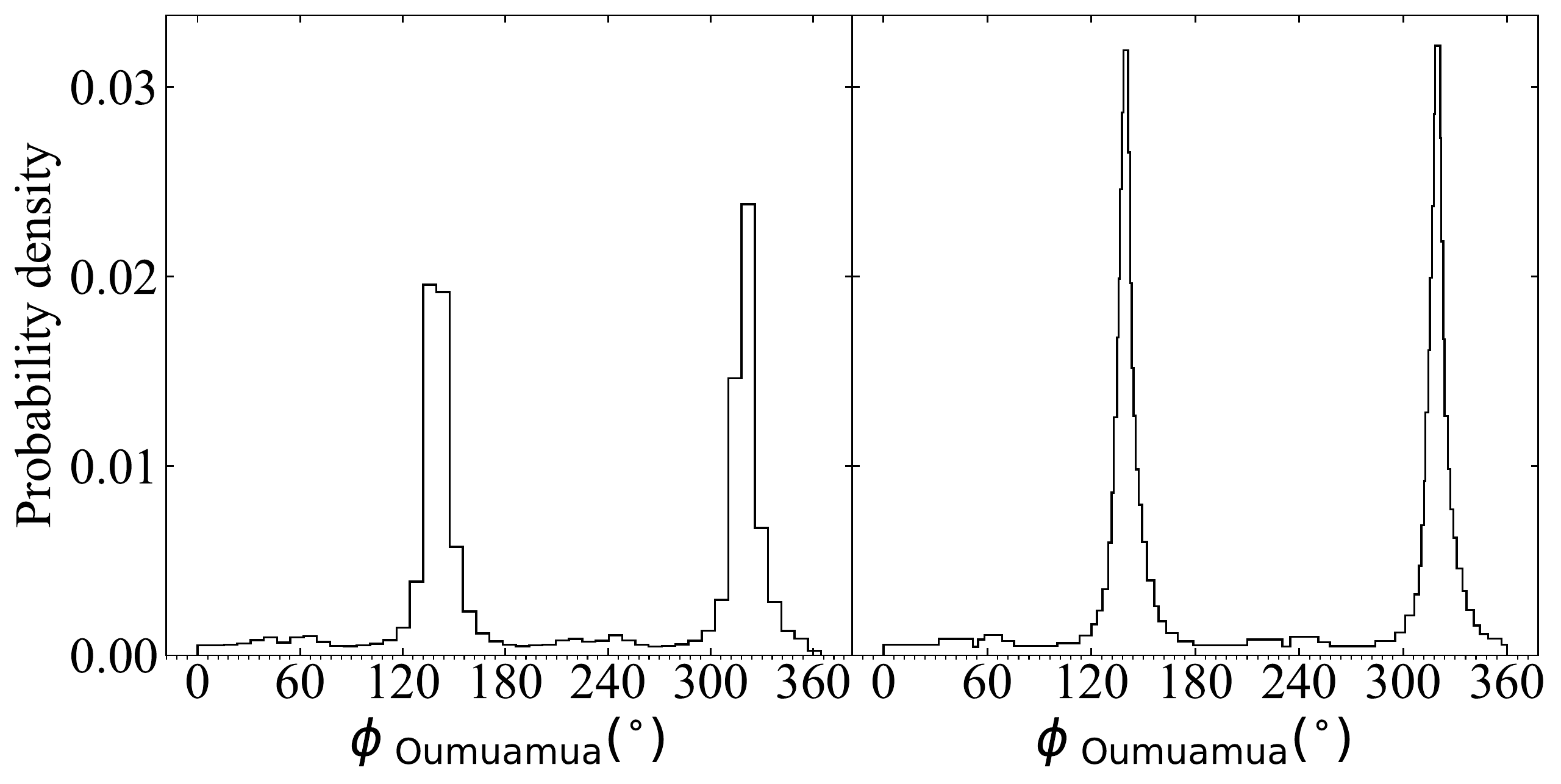}
         \caption{Distributions of equatorial obliquity and cometocentric longitude of the Sun at perihelion for 1I/2017~U1~(`Oumuamua).
                  The top left panel shows the histogram of equatorial obliquity, $\epsilon_{\rm \ `Oumuamua}$, with bins computed using the 
                  Freedman and Diaconis rule, while the top right panel uses the Bayesian Blocks technique. Similarly, the cometocentric 
                  longitude, $\phi_{\rm \ `Oumuamua}$, histograms displayed in the lower panels also use the Freedman and Diaconis rule 
                  (left) and Bayesian Blocks (right). Histograms are based on data from Table~\ref{elements1I} (see the text for details). 
                 }
         \label{oumuamuaSPIN}
      \end{figure}
%
%

      Figure~\ref{oumuamuaLAG} also shows a unimodal distribution for $\eta_{\rm \ `Oumuamua}$. The median and 16th and 84th percentiles of 
      the distribution are $0.08{\degr}_{-55{\degr}}^{+52{\degr}}$. This implies that the maximum outgassing was taking place mainly at 
      `Oumuamua's noon, when the Sun was at its highest as seen from the surface of the interstellar visitor.
%
%
      \begin{figure}
        \centering
         \includegraphics[width=0.99\linewidth]{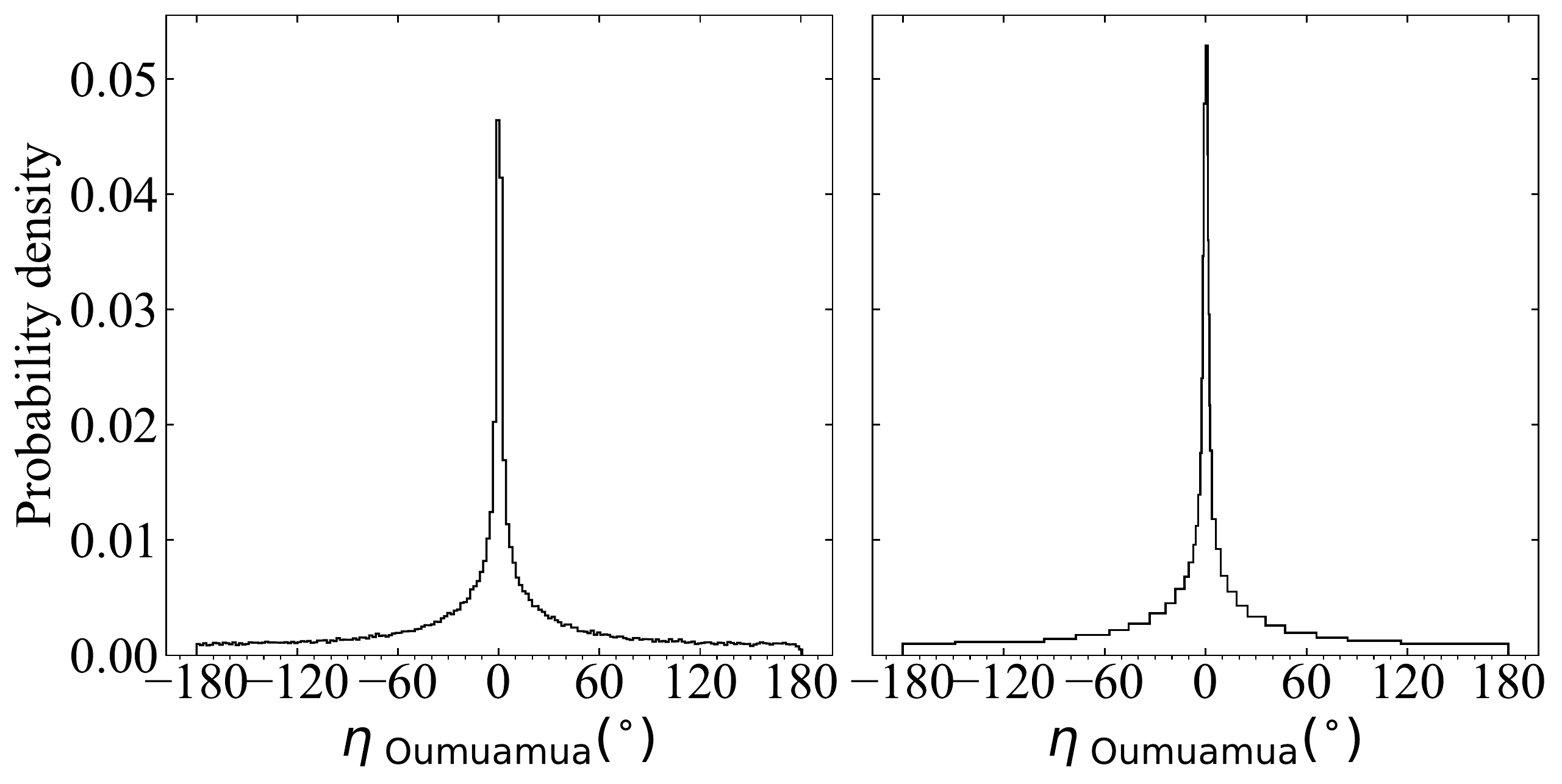}
         \caption{Distribution of thermal lag angle for 1I/2017~U1~(`Oumuamua).
                  The left panel shows the histogram with bins computed using the Freedman and Diaconis rule, while the right panel uses the 
                  Bayesian Blocks technique. Histograms are based on data from Table~\ref{elements1I}.
                 }
         \label{oumuamuaLAG}
      \end{figure}
%
%

      Figure~\ref{oumuamuaOBLA} shows a unimodal distribution for the oblateness of `Oumuamua with median and 16th and 84th percentiles of 
      $-0.13_{-1.56}^{+0.22}$. This result slightly favors a fusiform shape for `Oumuamua over a disk-like one. The actual shape of 
      `Oumuamua remains a controversial subject (see for example the discussion in \citealt{2020MNRAS.493.1546V}). Although multiple studies 
      have pointed out that the available observational data strongly favor that `Oumuamua has a very elongated shape (see for example 
      \citealt{2017ApJ...851L..31K,2018ApJ...852L...2B,2018NatAs...2..407D,2018NatAs...2..383F}), \citet{2018ApJ...856L..21B} stated that 
      its shape might be fusiform, but a disk-like appearance could not be ruled out. 

      \citet{2019MNRAS.489.3003M} presented a very detailed modeling of the light curve of `Oumuamua to show that his best-fitting model, 
      with a probability of 91\%, was a thin disk with $p$=0.836; his second choice, with a probability of 16\%, was a thin spindle with 
      $p$=$-$6.7. The thin-disk model was significantly more successful at reproducing the observed variations in apparent magnitude over 
      time. However, the results presented by \citet{2019MNRAS.489.3003M} have been contested by \citet{2020NatAs...4..852Z} by arguing that 
      multiple lines of evidence favor the very prolate or cigar-like shape against the very oblate or disk-like shape. In particular, they
      pointed out that the fusiform case is energetically more stable. In the following, we test the impact of both models on the values of 
      the rotational parameters of `Oumuamua. We first assume that $p$=$-$6.7 together with the data in Table~\ref{elements1I}, then we 
      consider $p$=0.836.   
%
%
      \begin{figure}
        \centering
         \includegraphics[width=0.99\linewidth]{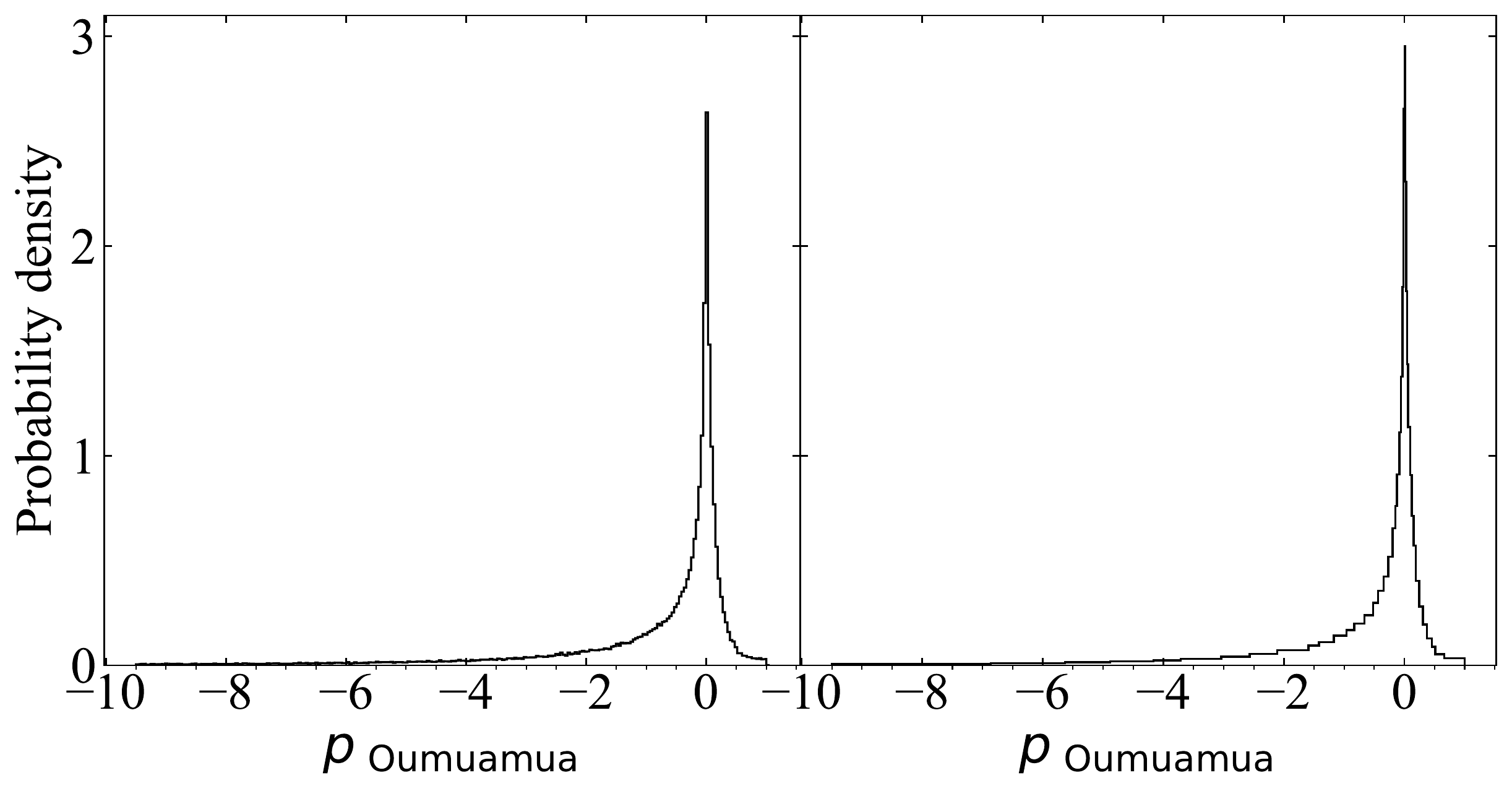}
         \caption{Distribution of values of the oblateness of interstellar object 1I/2017~U1~(`Oumuamua). The left panel shows the histogram 
                  with bins computed using the Freedman and Diaconis rule, while the right panel uses the Bayesian Blocks technique. 
                  Histograms are based on data from Table~\ref{elements1I}.
                 }
         \label{oumuamuaOBLA}
      \end{figure}
%
%

      Figures~\ref{oumuamuaSPINcigar} and \ref{oumuamuaLAGcigar}, show the distributions of $\epsilon_{\rm \ `Oumuamua}, 
      \phi_{\rm \ `Oumuamua}$ and $\eta_{\rm \ `Oumuamua}$ under the assumption of $p$=$-$6.7; we generated $10^{4}$ instances of a given 
      orbit. The results for the orientation of the pole are similar to those in Fig.~\ref{oumuamuaSPIN}: The equatorial obliquity becomes 
      93{\degr}$\pm$3{\degr} and $\phi_{\rm \ `Oumuamua}$$\sim$140{\degr} or $\sim$320{\degr}. However, Fig.~\ref{oumuamuaLAGcigar} shows a 
      relatively flat distribution for $\eta_{\rm \ `Oumuamua}$; in other words, outgassing takes place evenly throughout a rotation period. 
      The actual location of the pole is studied in Figs.~\ref{1Ipolecigar} and \ref{1Ipolecoorcigar}: Under the thin-spindle scenario, the 
      most probable spin-axis direction of `Oumuamua in equatorial coordinates could be $(280{\degr},~+46{\degr})$, with approximate 
      Galactic coordinates $l=75{\degr}$, $b=+21{\degr}$, that point slightly above the Galactic disk, toward the constellation of Lyra.
%
%
      \begin{figure}
        \centering
         \includegraphics[width=0.99\linewidth]{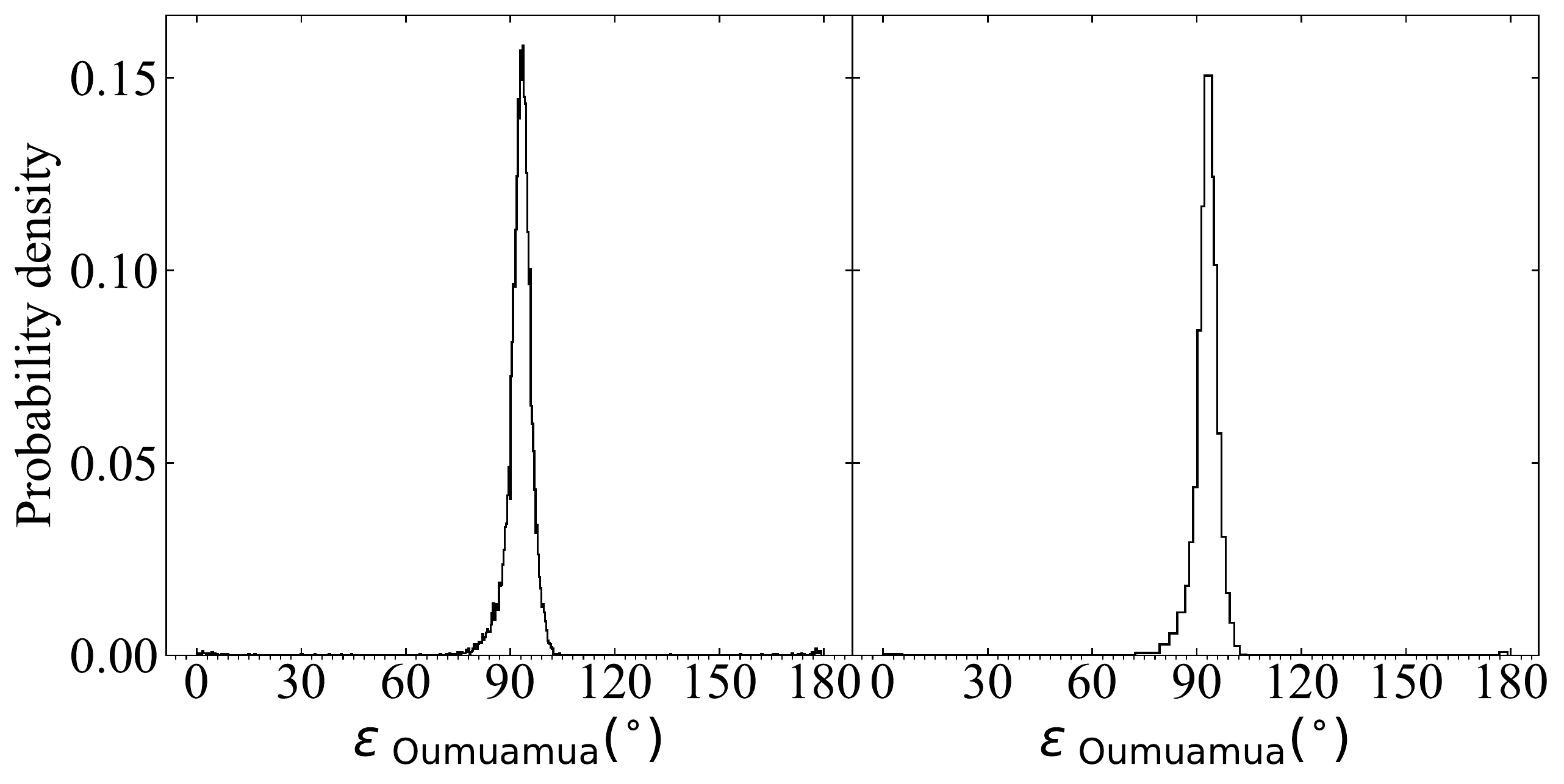}
         \includegraphics[width=0.99\linewidth]{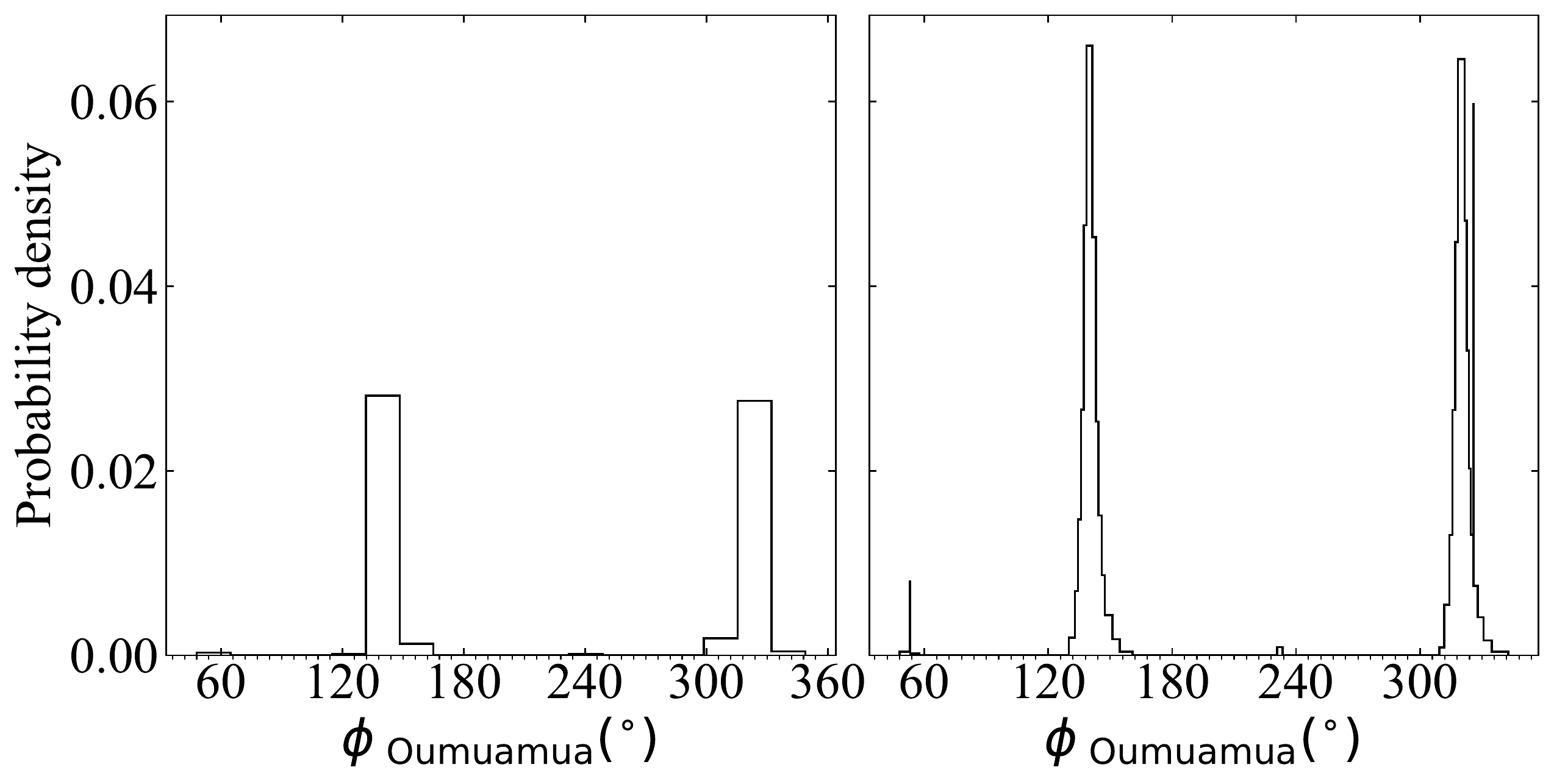}
         \caption{Same as Fig.~\ref{oumuamuaSPIN} but assuming a value for the oblateness of interstellar object 1I/2017~U1~(`Oumuamua) 
                  equal to $-$6.7 as suggested by \citet{2019MNRAS.489.3003M}.
                 }
         \label{oumuamuaSPINcigar}
      \end{figure}
%
%
%
%
      \begin{figure}
        \centering
         \includegraphics[width=0.99\linewidth]{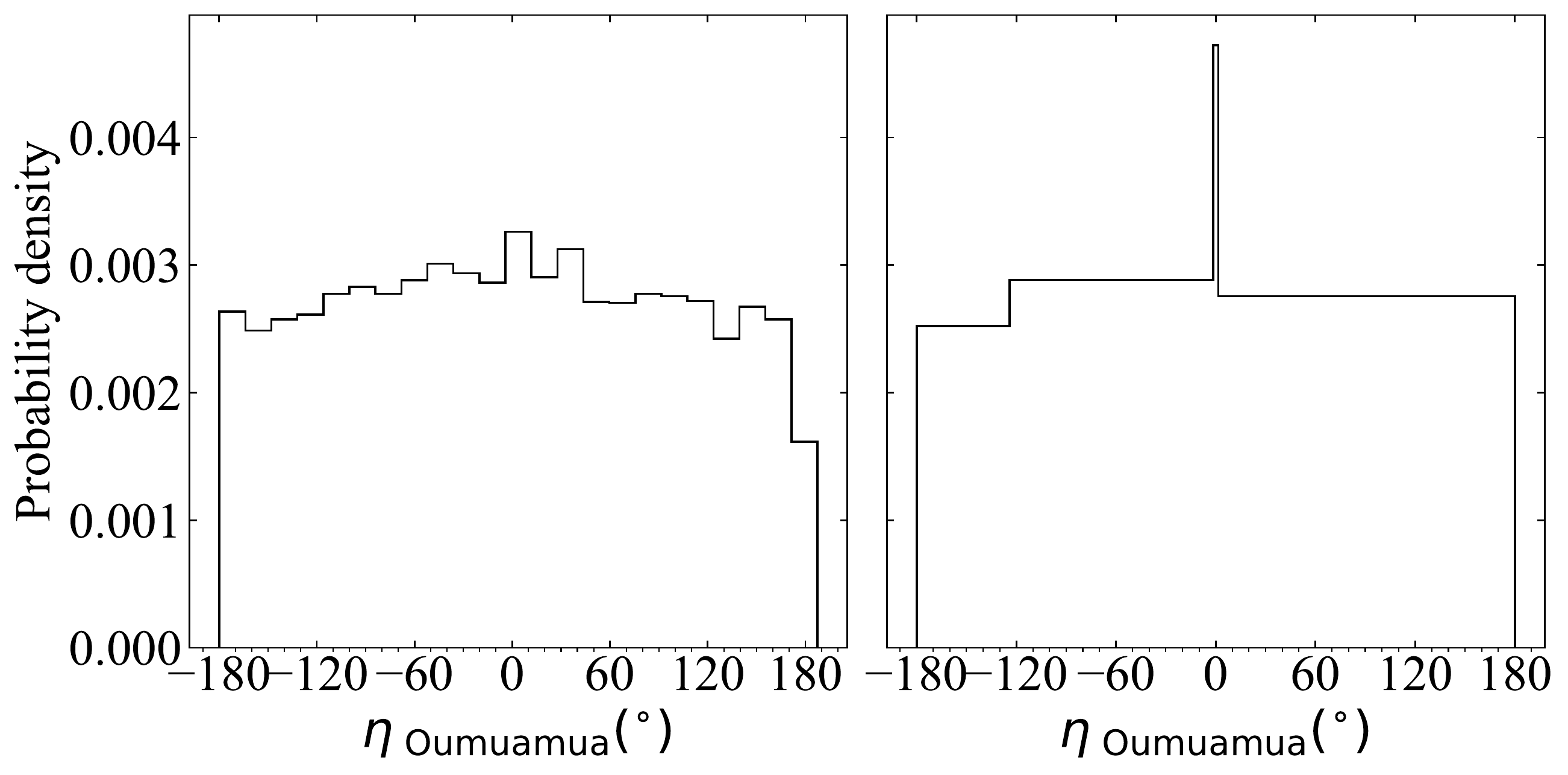}
         \caption{Same as Fig.~\ref{oumuamuaLAG} but assuming a value for the oblateness of interstellar object 1I/2017~U1~(`Oumuamua) equal 
                  to $-$6.7 as suggested by \citet{2019MNRAS.489.3003M}.
                 }
         \label{oumuamuaLAGcigar}
      \end{figure}
%
%
%
%
      \begin{figure}
        \centering
         \includegraphics[width=0.99\linewidth]{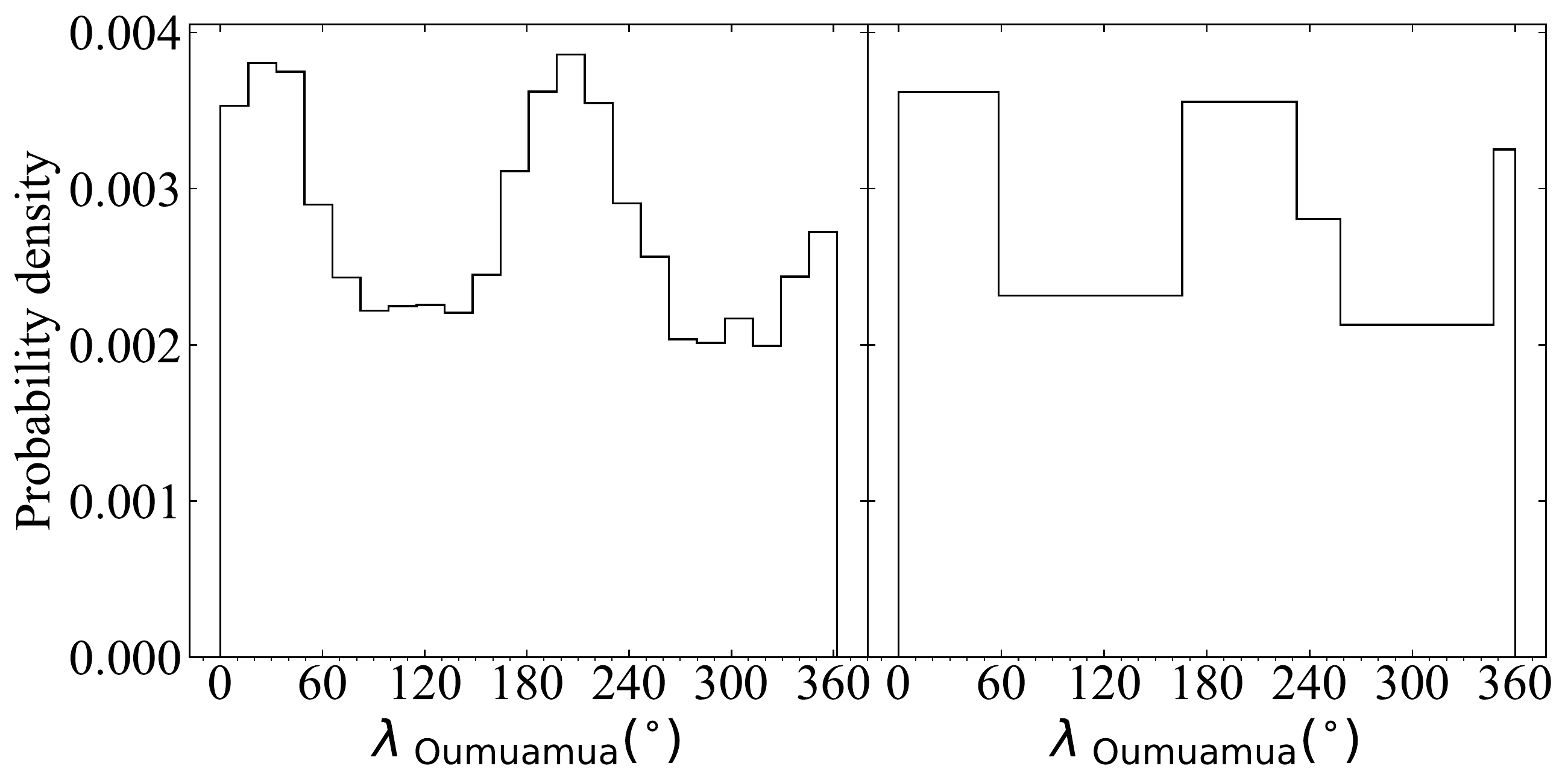}
         \includegraphics[width=0.99\linewidth]{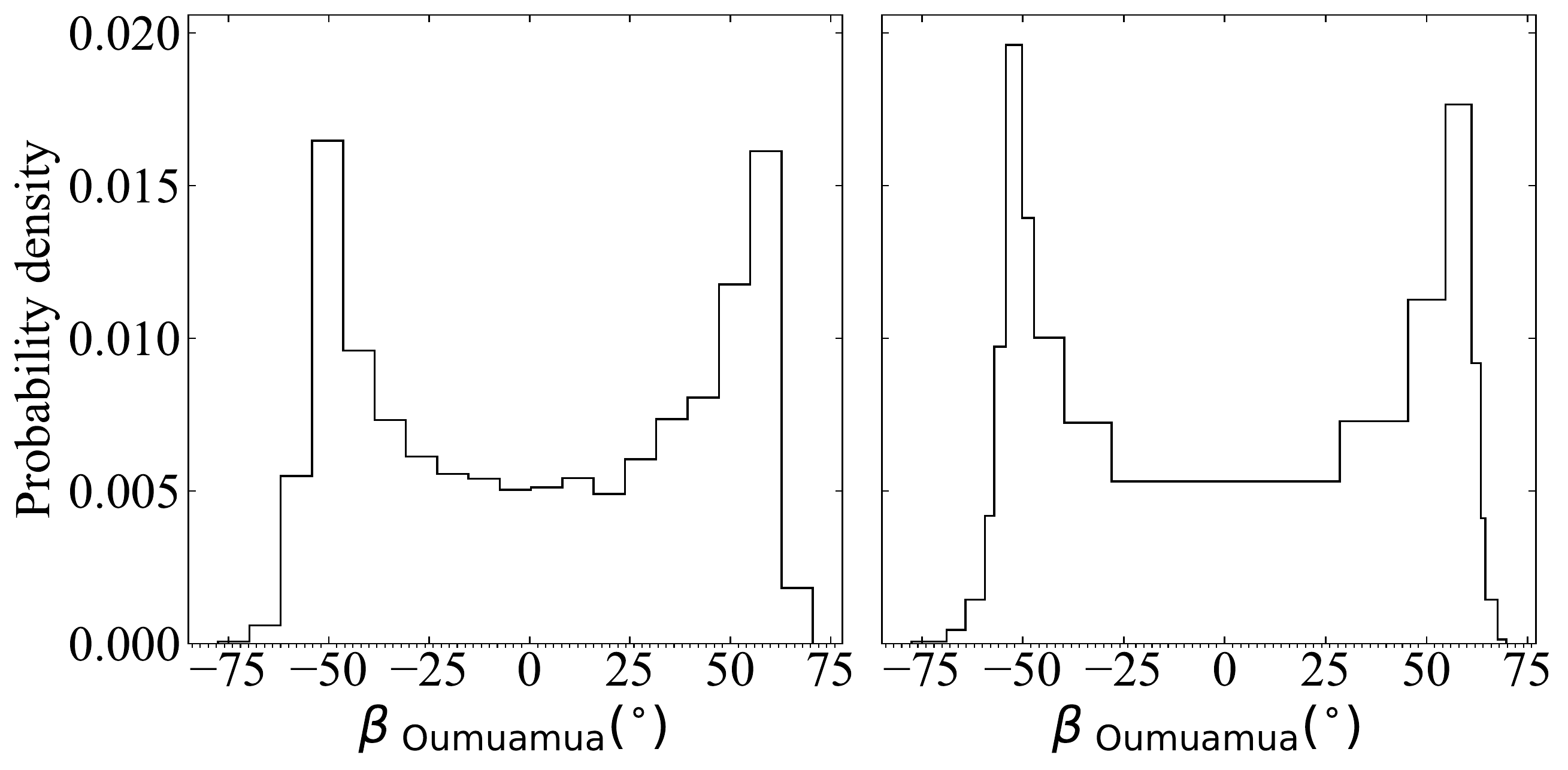}
         \caption{Distributions of spin-axis orientations, $(\lambda_{\rm p}, \beta_{\rm p})$, for 1I/2017~U1~(`Oumuamua) assuming a very
                  prolate shape, with $p=-6.7$. The left panel shows the histogram with bins computed using the Freedman and Diaconis rule, 
                  while the right panel uses the Bayesian Blocks technique. Histograms are based on data from Table~\ref{elements1I}.
                 }
         \label{1Ipolecigar}
      \end{figure}
%
%
%
%
      \begin{figure}
        \centering
         \includegraphics[width=\linewidth]{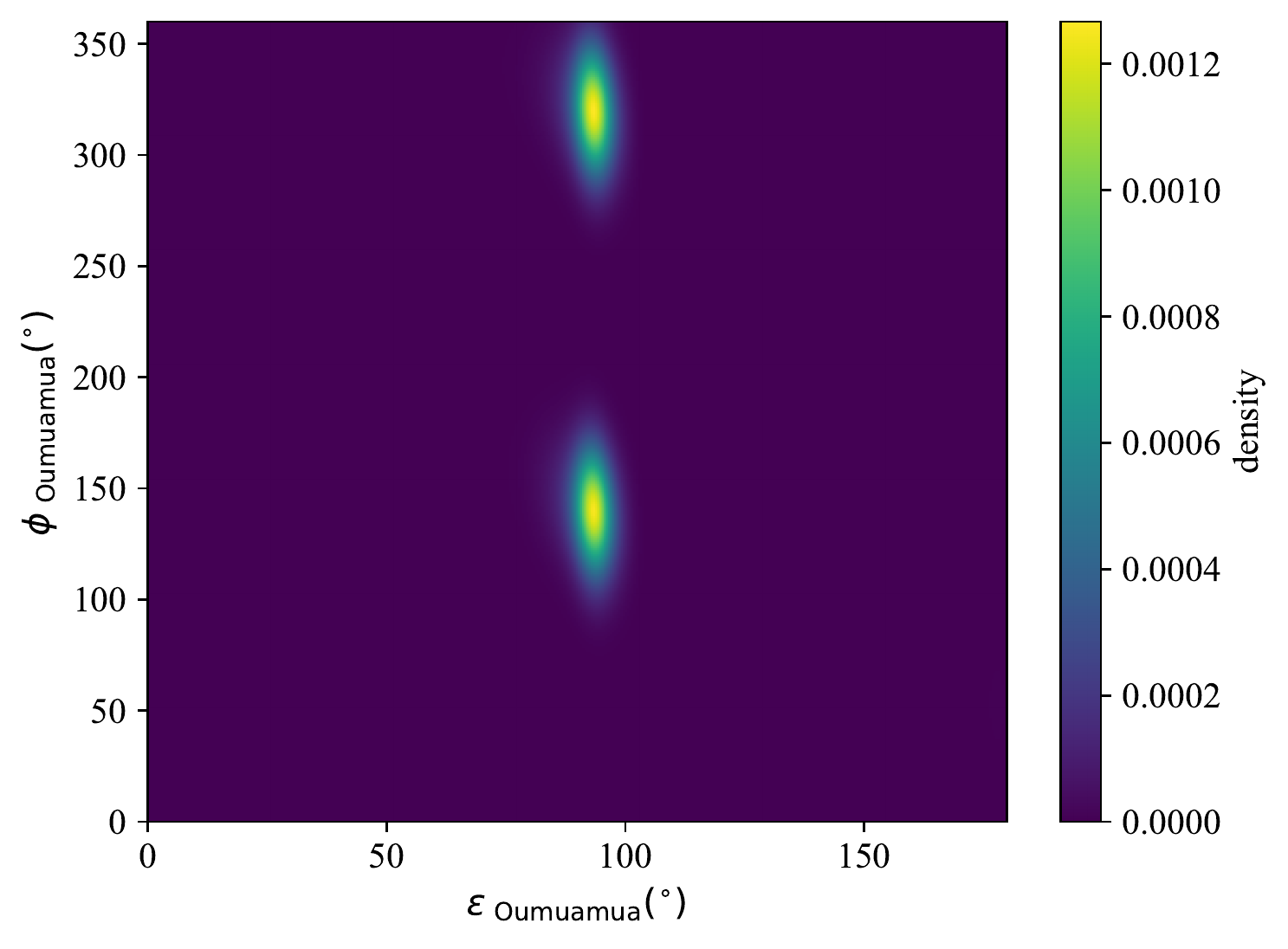}
         \includegraphics[width=\linewidth]{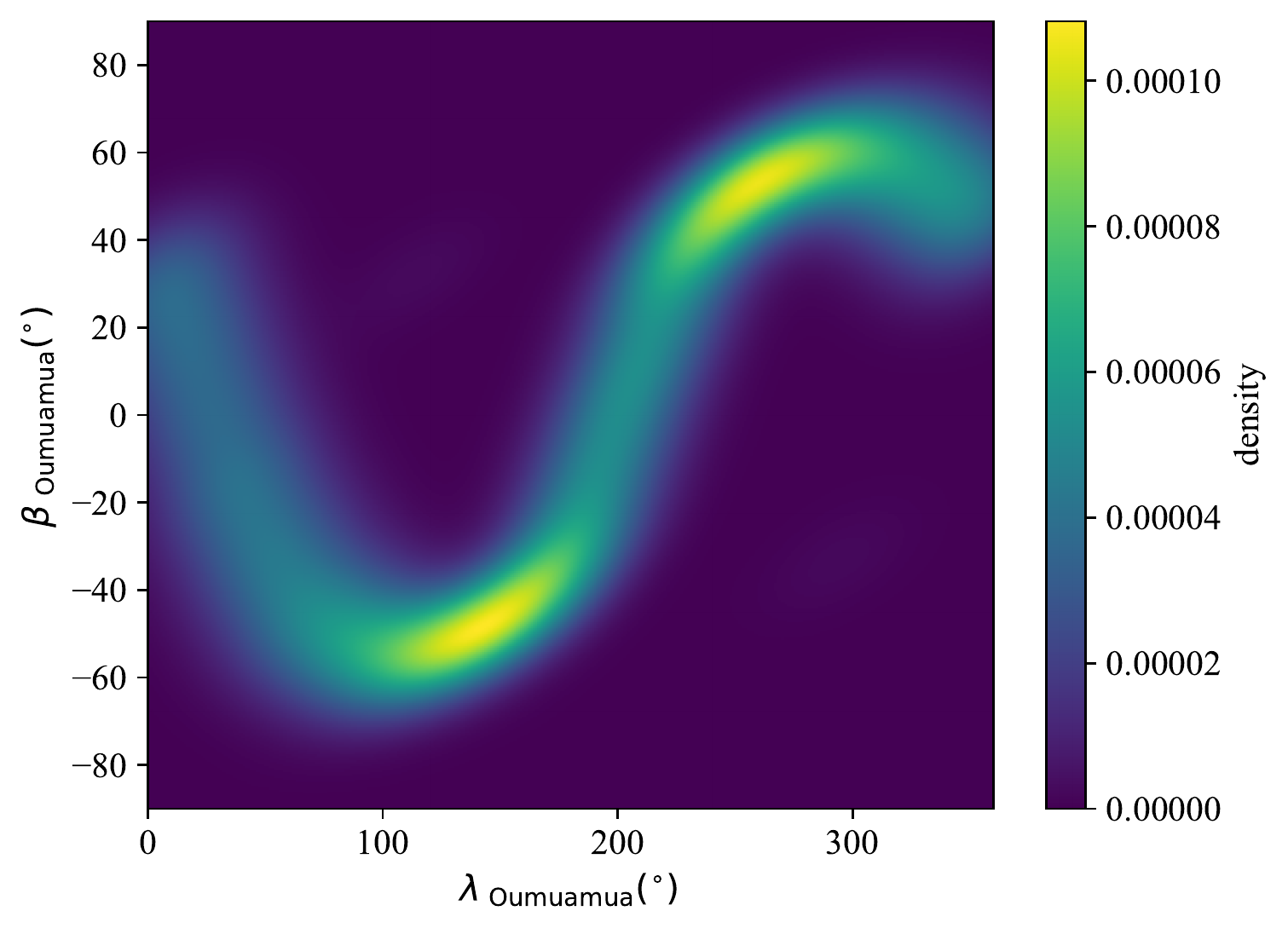}
         \includegraphics[width=\linewidth]{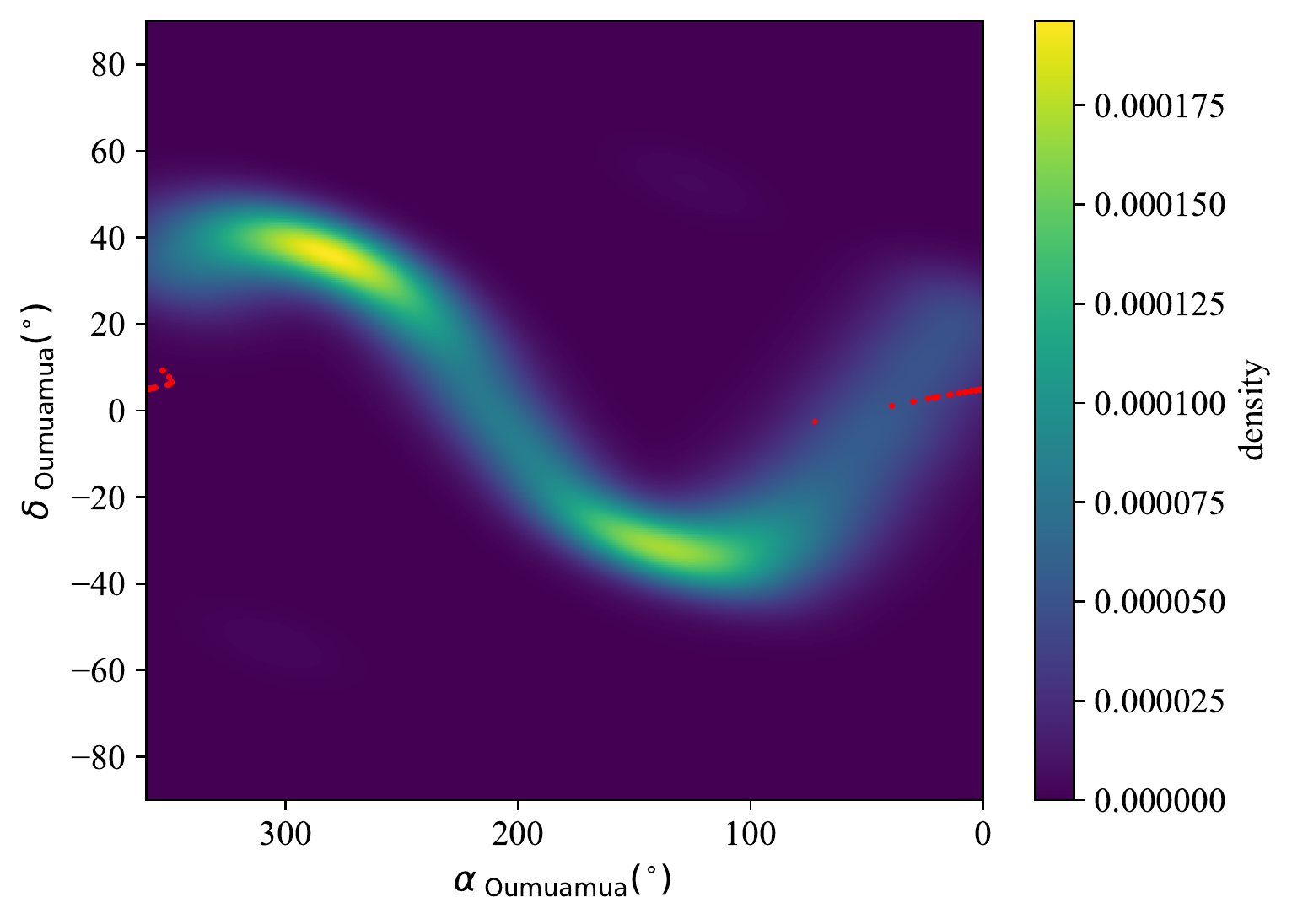}
         \caption{Gaussian kernel density estimation of the spin-axis orientations computed using the SciPy library
                  \citep{2020NatMe..17..261V} for a very prolate case ($p=-6.7$). The top panel shows the $(\epsilon_{\rm c}, \phi_{\rm c})$ 
                  map. The results in ecliptic coordinates, $(\lambda_{\rm p}, \beta_{\rm p})$, are shown in the middle panel and those in 
                  equatorial coordinates, $(\alpha_{\rm p}, \delta_{\rm p})$, are displayed in the bottom panel that includes the actual 
                  observations (in red) of interstellar object 1I/2017~U1~(`Oumuamua) available from the MPC.
                 }
         \label{1Ipolecoorcigar}
      \end{figure}
%
%

      Figures~\ref{oumuamuaSPINdisc} and \ref{oumuamuaLAGdisc}, show the distributions of $\epsilon_{\rm \ `Oumuamua}, 
      \phi_{\rm \ `Oumuamua}$, and $\eta_{\rm \ `Oumuamua}$ for $10^{4}$ instances of the orbit, under the assumption of $p$=0.836. The 
      results for the orientation of the pole are very different from those in Fig.~\ref{oumuamuaSPIN}: The equatorial obliquity is now 
      16{\degr}$_{-7\degr}^{+80\degr}$ and $\phi_{\rm \ `Oumuamua}$$\sim$67{\degr} or $\sim$247{\degr}. However, the distribution of 
      $\eta_{\rm \ `Oumuamua}$ in Fig.~\ref{oumuamuaLAGdisc} resembles that in Fig.~\ref{oumuamuaLAG}; maximum outgassing would take 
      place at `Oumuamua's noon with $\eta_{\rm \ `Oumuamua}$=3{\degr}$\pm$2{\degr}. The location of the pole under the thin-disk scenario 
      is shown in Figs.~\ref{1Ipoledisc} and \ref{1Ipolecoordisc}; the most probable spin-axis direction of `Oumuamua in equatorial 
      coordinates could be $(312{\degr},~-50{\degr})$. This is consistent with the estimate of the orbital pole of `Oumuamua in 
      \citet{2017RNAAS...1....5D,2017RNAAS...1....9D}: Compare Fig.~\ref{1Ipolecoordisc}, middle panel, with the middle panel in fig.~1 of
      \citet{2017RNAAS...1....9D}, both in ecliptic coordinates. The approximate Galactic coordinates of the rotational pole are 
      $l=349{\degr}$, $b=-39{\degr}$, that point slightly below the Galactic bulge, toward the constellation Indus. 
%
%
      \begin{figure}
        \centering
         \includegraphics[width=0.99\linewidth]{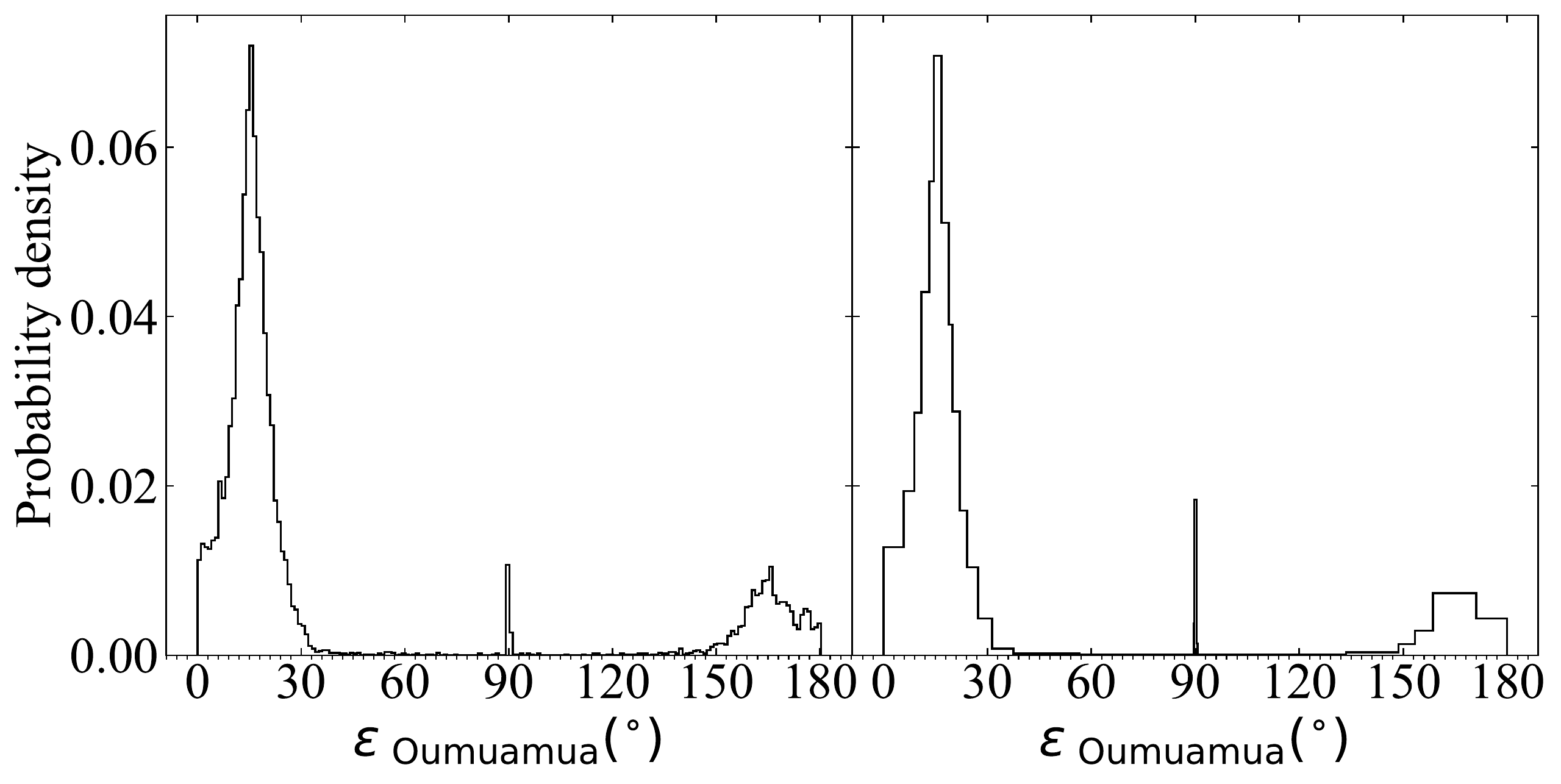}
         \includegraphics[width=0.99\linewidth]{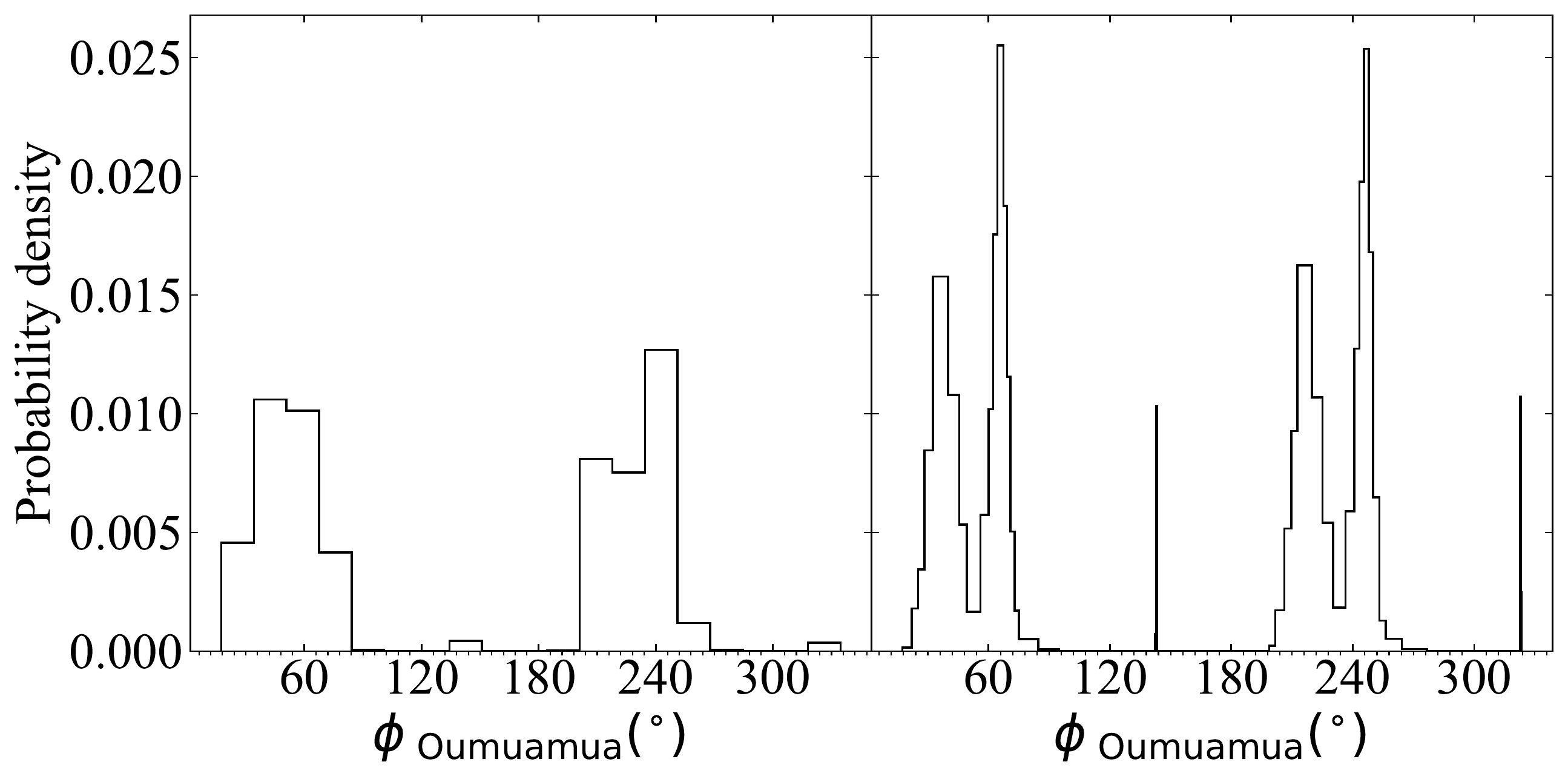}
         \caption{Same as Fig.~\ref{oumuamuaSPIN} but assuming a value for the oblateness of interstellar object 1I/2017~U1~(`Oumuamua) 
                  equal to 0.836 as suggested by \citet{2019MNRAS.489.3003M}.
                 }
         \label{oumuamuaSPINdisc}
      \end{figure}
%
%
%
%
      \begin{figure}
        \centering
         \includegraphics[width=0.99\linewidth]{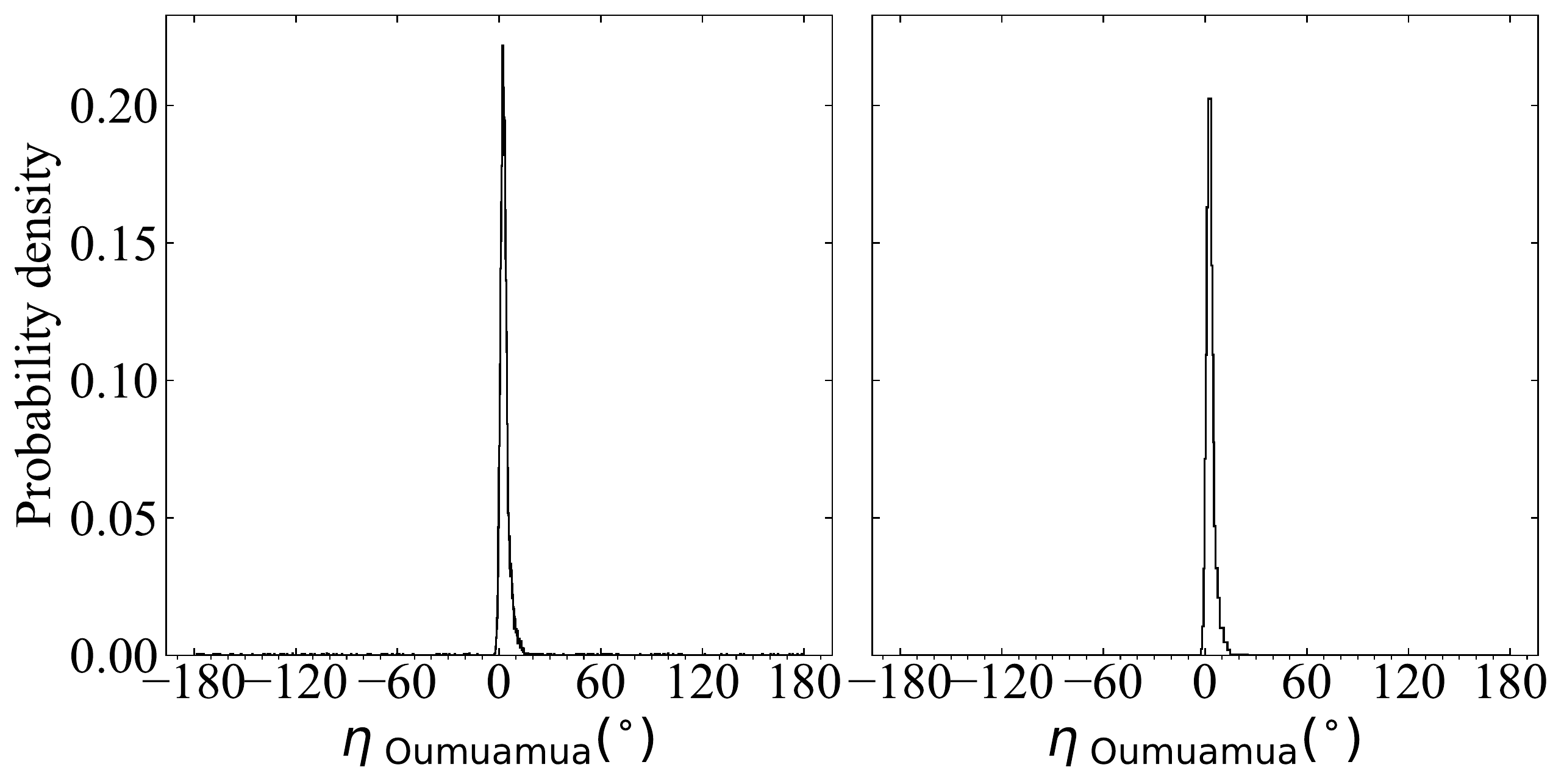}
         \caption{Same as Fig.~\ref{oumuamuaLAG} but assuming a value for the oblateness of interstellar object 1I/2017~U1~(`Oumuamua) equal 
                  to 0.836 as suggested by \citet{2019MNRAS.489.3003M}.
                 }
         \label{oumuamuaLAGdisc}
      \end{figure}
%
%
%
%
      \begin{figure}
        \centering
         \includegraphics[width=0.99\linewidth]{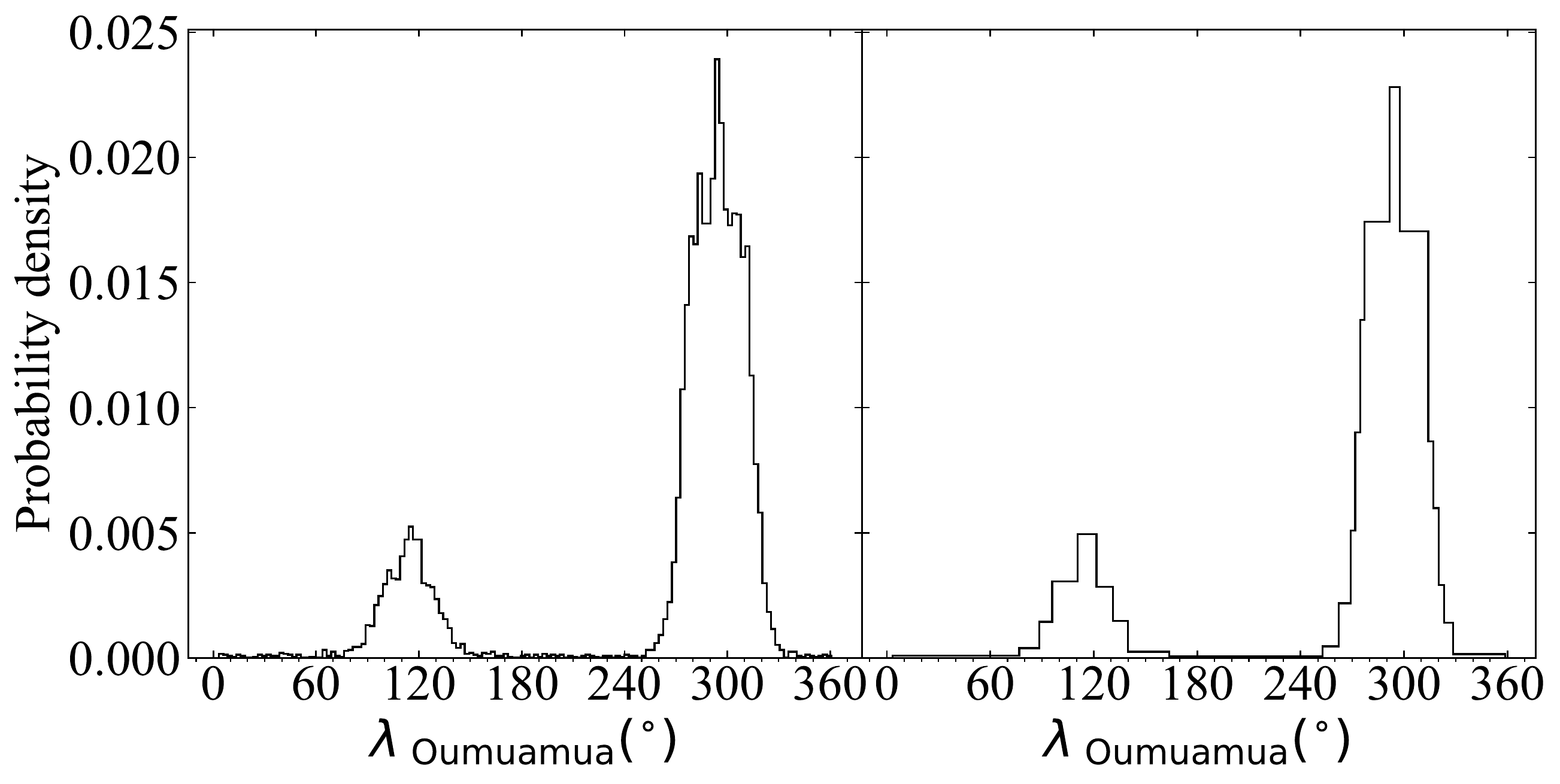}
         \includegraphics[width=0.99\linewidth]{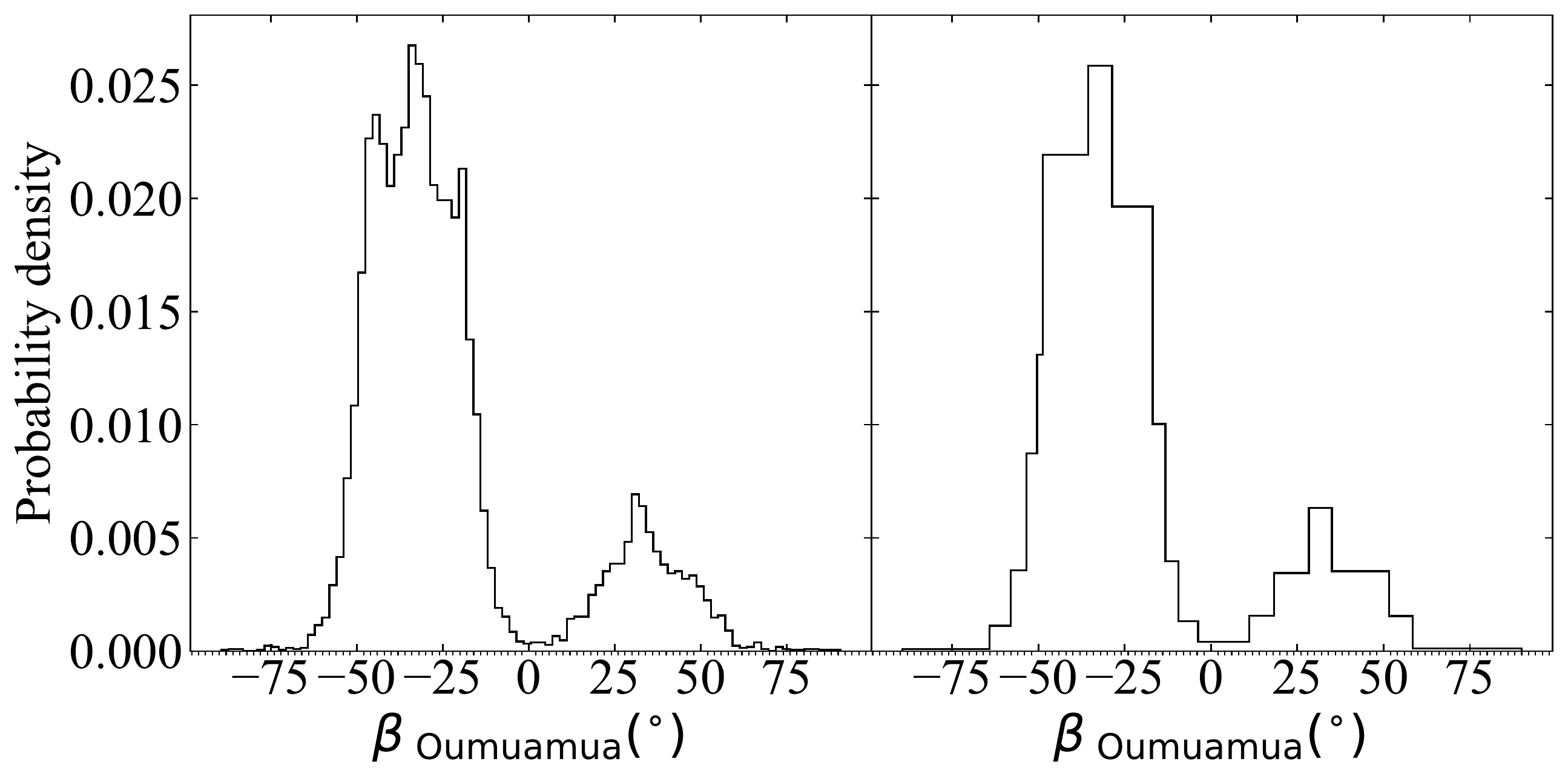}
         \caption{Distributions of spin-axis orientations, $(\lambda_{\rm p}, \beta_{\rm p})$, for 1I/2017~U1~(`Oumuamua) assuming a very
                  oblate shape, with $p$=0.836. The left panel shows the histogram with bins computed using the Freedman and Diaconis rule,
                  while the right panel uses the Bayesian Blocks technique. Histograms are based on data from Table~\ref{elements1I}.
                 }
         \label{1Ipoledisc}
      \end{figure}
%
%
%
%
      \begin{figure}
        \centering
         \includegraphics[width=\linewidth]{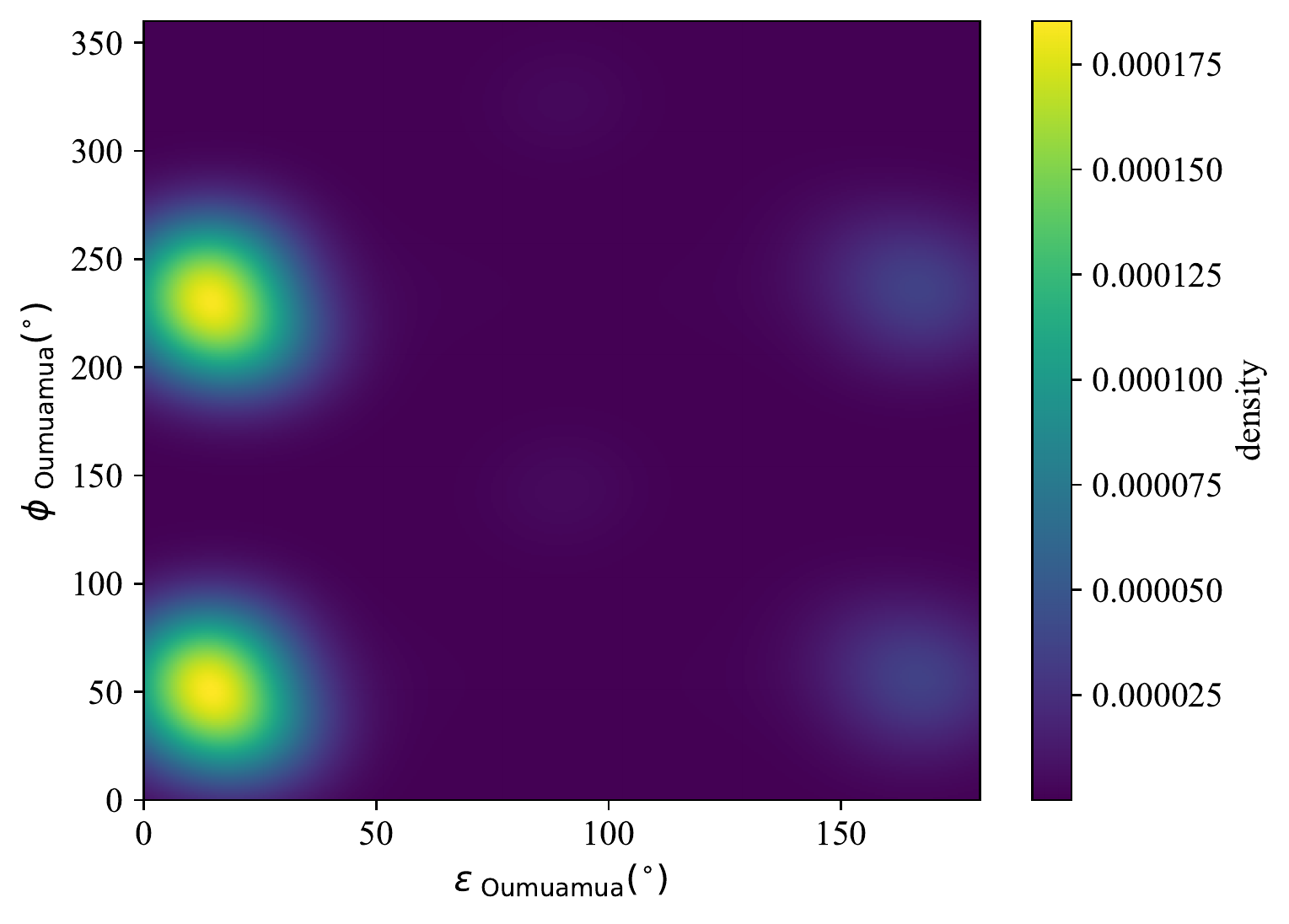}
         \includegraphics[width=\linewidth]{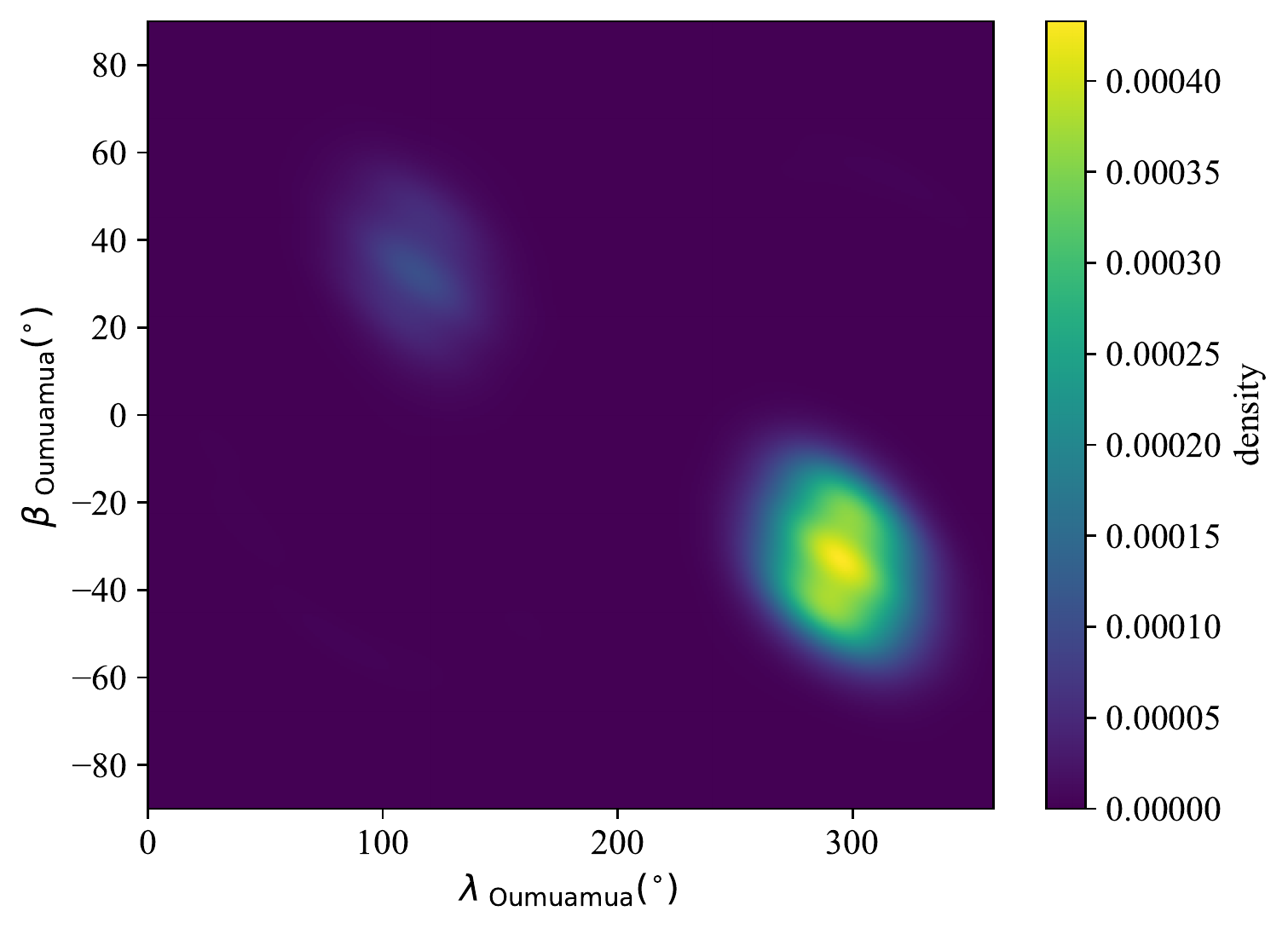}
         \includegraphics[width=\linewidth]{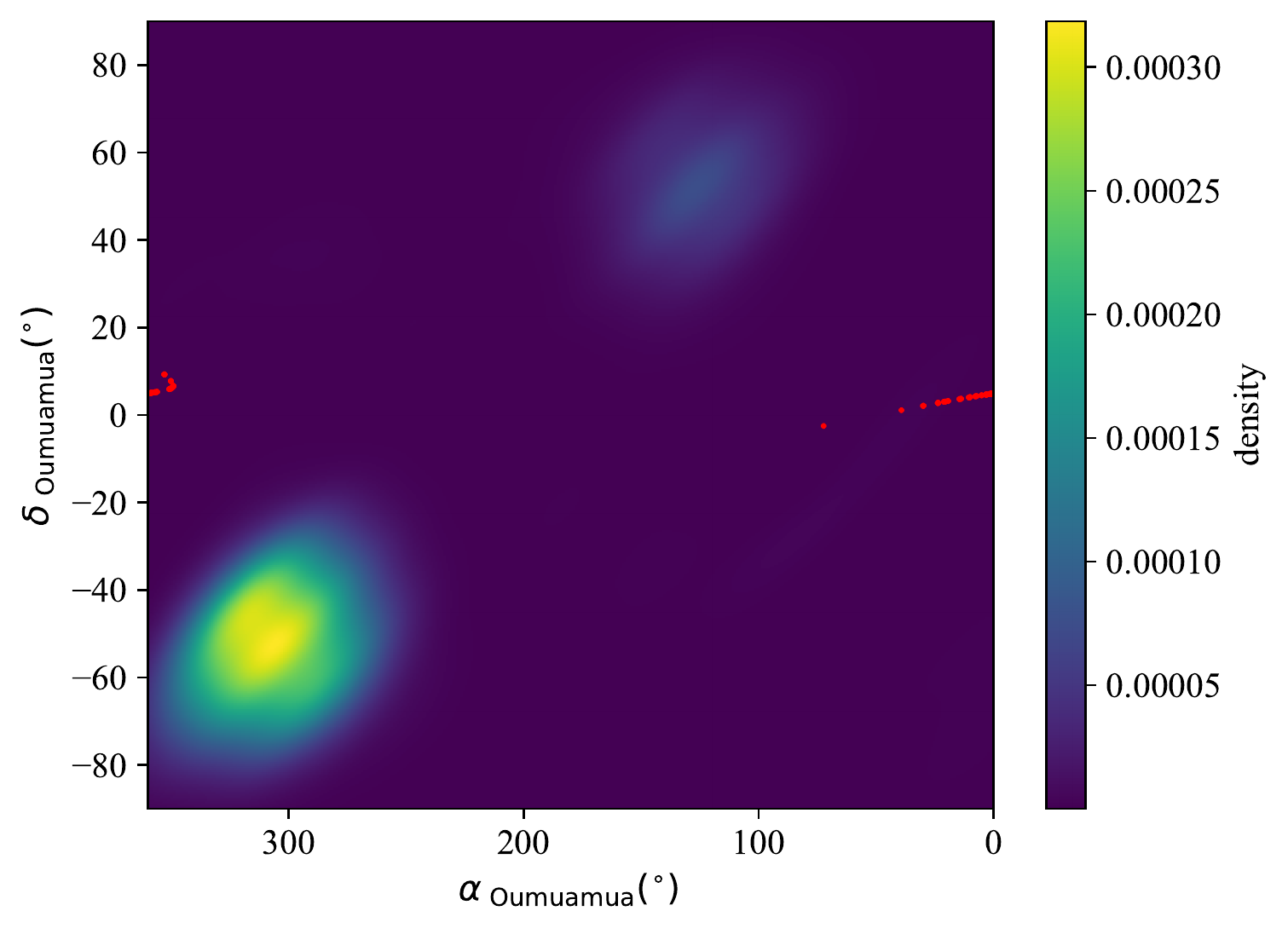}
         \caption{Gaussian kernel density estimation of the spin-axis orientations computed using the SciPy library
                  \citep{2020NatMe..17..261V} for a very oblate case ($p=0.836$). The top panel shows the $(\epsilon_{\rm c}, \phi_{\rm c})$
                  map. The results in ecliptic coordinates, $(\lambda_{\rm p}, \beta_{\rm p})$, are shown in the middle panel and those in
                  equatorial coordinates, $(\alpha_{\rm p}, \delta_{\rm p})$, are displayed in the bottom panel that includes the actual
                  observations (in red) of interstellar object 1I/2017~U1~(`Oumuamua) available from the MPC.
                 }
         \label{1Ipolecoordisc}
      \end{figure}
%
%

      The rotational properties provided by our modeling are very different depending on how extreme the shape of `Oumuamua is; this 
      suggests that the spin-axis orientation of `Oumuamua is tightly controlled by its shape. In the Solar System, most cometary nuclei are
      closer to a prolate configuration, not oblate \citep{2004come.book..281S}. Although the work by \citet{2019MNRAS.489.3003M} is widely 
      regarded as the best available at the moment, the arguments put forward by \citet{2020NatAs...4..852Z} may indicate that a final 
      answer on `Oumuamua's shape still remains elusive.              

   \section{Interstellar comet 2I/Borisov\label{2I}}
      It is unlikely that additional data on 1I/2017~U1~(`Oumuamua) may eventually emerge; in sharp contrast, interstellar comet 2I/Borisov 
      is still being actively observed. The two most recent orbit determinations of 2I/Borisov are shown in Tables~\ref{elements2Ia} and 
      \ref{elements2Ib} and both consider nongravitational effects driven by sublimation of water ice (see Sect.~\ref{Intro}). However,
      \citet{2020AJ....159...77Y} argued that a model based on sublimation of CO could be more consistent with the observations compared to 
      a H$_2$O model. In addition, \citet{2020NatAs...4..867B} have pointed out that the coma of 2I/Borisov contains significantly more CO 
      than H$_2$O gas, with abundances of at least 173\%, more than three times higher than previously measured for any comet in the inner 
      ($<$2.5~AU) Solar System. However, the water production rate prior to perihelion was found to increase faster than for most known
      dynamically new comets \citep{2020ApJ...893L..48X}; a hyperactive nucleus was also favored by \citet{2020ApJ...889L..10M}. Again and 
      as considered above, an orbit determination may produce numerically correct results (in the sense of correct ephemeris predictions) 
      even if the hypotheses used to carry out the calculations are physically questionable or even invalid.

      The orbit determination in Table~\ref{elements2Ia} corresponds to an epoch nearly a month past perihelion and shows that only the 
      normal acceleration may be playing a significant role on the dynamical evolution of 2I/Borisov as the other components have values 
      statistically compatible with zero. Using data from Table~\ref{elements2Ia} as input, we obtained the distributions in 
      Figs.~\ref{borisovSPINa} and \ref{borisovLAGa}. The distribution of $\epsilon_{\rm \ 2I/Borisov}$ is unimodal but wide (see 
      Fig.~\ref{borisovSPINa}, top panels); there is one statistically significant maximum and the median and 16th and 84th percentiles are 
      $59{\degr}_{-18{\degr}}^{+15{\degr}}$. Figure~\ref{borisovSPINa}, bottom panels, shows a multimodal distribution for 
      $\phi_{\rm \ 2I/Borisov}$; the absolute maxima are found at about 90{\degr} and 270{\degr}, with secondary maxima at 40{\degr} and 
      220{\degr} (therefore, each pair separated by 180{\degr}), and the associated dispersions are relatively small. 
      Figure~\ref{borisovLAGa} displays a multimodal distribution for $\eta_{\rm \ 2I/Borisov}$, with a dominant maximum at about 
      $-$130{\degr}, which is on the nightside of the comet. 
%
%
      \begin{figure}
        \centering
         \includegraphics[width=0.99\linewidth]{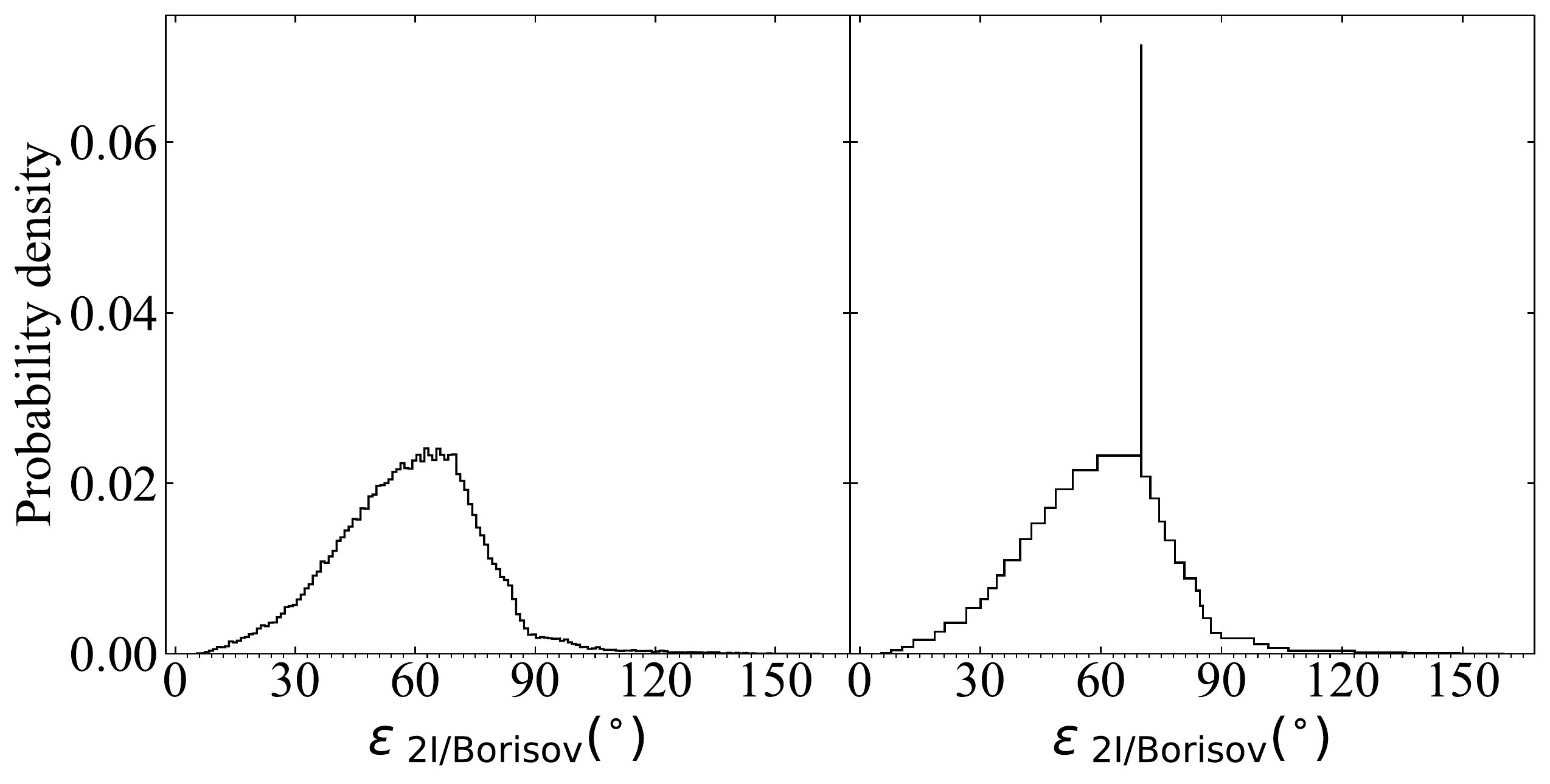}
         \includegraphics[width=0.99\linewidth]{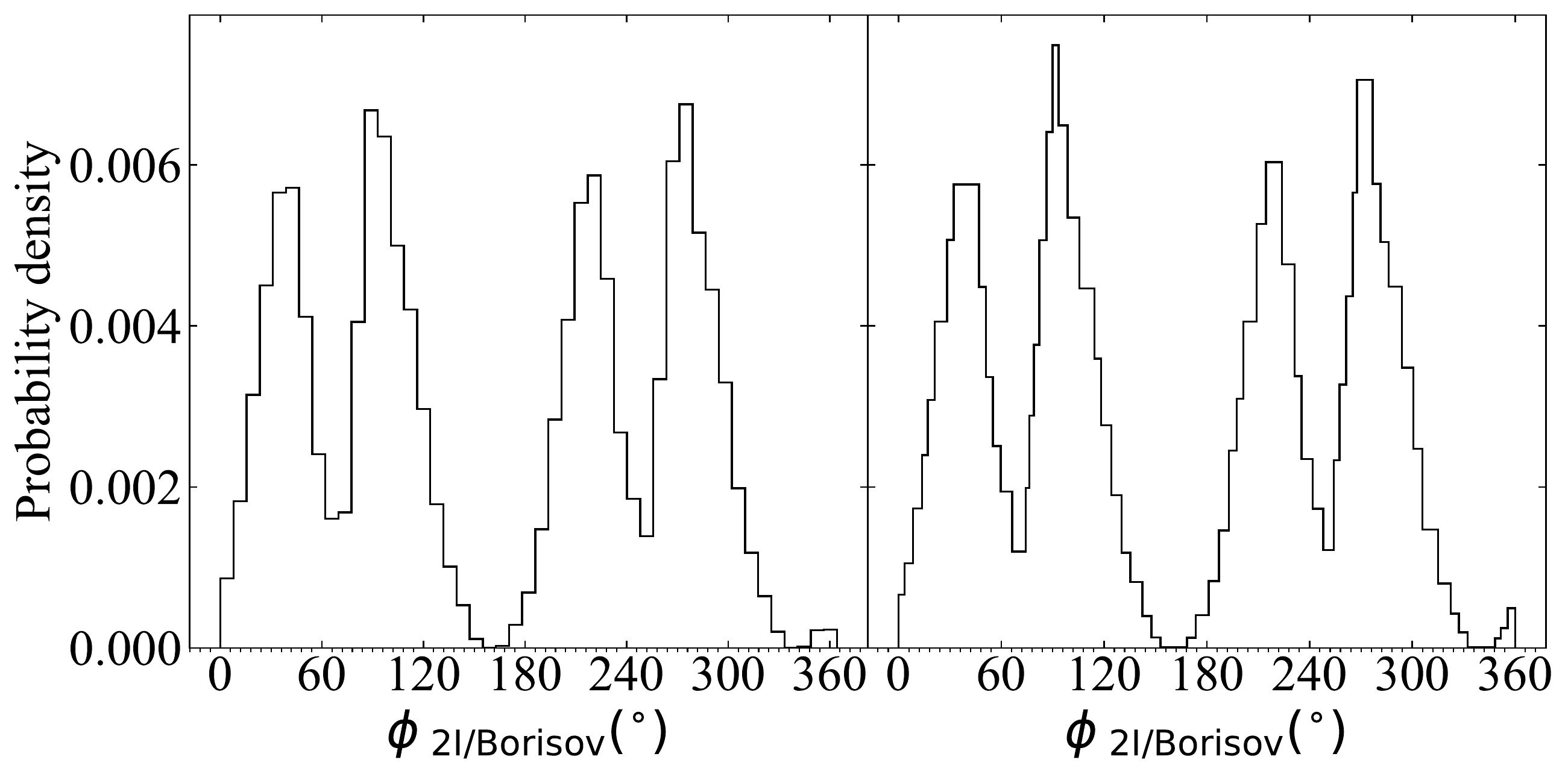}
         \caption{Distributions of equatorial obliquity and cometocentric longitude of the Sun at perihelion for 2I/Borisov.
                  These results correspond to the orbit determination of comet 2I/Borisov in Table~\ref{elements2Ia}.
                 }
         \label{borisovSPINa}
      \end{figure}
%
%
%
%
      \begin{figure}
        \centering
         \includegraphics[width=0.99\linewidth]{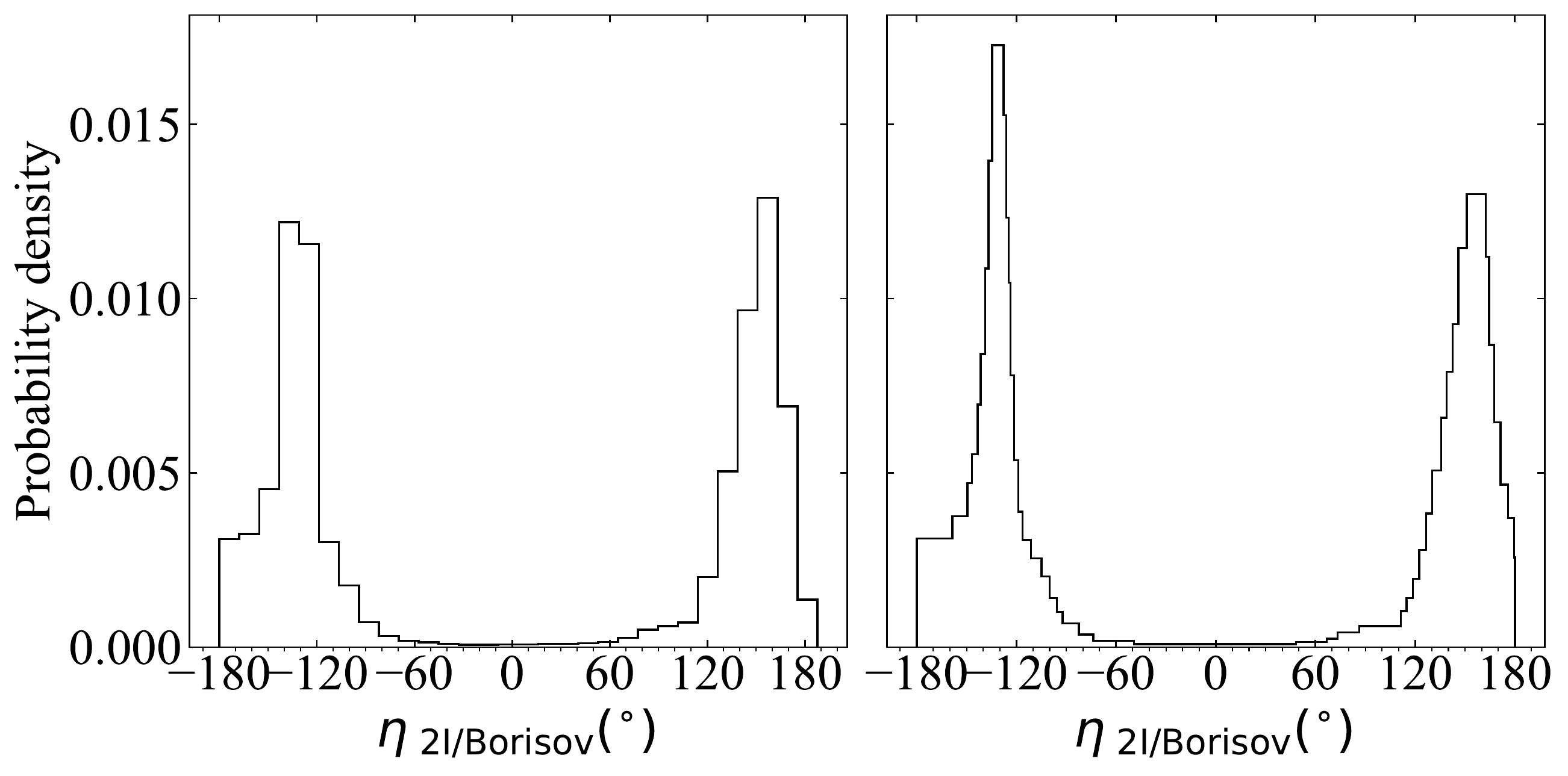}
         \caption{Distribution of thermal lag angle for 2I/Borisov.
                  These results correspond to the orbit determination of comet 2I/Borisov in Table~\ref{elements2Ia}.
                 }
         \label{borisovLAGa}
      \end{figure}
%
%

      The most probable value of the equatorial obliquity of 2I/Borisov is sometimes found for periodic comets in the Solar System, when 
      outgassing comes from a single active region located at a rotation pole (see sect. 3.1 in \citealt{2004come.book..137Y}). Surprisingly, 
      outgassing appears to be restricted to the nightside of the comet. Most comets tend to exhibit afternoon activity 
      \citep{1981AREPS...9..113S}. The lack of dayside activity seems problematic when considering that `Oumuamua's activity was restricted 
      to the dayside for the very oblate case; however, outbursts from the nightside of comets 9P/Tempel~1 \citep{2007Icar..187...26F,
      2013Icar..222..540F}, 103P/Hartley~2 \citep{2011Sci...332.1396A,2013Icar..222..610B}, and 67P/Churyumov-Gerasimenko 
      \citep{2016A&A...596A..89K,2019A&A...630A..21R} have been observed.

%
%
      \begin{figure}
        \centering
         \includegraphics[width=0.99\linewidth]{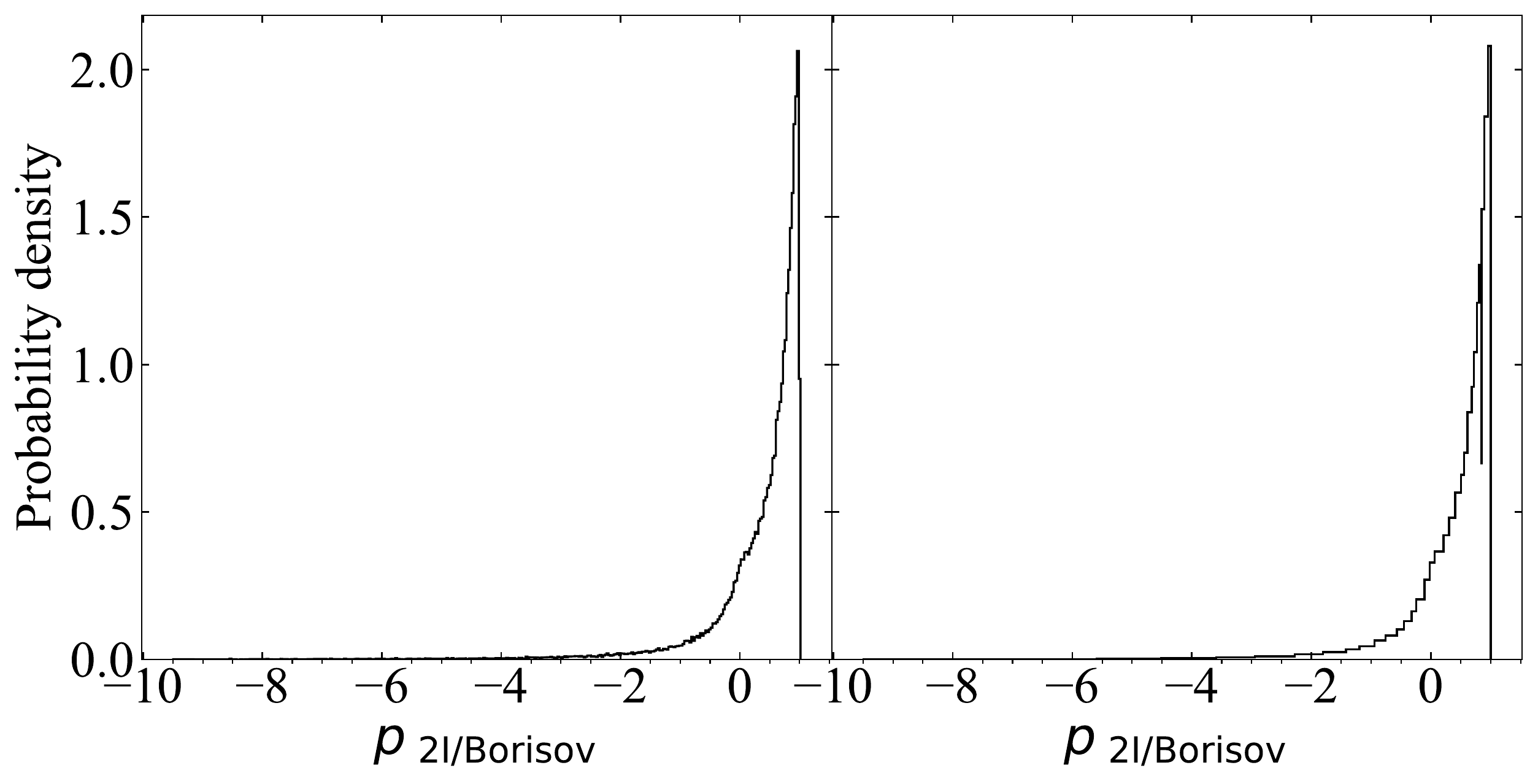}
         \caption{Distribution of the oblateness parameter for 2I/Borisov.
                  These results correspond to the orbit determination of comet 2I/Borisov in Table~\ref{elements2Ia}.
                 }
         \label{borisovOBLAa}
      \end{figure}
%
%
      Figure~\ref{borisovOBLAa} shows that the most probable value for the oblateness parameter for 2I/Borisov corresponds to that of a very
      oblate body with median and 16th and 84th percentiles of $0.6_{-0.7}^{+0.3}$; the peak of the distribution is found at $p=0.95$. This 
      is even more extreme than the value of $p$=0.836 favored by \citet{2019MNRAS.489.3003M} for 1I/2017~U1~(`Oumuamua).  

      The location of the pole is shown in Figs.~\ref{2Ipole09Jan} and \ref{2Ipolecoor09Jan}. The most probable spin-axis direction of 
      2I/Borisov in equatorial coordinates could be $(275{\degr},~+65{\degr})$, see Fig.\ref{2Ipolecoor09Jan}, bottom panel, with 
      approximate Galactic coordinates $l=95{\degr}$, $b=+28{\degr}$ that point slightly above the Galactic disk, toward the constellation 
      Draco.
%
%
      \begin{figure}
        \centering
         \includegraphics[width=0.99\linewidth]{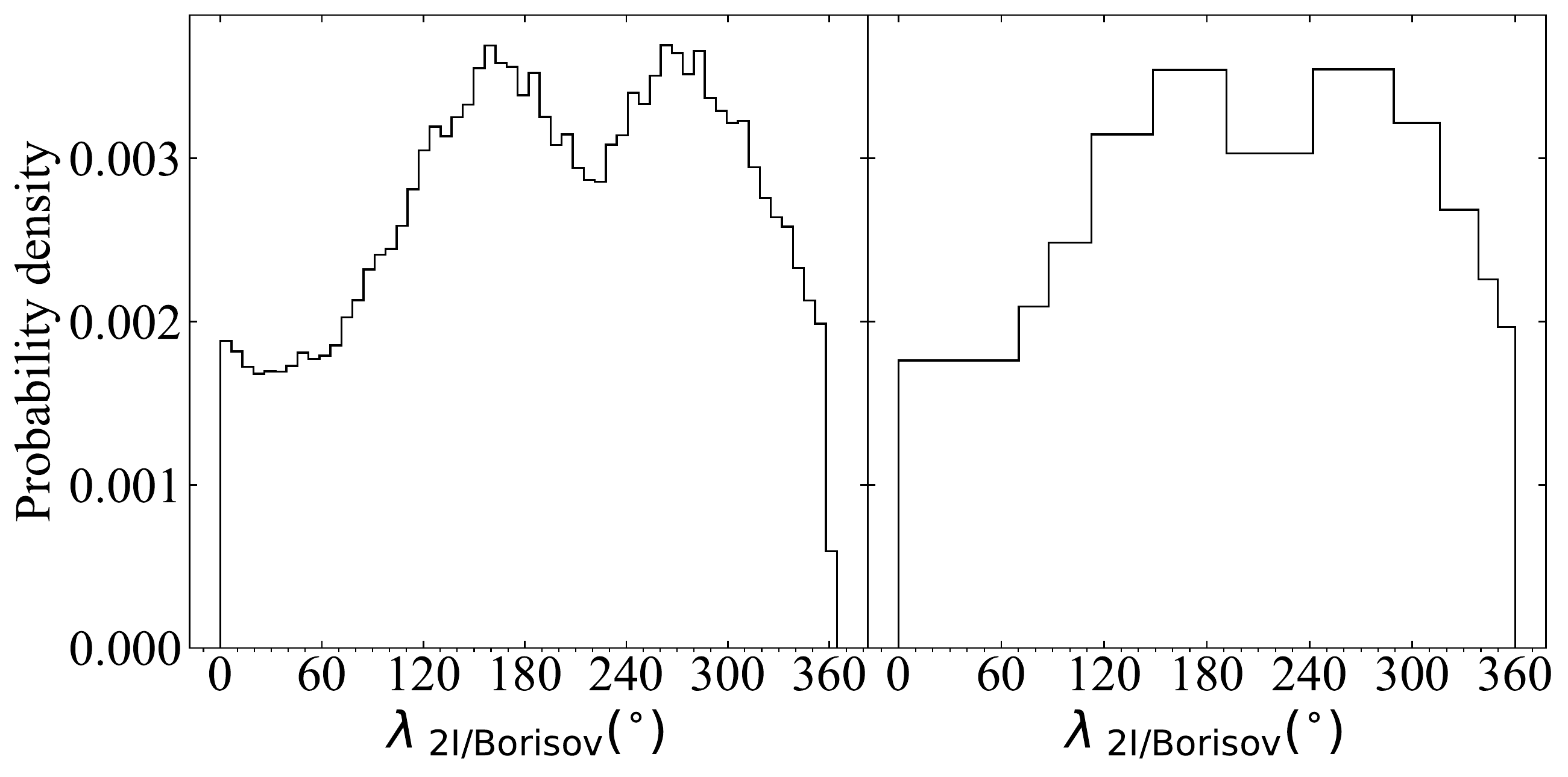}
         \includegraphics[width=0.99\linewidth]{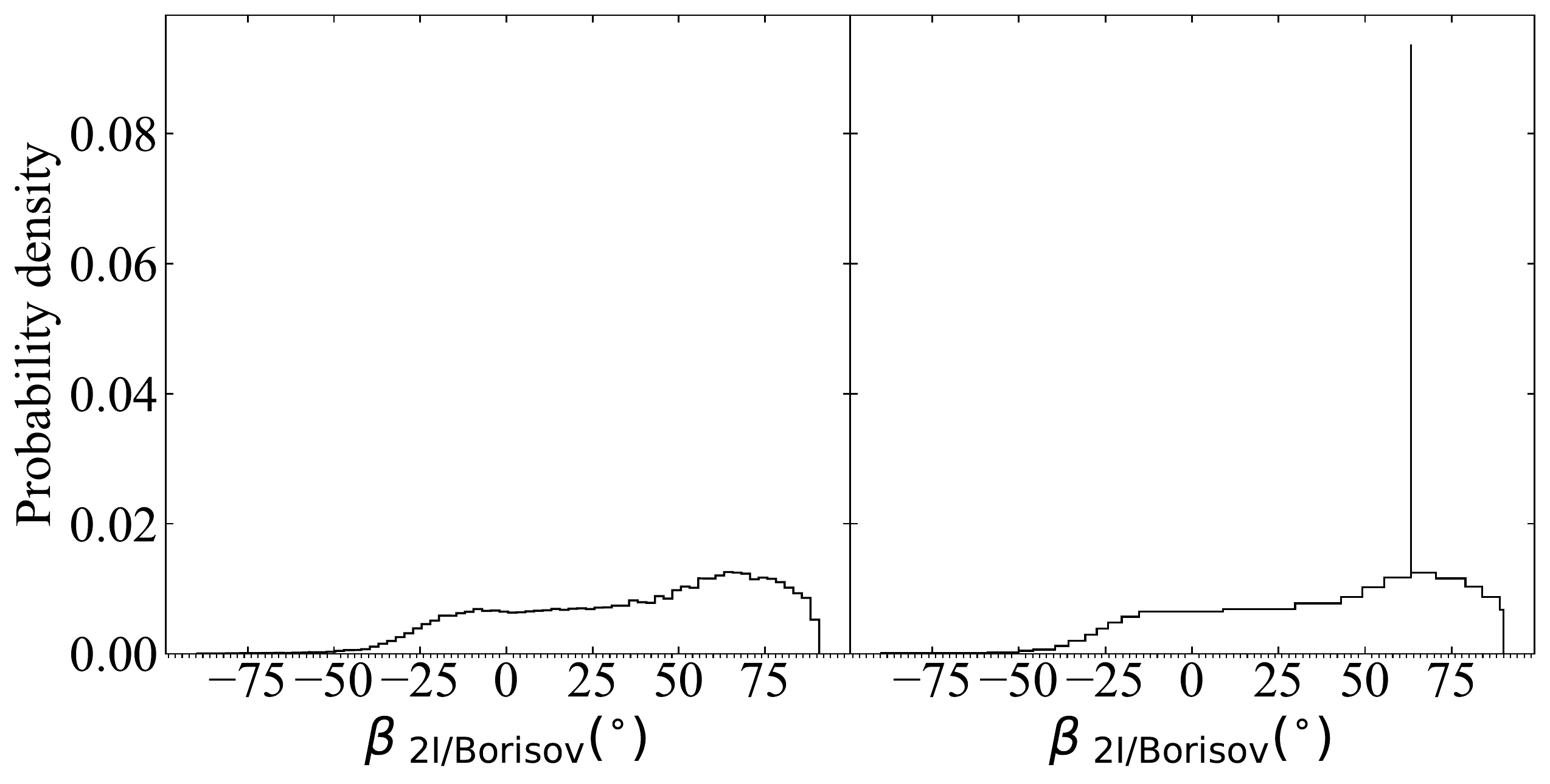}
         \caption{Distributions of spin-axis orientations, $(\lambda_{\rm p}, \beta_{\rm p})$, for 2I/Borisov. The left panel shows the
                  histogram with bins computed using the Freedman and Diaconis rule, while the right panel uses the Bayesian Blocks
                  technique. Histograms are based on data from Table~\ref{elements2Ia}.
                 }
         \label{2Ipole09Jan}
      \end{figure}
%
%
%
%
      \begin{figure}
        \centering
         \includegraphics[width=\linewidth]{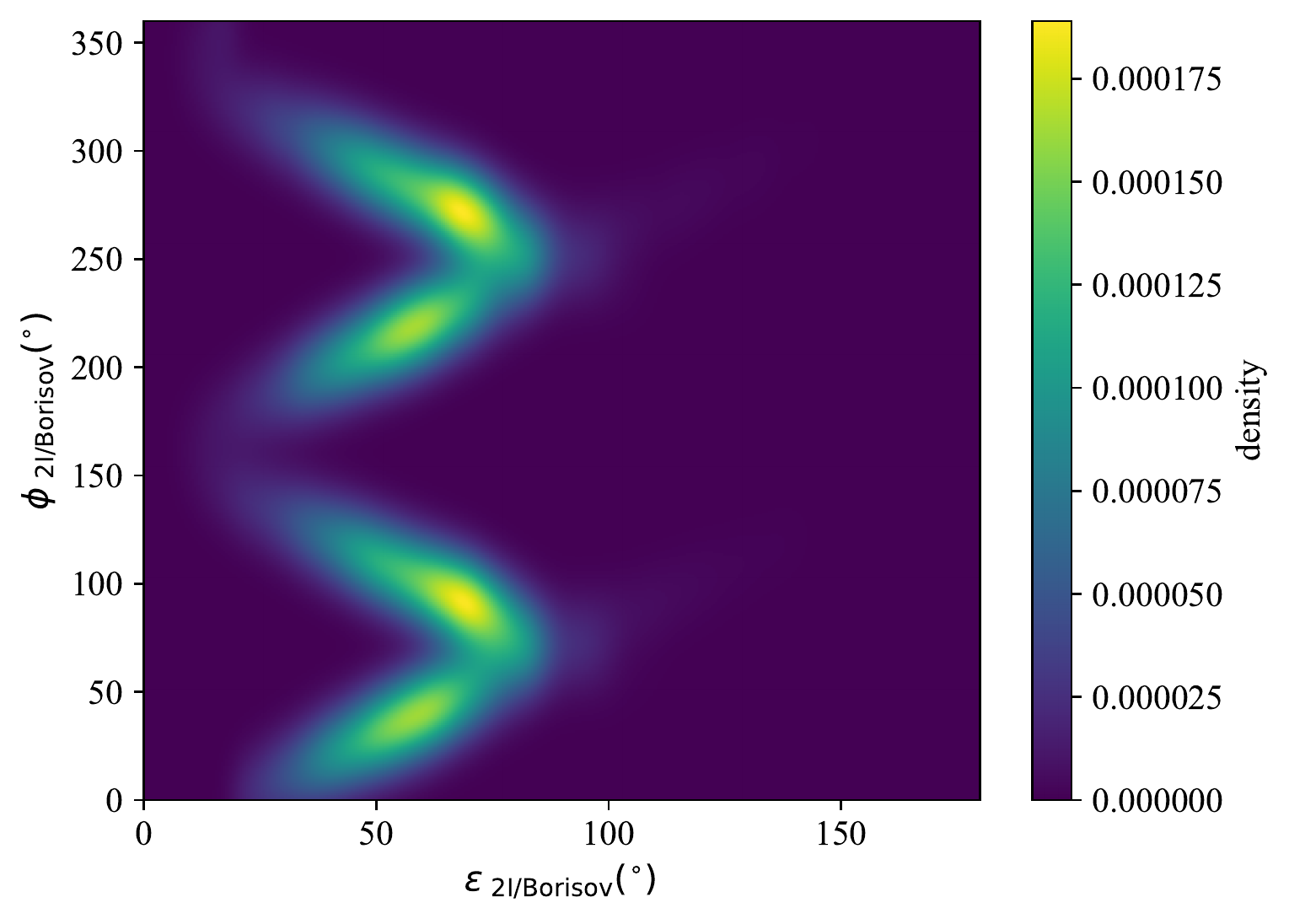}
         \includegraphics[width=\linewidth]{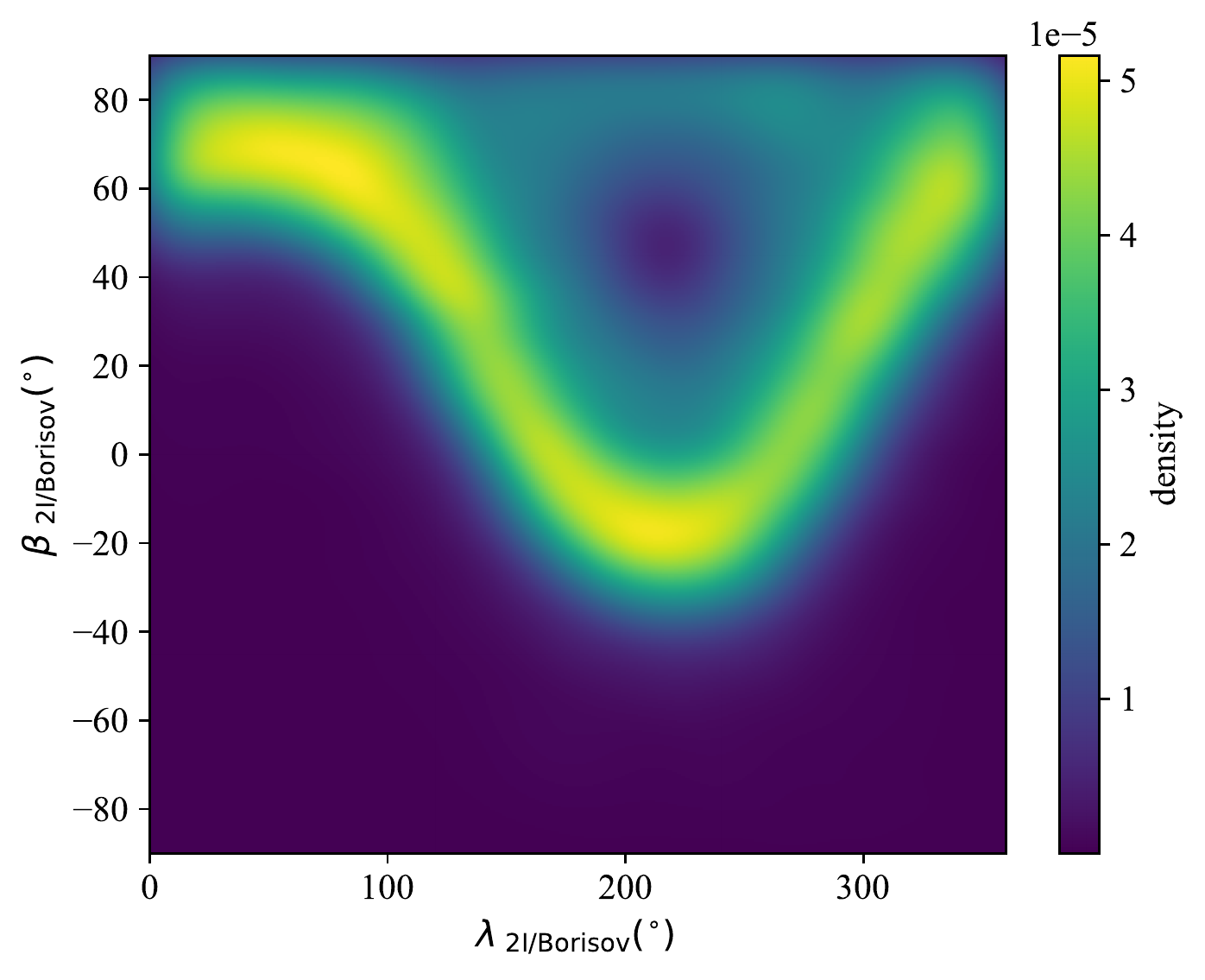}
         \includegraphics[width=\linewidth]{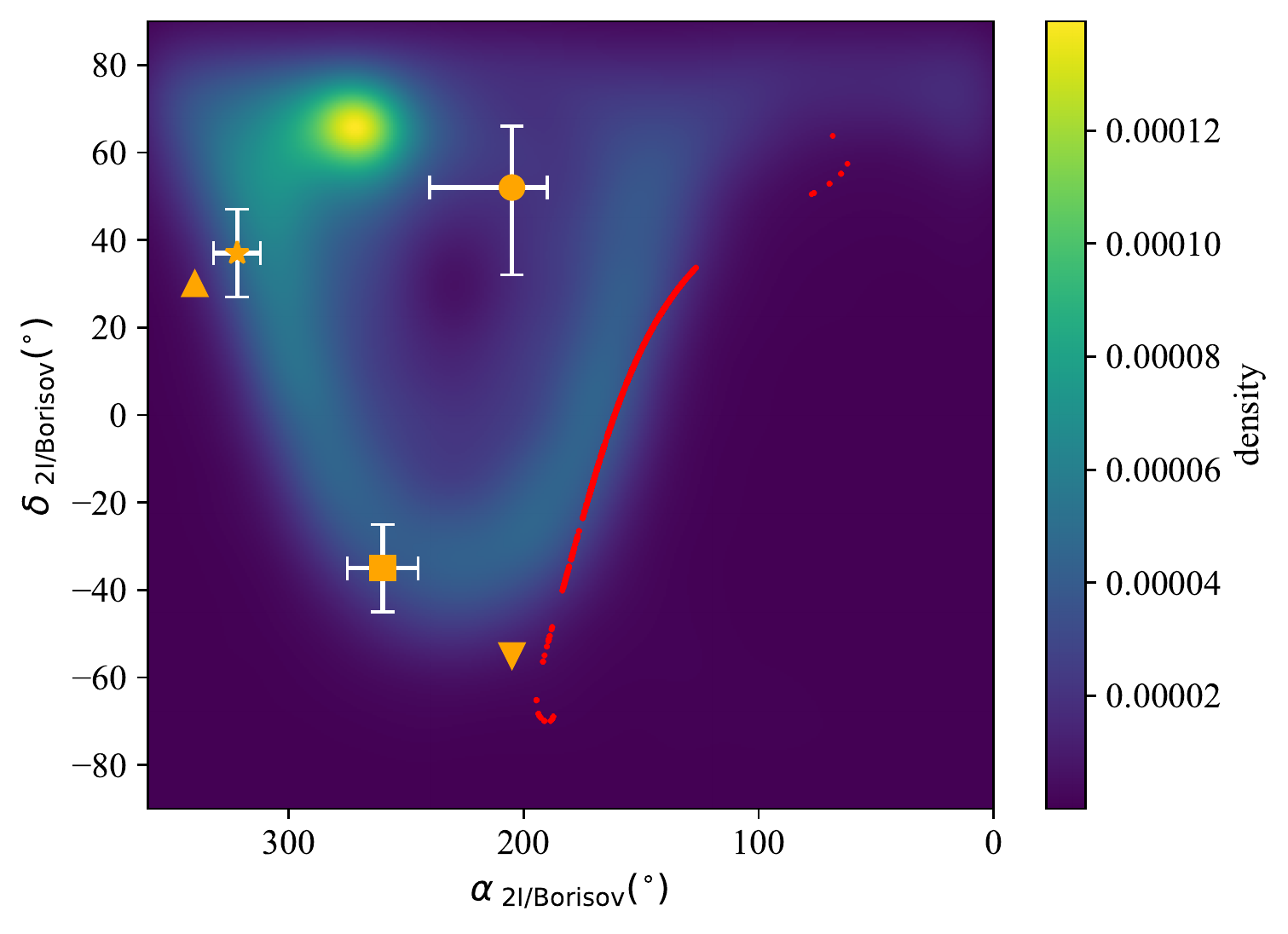}
         \caption{Gaussian kernel density estimation of the spin-axis orientations of 2I/Borisov computed using the SciPy library
                  \citep{2020NatMe..17..261V} and based on data from Table~\ref{elements2Ia}. The top panel shows the $(\epsilon_{\rm c}, 
                  \phi_{\rm c})$ map. The results in ecliptic coordinates, $(\lambda_{\rm p}, \beta_{\rm p})$, are shown in the middle panel 
                  and those in equatorial coordinates, $(\alpha_{\rm p}, \delta_{\rm p})$, are displayed in the bottom panel that includes 
                  the actual observations (in red) of 2I/Borisov available from the MPC. Already published spin-axis directions are plotted
                  as filled orange symbols with white error bars (when available): triangles \citep{2020AJ....159...77Y}, square 
                  \citep{2020MNRAS.495L..92M}, star \citep{2020AJ....160...26B}, and circle \citep{2020ApJ...895L..34K}.
                 }
         \label{2Ipolecoor09Jan}
      \end{figure}
%
%

      Our results are quite different when considering the most recent orbit determination of 2I/Borisov in Table~\ref{elements2Ib}. The 
      distribution of $\epsilon_{\rm \ 2I/Borisov}$ in Fig.~\ref{borisovSPINb}, top panels, is still unimodal but now the median and 16th 
      and 84th percentiles are 90{\degr}$\pm$52{\degr}. The distribution of $\phi_{\rm \ 2I/Borisov}$ in Fig.~\ref{borisovSPINb}, bottom 
      panels, is multimodal as the one in Fig.~\ref{borisovSPINa}, bottom panels, but now the dominant maxima are found at about 100{\degr} 
      and 280{\degr}. The distribution of $\eta_{\rm \ 2I/Borisov}$ in Fig.~\ref{borisovLAGb} is very different from that in 
      Fig.~\ref{borisovLAGa} and shows peaks at $-$20{\degr} and 20{\degr}, with median and 16th and 84th percentiles of 
      0{\degr}$\pm$29{\degr}; the maximum of activity takes place around noon as seen from the comet.
%
%
      \begin{figure}
        \centering
         \includegraphics[width=0.99\linewidth]{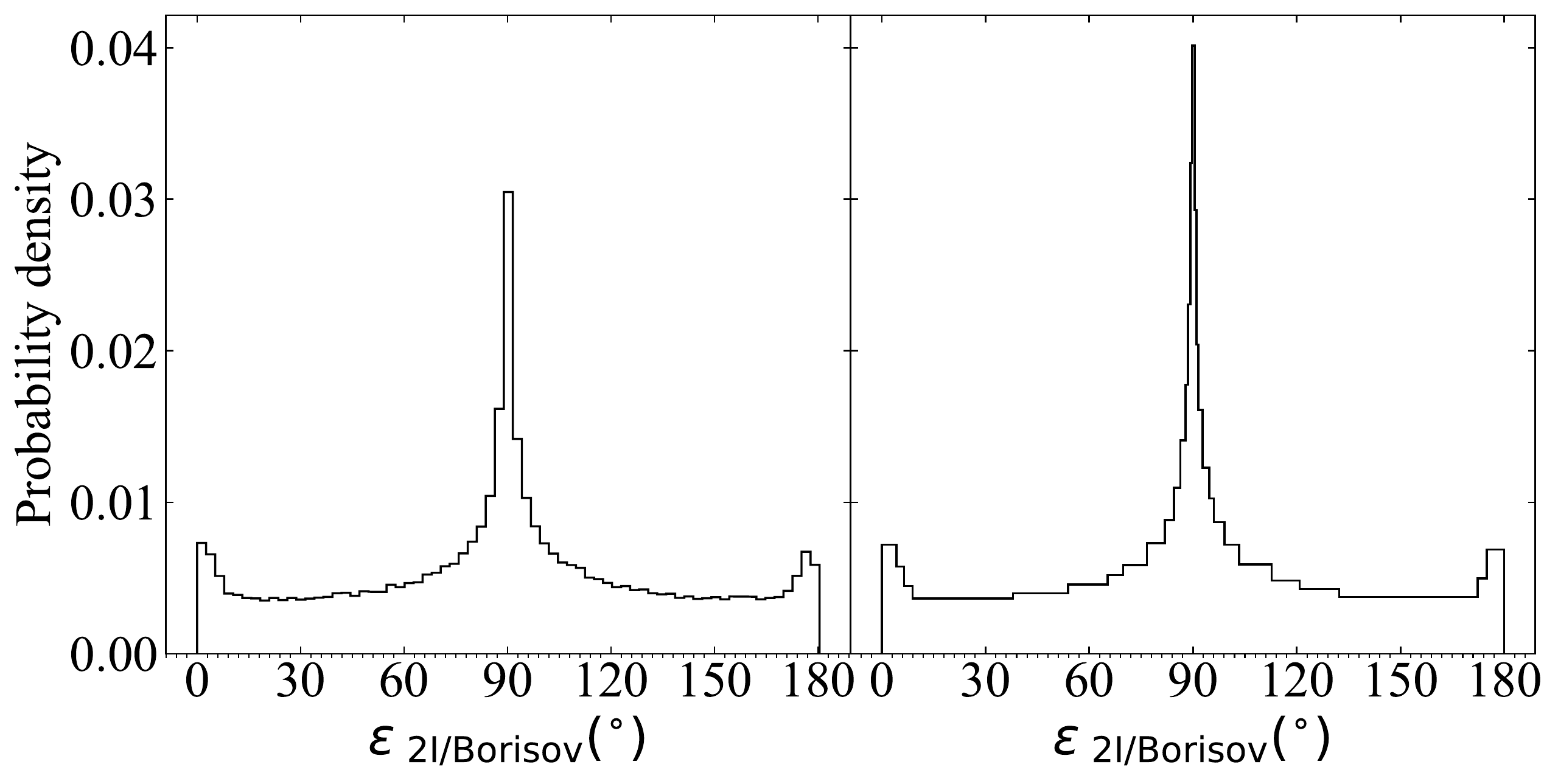}
         \includegraphics[width=0.99\linewidth]{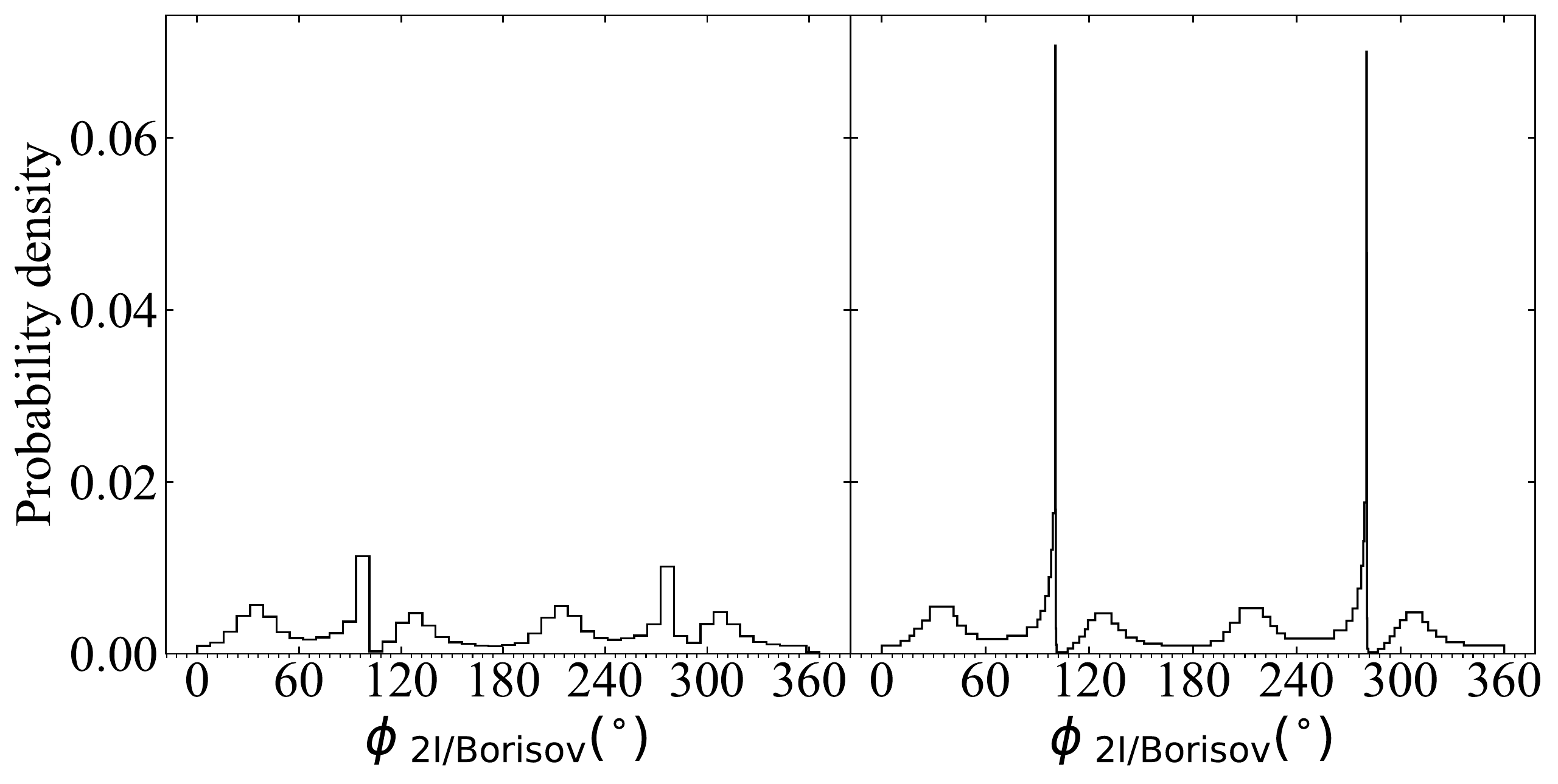}
         \caption{Distributions of equatorial obliquity and cometocentric longitude of the Sun at perihelion for 2I/Borisov.
                  Same as Fig.~\ref{borisovSPINa} but for the orbit determination of comet 2I/Borisov in Table~\ref{elements2Ib}.
                 }
         \label{borisovSPINb}
      \end{figure}
%
%
%
%
      \begin{figure}
        \centering
         \includegraphics[width=0.99\linewidth]{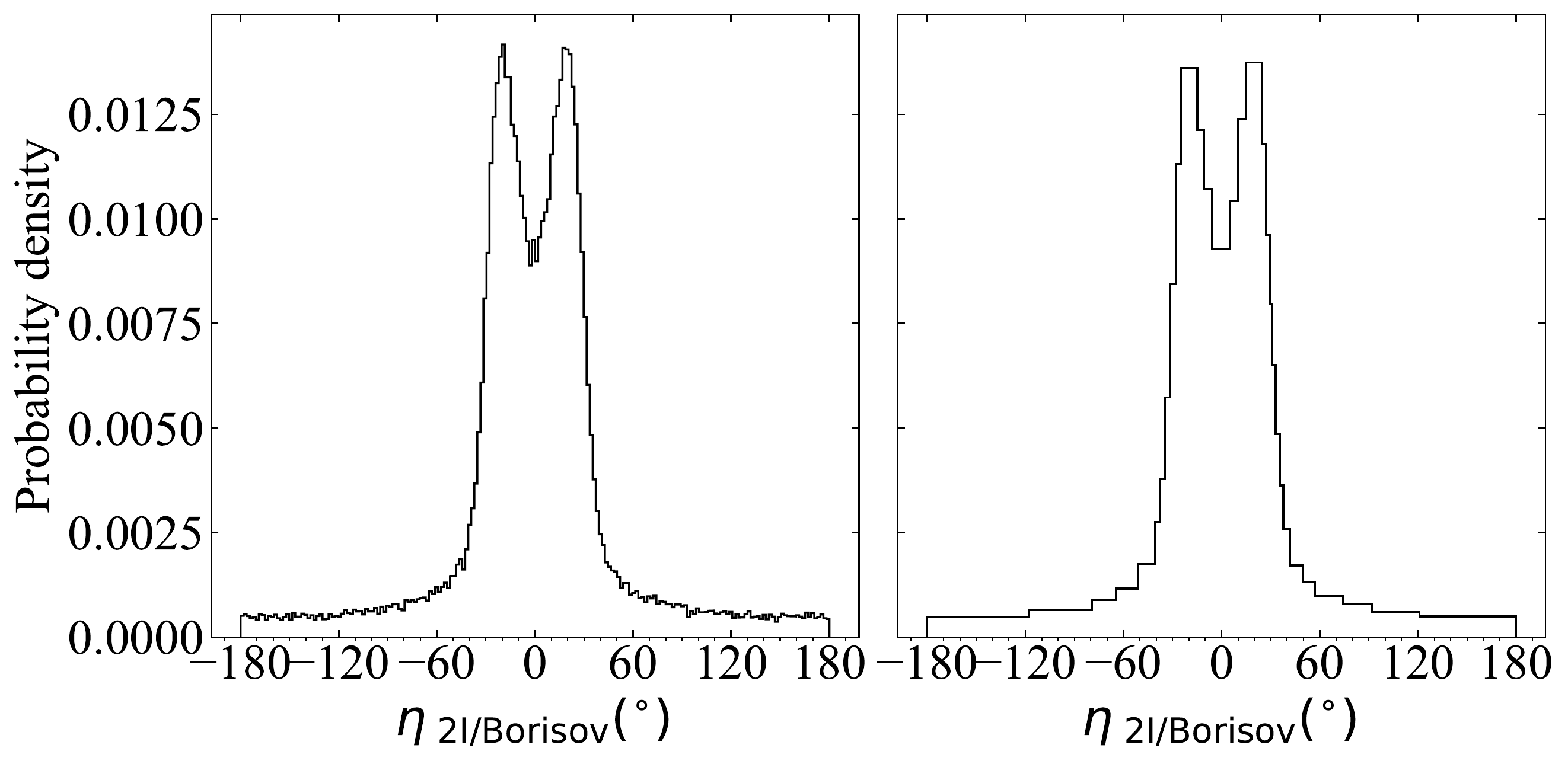}
         \caption{Distribution of thermal lag angle for 2I/Borisov.
                  Same as Fig.~\ref{borisovLAGa} but for the orbit determination of comet 2I/Borisov in Table~\ref{elements2Ib}.
                 }
         \label{borisovLAGb}
      \end{figure}
%
%

      Figure~\ref{borisovOBLAb} shows that the distribution of the values for the oblateness parameter of 2I/Borisov is bimodal with an
      absolute maximum of about 0.34 and a secondary peak at $-$0.36. Although our analysis favors an oblate shape for 2I/Borisov, a prolate
      one cannot be ruled out.

%
%
      \begin{figure}
        \centering
         \includegraphics[width=0.99\linewidth]{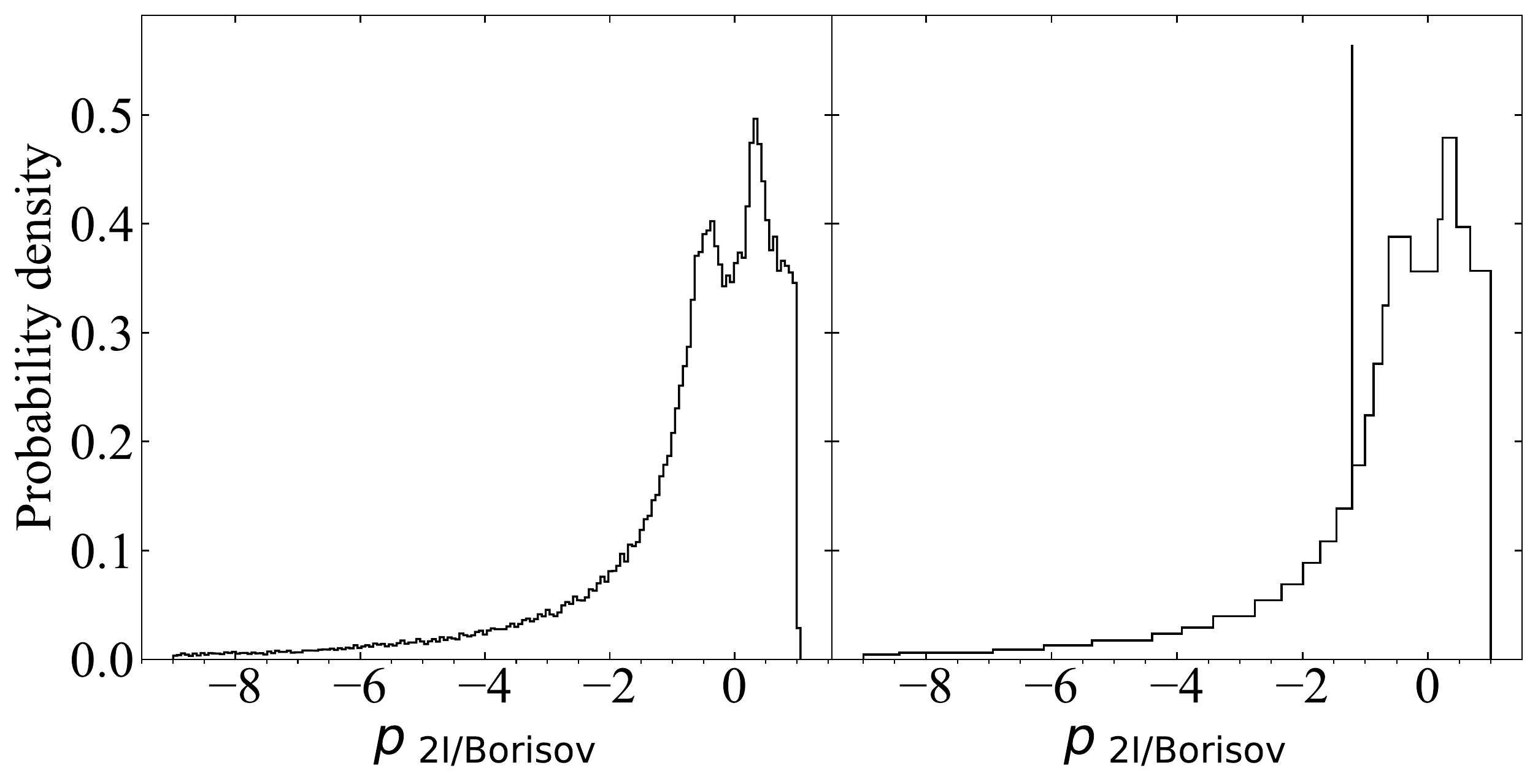}
         \caption{Distribution of the oblateness parameter for 2I/Borisov.
                  Same as Fig.~\ref{borisovOBLAa} but for the orbit determination of comet 2I/Borisov in Table~\ref{elements2Ib}.
                 }
         \label{borisovOBLAb}
      \end{figure}
%
%
      The location of the pole is shown in Figs.~\ref{2Ipole19Mar} and \ref{2Ipolecoor19Mar}. The most probable spin-axis direction of
      2I/Borisov in equatorial coordinates could be $(231{\degr},~+30{\degr})$, see Fig.\ref{2Ipolecoor19Mar}, bottom panel, with 
      approximate Galactic coordinates $l=47{\degr}$, $b=+56{\degr}$ that point well above the Galactic disk, toward the constellation 
      Corona Borealis.
%
%
      \begin{figure}
        \centering
         \includegraphics[width=0.99\linewidth]{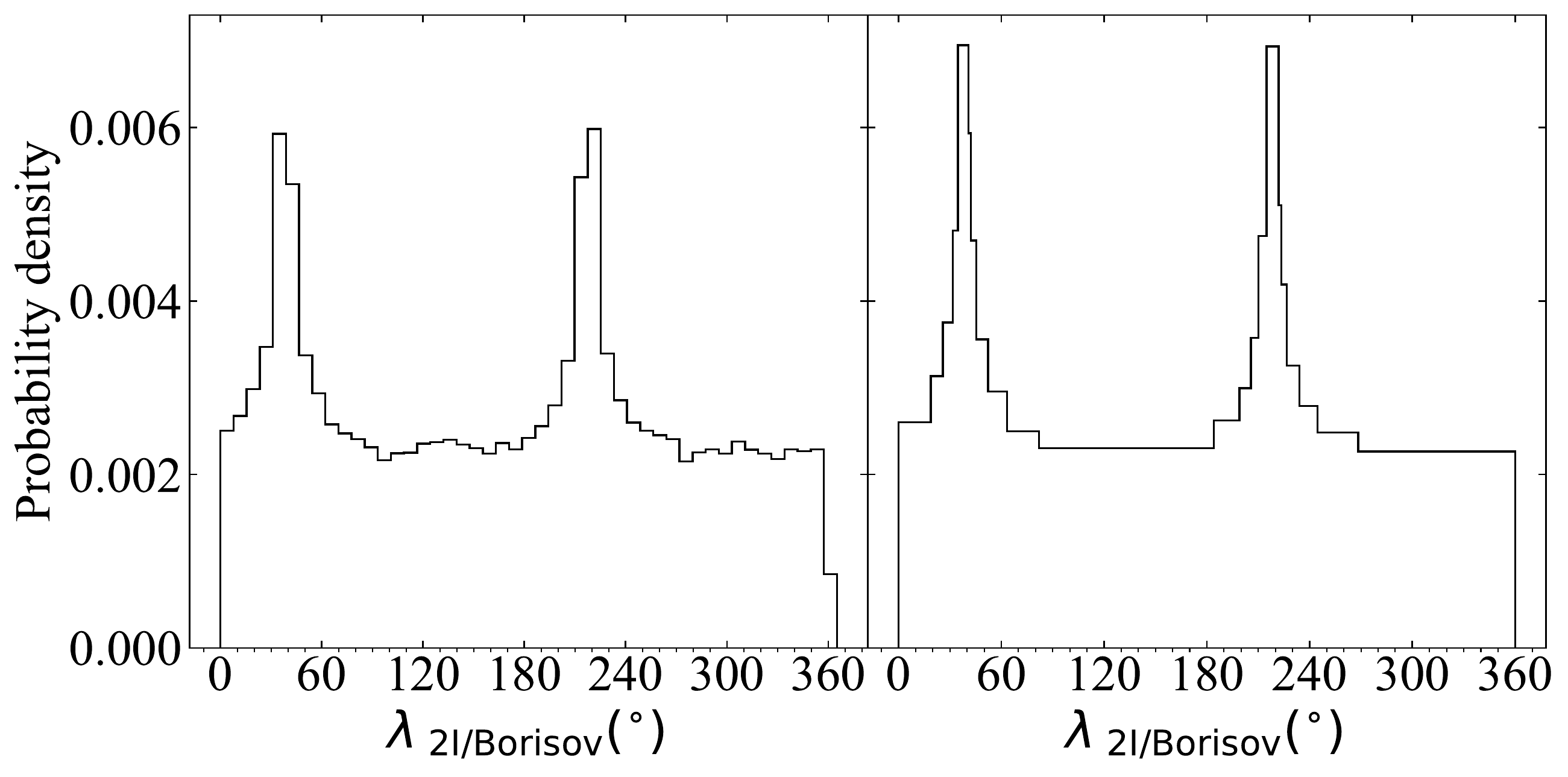}
         \includegraphics[width=0.99\linewidth]{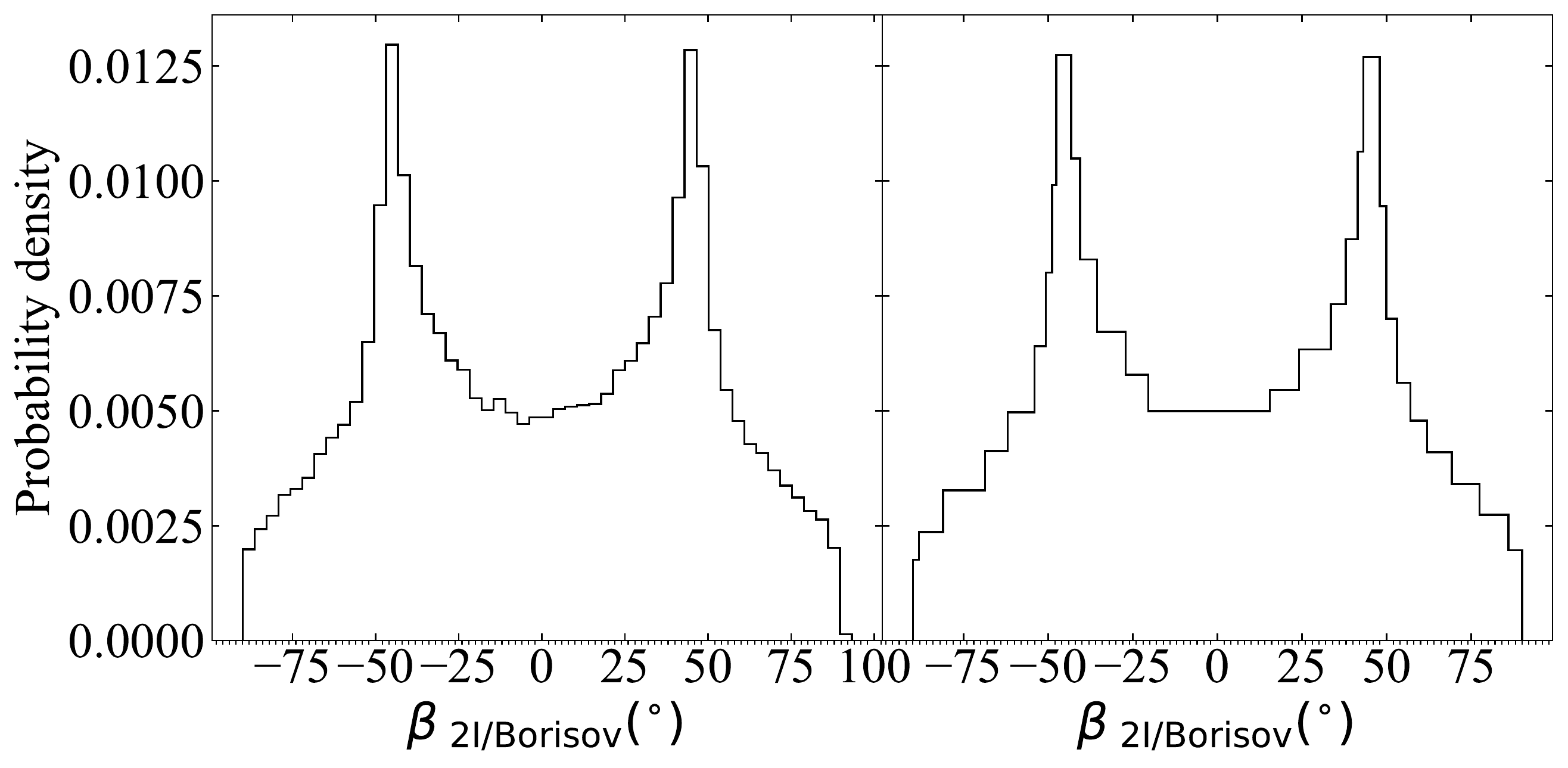}
         \caption{Distributions of spin-axis orientations, $(\lambda_{\rm p},~\beta_{\rm p})$, for 2I/Borisov. 
                  Same as Fig.~\ref{2Ipole09Jan} but based on data from Table~\ref{elements2Ib}.
                 }
         \label{2Ipole19Mar}
      \end{figure}
%
%
%
%
      \begin{figure}
        \centering
         \includegraphics[width=\linewidth]{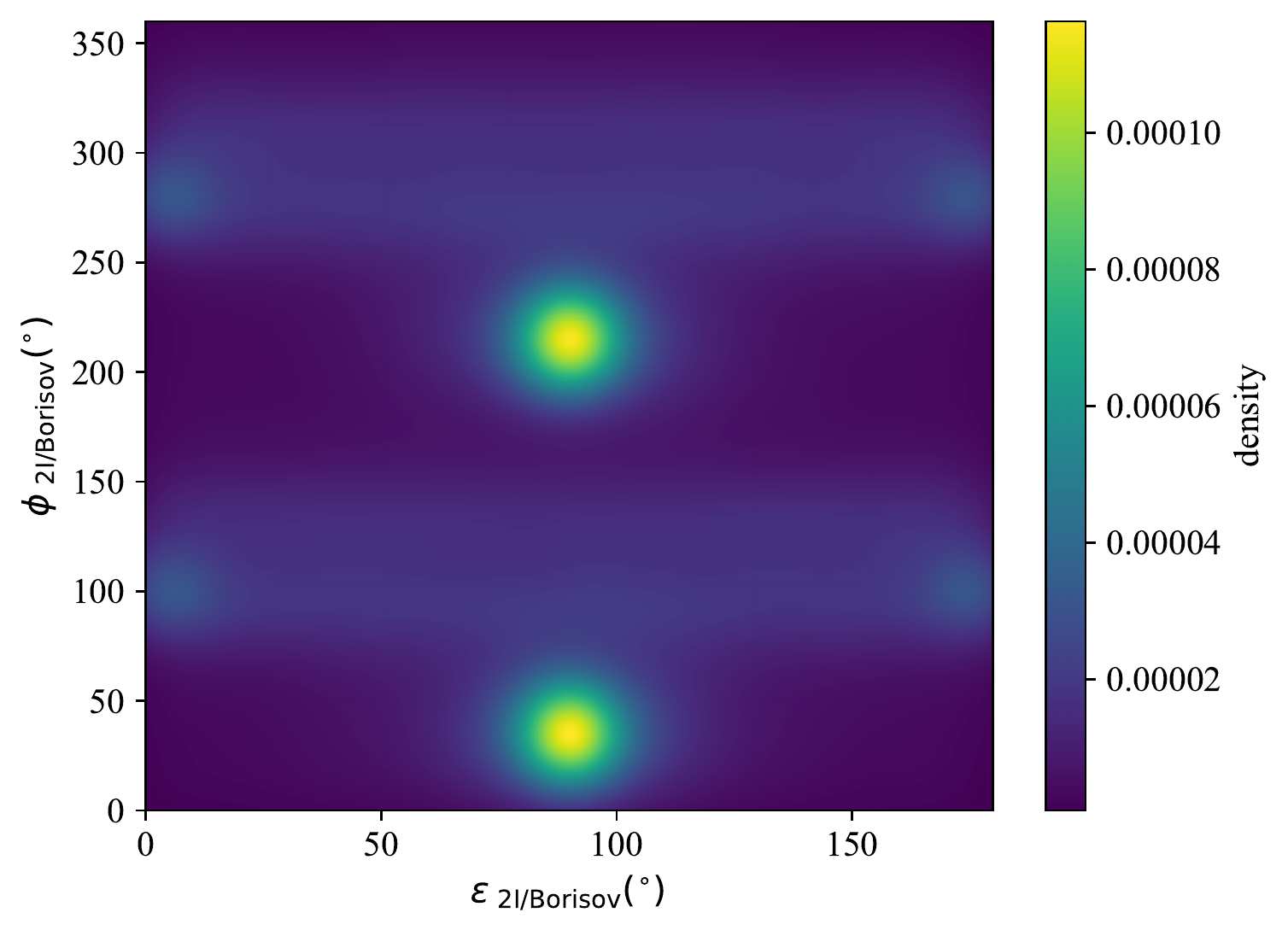}
         \includegraphics[width=\linewidth]{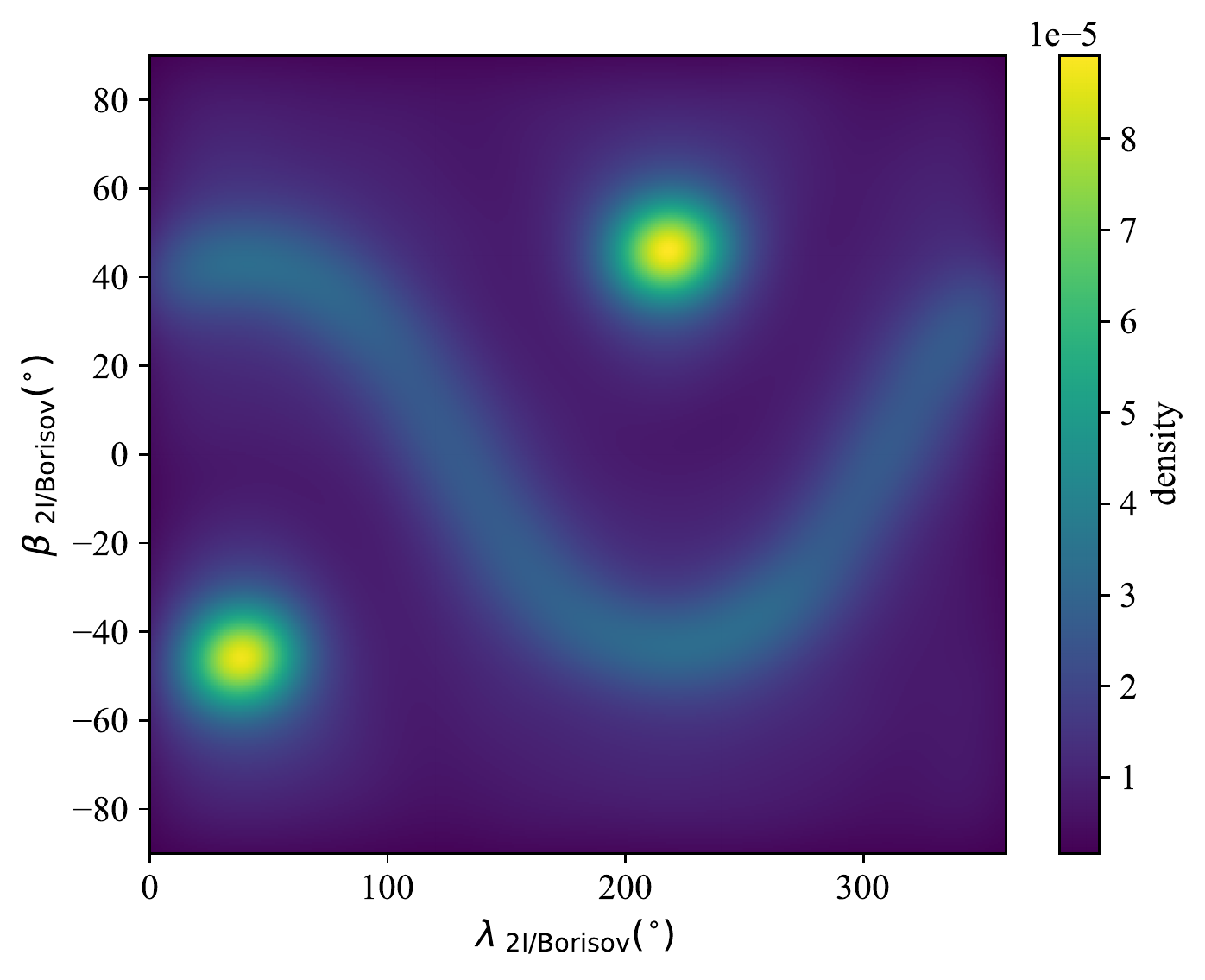}
         \includegraphics[width=\linewidth]{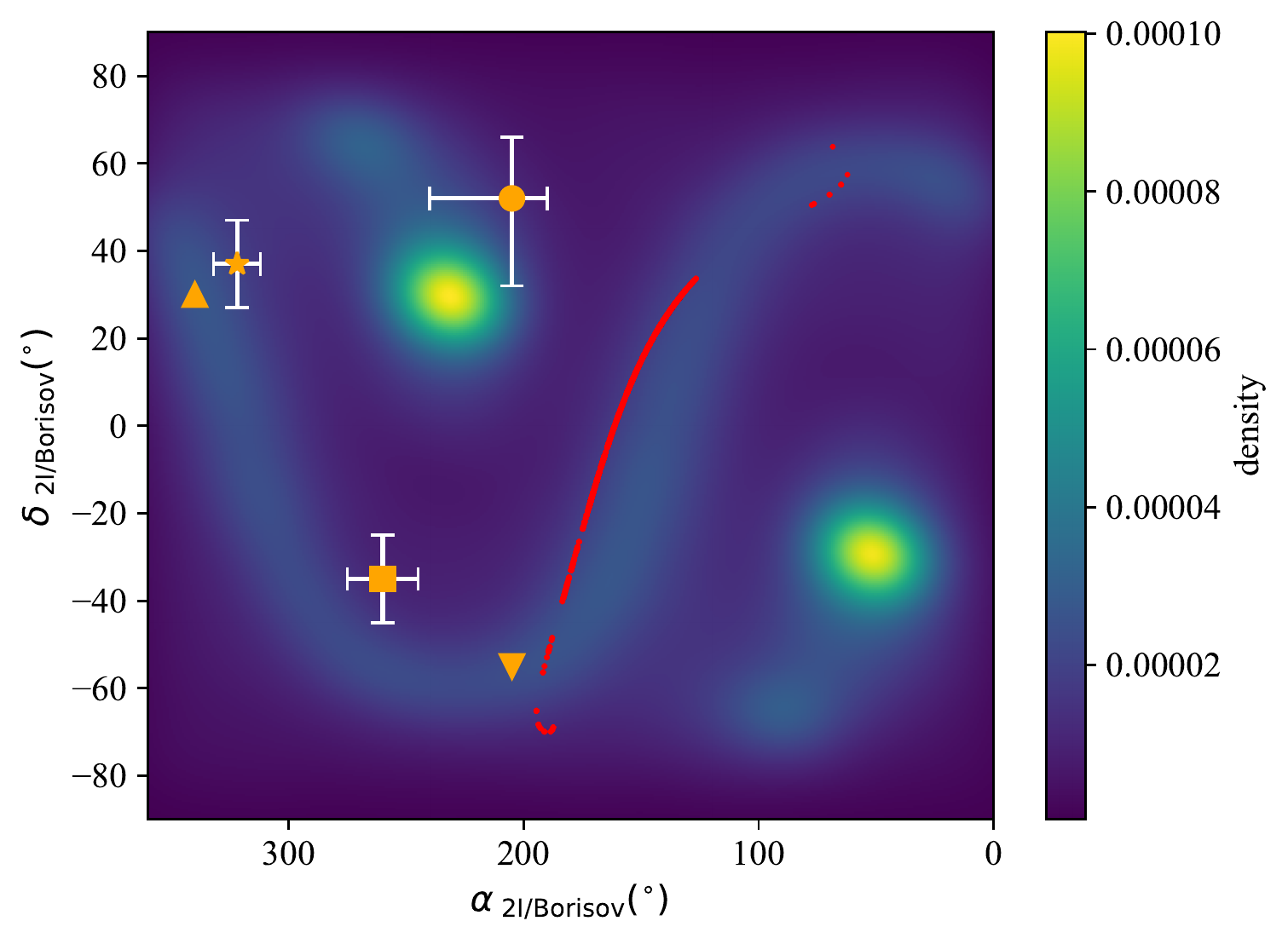}
         \caption{Same as Fig.~\ref{2Ipolecoor09Jan} but based on data from Table~\ref{elements2Ib}.
                 }
         \label{2Ipolecoor19Mar}
      \end{figure}
%
%

   \section{Discussion\label{Discussion}} 
      The main objective of the research presented here is not to argue against or in favor of the various orbit determinations (and their 
      underlying outgassing models) used to generate input data for our analysis, but to point out the consequences and implications that a 
      given nongravitational solution may have on certain comet properties such as the spin-axis orientation, the thermal lag angle, or the 
      value of the oblateness. On the other hand, both 1I/2017~U1~(`Oumuamua) and 2I/Borisov are single-apparition minor bodies and it is 
      unclear how many of the trends observed for well-studied periodic comets and discussed in the literature can be expected to be present 
      for these extrasolar visitors. \citet{1981AREPS...9..113S} argues that most periodic comets tend to have values of the obliquity close 
      to 90{\degr} and the spin axis tends to point toward the Sun at perihelion, which results in the most extreme insolation possible. In 
      addition, the forced precession model of a nonspherical cometary nucleus provides better results when applied to data from several 
      consecutive apparitions of the body under study \citep{1998A&A...335..757K}, but this is clearly not possible when considering 
      interstellar objects. 
   
      \subsection{1I/2017~U1~(`Oumuamua)}
         Although no estimates of the orientation of `Oumuamua's rotation pole have been made public yet, \citet{2018NatAs...2..133F} 
         assumed an obliquity of 0{\degr} in their analyses and \citet{2018Natur.559..223M} considered 45{\degr}, but additional modeling 
         by \citet{2019arXiv191112228Z} indicated that `Oumuamua's effective obliquity when it entered the Solar System may have been 
         93\fdg5 (this result has been removed from the final, published version of this work, \citealt{2020ApJ...899...42Z}). The value 
         assumed by \citet{2018NatAs...2..133F} is marginally consistent with the one computed here for a very oblate shape (see 
         Fig.~\ref{oumuamuaSPINdisc}, top panel); the value obtained by \citet{2019arXiv191112228Z} coincides with our estimate in the case 
         of a very prolate shape (see Fig.~\ref{oumuamuaSPINcigar}, top panel). 

         \citet{2019ApJ...876L..26S} argued that `Oumuamua's outgassing is made of H$_{2}$O molecules and that the venting of this material 
         is expected to be collimated toward the Sun. We found a small lag angle between the subsolar meridian and the direction of maximum 
         mass ejection for the case of a very oblate shape, but the very prolate case shows no statistically significant value of the lag 
         angle, which can be interpreted as lack of correlation between level of activity and elevation of the Sun over the local horizon. 
         On the other hand, the equatorial coordinates of the Sun when `Oumuamua reached perihelion on September 9, 2017 were approximately 
         $(167\fdg8,~+5\fdg2)$, therefore none of our predictions regarding the spin-axis direction are close to the position favored (see 
         above) by \citet{1981AREPS...9..113S}.

         \citet{2018ApJ...867L..17R} showed that under standard conditions, if `Oumuamua is an elongated body experiencing outgassing, it
         should have followed a rapid and violent rotational evolution leading to rotational fission and probably catastrophic disruption 
         prior to its discovery. The fact is that `Oumuamua was discovered and \citet{2019ApJ...876L..26S} explained that being long-term 
         structurally stable, outgassing, and having an extremely oblong appearance could be simultaneously possible if the active region
         tracks the subsolar point (maximal venting occurs when both are aligned). The modeling presented by \citet{2019MNRAS.489.3003M} is
         consistent with the discussion in \citet{2019ApJ...876L..26S}. 

         Our analysis of the rotational properties of `Oumuamua in Sect.~\ref{1I} shows that, for the same orbit determination in 
         Table~\ref{elements1I}, a very prolate or cigar-like shape with $p$=$-$6.7 leads to outgassing taking place evenly throughout the 
         rotation period (see Fig.~\ref{oumuamuaLAGcigar}); in sharp contrast, a very oblate or disk-like shape with $p$=0.836 leads to a 
         strong correlation between time of maximum activity or outgassing and alignment between active region and subsolar point (see 
         Fig.~\ref{oumuamuaLAGdisc}) as demanded by \citet{2019ApJ...876L..26S}. On the other hand, the Solar System is no strange to very 
         oblate or flattened bodies, 486958~Arrokoth (2014~MU$_{69}$), a trans-Neptunian contact binary, has been found to have an average 
         oblateness of $p$=0.695, with 0.808 for the large lobe and 0.582 for the small lobe \citep{2020MNRAS.496.4154A}. 

         It can be argued that the values of the parameters obtained when no value of $p$ is imposed ($\epsilon_{\rm \ `Oumuamua} = 
         93{\degr}_{-13{\degr}}^{+8{\degr}}$, $\eta_{\rm \ `Oumuamua} = 0.08{\degr}_{-55{\degr}}^{+52{\degr}}$) lead to a trivial solution 
         to the problem under study here. The effective equatorial plane of `Oumuamua was nearly perpendicular to its orbital plane and the 
         maximum outgassing was taking place at `Oumuamua's noon, when the Sun was at its highest as seen from the surface of the 
         interstellar visitor. In this case and if `Oumuamua is fusiform, its body would be parallel to its orbital plane and the thermal 
         lag could be effectively zero, then the outgassing is confined only to the radial direction and the two poles are indistinct. This 
         is caused by the fact that both the transverse and normal components of the force are statistically compatible with zero. The 
         inclusion of information on $p$ leads to more complex solutions. 

         The orbit determination in Table~\ref{elements1I} corresponds to 74~d past perihelion and shows that only the radial acceleration 
         may have played a significant role on the dynamical evolution of `Oumuamua as it visited the inner Solar System because the other 
         components have rather uncertain values, statistically compatible with zero \citep{2018Natur.559..223M}. `Oumuamua is in a tumbling 
         rotational state \citep{2018ApJ...856L..21B,2018NatAs...2..407D,2018NatAs...2..383F}. \citet{2018ApJ...856L..21B} argued for two 
         periods ---a short one of 8.7~h and a long one (slow precession) of 54.5~h that are the result of tumbling--- and a long, narrow 
         rod-like shape. \citet{2018NatAs...2..407D} found a variable light curve with a periodicity of 7.56~h consistent with a very 
         elongated shape resulting from a catastrophic collision in the distant past. Extensive reviews on `Oumuamua's properties can be 
         found in \citet{2018Msngr.173...13H} and \citet{2019NatAs...3..594O}.
      
      \subsection{2I/Borisov}
         Although no estimates on the spin-axis orientation of `Oumuamua have been presented in the literature, there are a number of 
         released results for 2I/Borisov although they are all mutually exclusive. \citet{2020AJ....159...77Y} found two equally preferred 
         locations (see fig.~7 in their paper) for the pole's right ascension and declination: (340{\degr},~+30{\degr}) and 
         (205{\degr},~$-$55{\degr}). \citet{2020ApJ...895L..34K} discussed a persistent asymmetry in the dust coma that could be best 
         explained by a thermal lag on the rotating nucleus causing peak mass loss to take place during the comet nucleus afternoon; these 
         authors calculated a value for the obliquity of 2I/Borisov of 30{\degr} and estimated a pole direction of 
         ($\alpha$,~$\delta$)=(205{\degr},~+52{\degr}). \citet{2020MNRAS.495L..92M} computed a spin-axis direction 
         ($\alpha$,~$\delta$)=(260{\degr}$\pm$15{\degr},~$-$35{\degr}$\pm$10{\degr}). \citet{2020MNRAS.497.4031B} estimated a value of the 
         oblateness $\leq$$-$0.5 and argued that a jet is close to the rotation pole that points toward 
         ($\alpha$,~$\delta$)=(322{\degr}$\pm$10{\degr},~+37{\degr}$\pm$10{\degr}), mentioning that the rotation of the nucleus of comet 
         2I/Borisov seems to occur on a single axis and that it appears to be not chaotic. 

         Together with our own estimates in the form of maps of statistical significance, the various estimates for the spin-axis 
         orientation of 2I/Borisov are plotted in Figs.~\ref{2Ipolecoor09Jan} and \ref{2Ipolecoor19Mar}, bottom panels, as filled orange 
         symbols with white error bars (when available): triangles \citep{2020AJ....159...77Y}, square \citep{2020MNRAS.495L..92M}, star 
         \citep{2020MNRAS.497.4031B}, and circle \citep{2020ApJ...895L..34K}. The published solutions for the spin-axis orientation of 
         2I/Borisov are quite different and they may signal actual short-term changes in the values of the rotational properties of this 
         object or problems with the interpretation of some of the observed features. Our estimate based on Table~\ref{elements2Ia} (see 
         Fig.~\ref{2Ipolecoor09Jan}, bottom panel) is inconsistent with any of the other four, but the one based on the orbit determination 
         in Table~\ref{elements2Ib} (see Fig.~\ref{2Ipolecoor19Mar}, bottom panel) is marginally consistent with the result obtained by 
         \citet{2020ApJ...895L..34K} although the associated values of the obliquity are quite different, 30{\degr} versus 90{\degr}. Our 
         value of the oblateness for the solution in Table~\ref{elements2Ib} is inconsistent with the one computed by 
         \citet{2020MNRAS.497.4031B}. On the other hand, the equatorial coordinates of the Sun when 2I/Borisov reached perihelion on 
         December 8, 2019 were approximately $(254\fdg6,~-22\fdg7)$, so none of our predictions are close to the position favored by 
         \citet{1981AREPS...9..113S}. This location is also inconsistent with the estimates provided by \citet{2020AJ....159...77Y}, 
         \citet{2020MNRAS.497.4031B}, and \citet{2020ApJ...895L..34K}; it is, however, somewhat consistent with the spin-axis direction 
         computed by \citet{2020MNRAS.495L..92M}.  

         Most of the observed properties of interstellar comet 2I/Borisov have been found to be remarkably similar to those of known Solar 
         System comets (see for example \citealt{2019RNAAS...3..131D,2020MNRAS.495.2053D,2019ApJ...885L...9F,2019A&A...631L...8O,
         2020ApJ...893L..12C,2020NatAs...4...53G,2020ApJ...888L..23J,2020ApJ...889L..38K,2020A&A...634L...6Y}). Recent estimates, placed the 
         rotation period of 2I/Borisov at about 5.3~h or perhaps 10.6~h \citep{2020MNRAS.497.4031B}. \citet{2019RNAAS...3..187G} had found a 
         very slow rotation rate with a period of about 13~d. If confirmed, this would be the longest rotation period ever observed in a 
         comet. Slowly rotating Solar System comets, such as 1P/Halley and 109P/Swift-Tuttle, have rotation periods of about 2.8~d (see for 
         example table~1 in \citealt{2004come.book..281S}); comet 29P/Schwassmann-Wachmann~1 ---well known for its enormous and random 
         outbursts--- was reported to have a rotation period of about 5~d (see for example \citealt{1981AREPS...9..113S}), but recent 
         determinations place its spin rate close to 1.4~d (see for example table~1 in \citealt{2004come.book..281S}). JPL's SBDB indicates 
         that, among known comets, P/2006~HR30 (Siding Spring) has the longest rotation period with 2.9~d. 

         The thermal lag angle is relatively small for our results based on Table~\ref{elements2Ib} and this suggests that outgassing for 
         2I/Borisov appears to have been active only when the Sun was rather high over the local horizon. As for the large lag angle 
         obtained in the case of the results based on Table~\ref{elements2Ia}, it suggests that an insulating crust of nonvolatile material 
         might have been present over most of the surface of this comet prior to the sequence of outbursts pointed out above; no significant 
         change in the slightly reddish color of the comet was observed from late September 2019 through late January 2020 
         \citep{2020AJ....160...92H}. Alternatively, this could be the result of its putative long period as the effects of the outgassing 
         from the comet's dayside may cancel out and only discrete nightside outbursts may result in a net contribution to the value of 
         nongravitational normal acceleration parameter. Yet another scenario may involve an active vent close to the south pole of the 
         comet that, due to the viewing geometry, remained on the nightside of the comet during the time interval covered by the 
         observations. Both \citet{2020MNRAS.495L..92M} and \citet{2020MNRAS.497.4031B} argued that such a scenario may have been possible.

         An improved orbit determination of 2I/Borisov was published on August 21, 2020 (see Table~\ref{elements2Ic}), when this work was 
         under review, that also considers nongravitational effects driven by sublimation of water ice (see Sect.~\ref{Intro}). This 
         solution is close to that in Table~\ref{elements2Ib} although it assummes asymmetric nongravitational forces including an 
         additional parameter, $d\tau$, to account for an asymmetry of the comet's outgassing (see for example 
         \citealt{2004come.book..137Y}). We have repeated the calculations discussed in Sect.~\ref{2I} using the data in 
         Table~\ref{elements2Ic} as input and found results that are similar (see Fig.~\ref{2Ipolecoor21Aug}) to those obtained for the 
         orbit determination in Table~\ref{elements2Ib}. The inclusion of the nongravitational perihelion offset has no effects on the 
         mathematical expressions used to perform our calculations (see Sect.~\ref{Intro}). 
%
%
     \begin{table*}
        \fontsize{8}{12pt}\selectfont
        \tabcolsep 0.15truecm
        \caption{\label{elements2Ic}Heliocentric and barycentric orbital elements, and 1$\sigma$ uncertainties of interstellar comet 2I/Borisov (III).
                }
        \centering
        \begin{tabular}{lccc}
           \hline\hline
            Orbital parameter                                                        &   & Heliocentric                           & Barycentric  \\
           \hline
            Eccentricity, $e$                                                        & = &    3.356215$\pm$0.000012               &    3.358837  \\
            Perihelion distance, $q$ (AU)                                            & = &    2.006582$\pm$0.000005               &    2.011824  \\
            Inclination, $i$ (\degr)                                                 & = &   44.052571$\pm$0.000006               &   44.062173  \\
            Longitude of the ascending node, $\Omega$ (\degr)                        & = &  308.14873$\pm$0.00004                 &  308.10019   \\
            Argument of perihelion, $\omega$ (\degr)                                 & = &  209.12368$\pm$0.00012                 &  209.16653   \\
            Time of perihelion passage, $\tau$ (TDB)                                 & = & 2458826.0451$\pm$0.0003                & 2458826.2622 \\
           \hline
            Nongravitational radial acceleration parameter, $A_1$ (AU d$^{-2}$)      & = & $7.09\times10^{-8}\pm7.6\times10^{-9}$ &              \\
            Nongravitational transverse acceleration parameter, $A_2$ (AU d$^{-2}$)  & = & $-1.4\times10^{-8}\pm2.7\times10^{-9}$ &              \\
            Nongravitational normal acceleration parameter, $A_3$ (AU d$^{-2}$)      & = & $6.5\times10^{-10}\pm1.4\times10^{-9}$ &              \\
            Nongravitational perihelion offset, $d\tau$ (d)                          & = & $87.3\pm4.6$                           &              \\
           \hline
        \end{tabular}
        \tablefoot{This solution is hyperbolic with a statistical significance of 189938-$\sigma$ (barycentric, using all the decimal figures provided
                   by the data source) and it is based on 1428 observations that span a data-arc of 311~d. The last observation used in these 
                   calculations was acquired on July 6, 2020; the first one, on August 30, 2019. The orbit determination has been computed by 
                   D. Farnocchia at epoch JD 2459062.5 that corresponds to 00:00:00.000 TDB on 2020 August 1, J2000.0 ecliptic and equinox. Source: 
                   JPL's SSDG SBDB (solution date, 2020-Aug-21 09:32:58 PDT).
                  }
     \end{table*}
%
%
%
%
      \begin{figure}
        \centering
         \includegraphics[width=\linewidth]{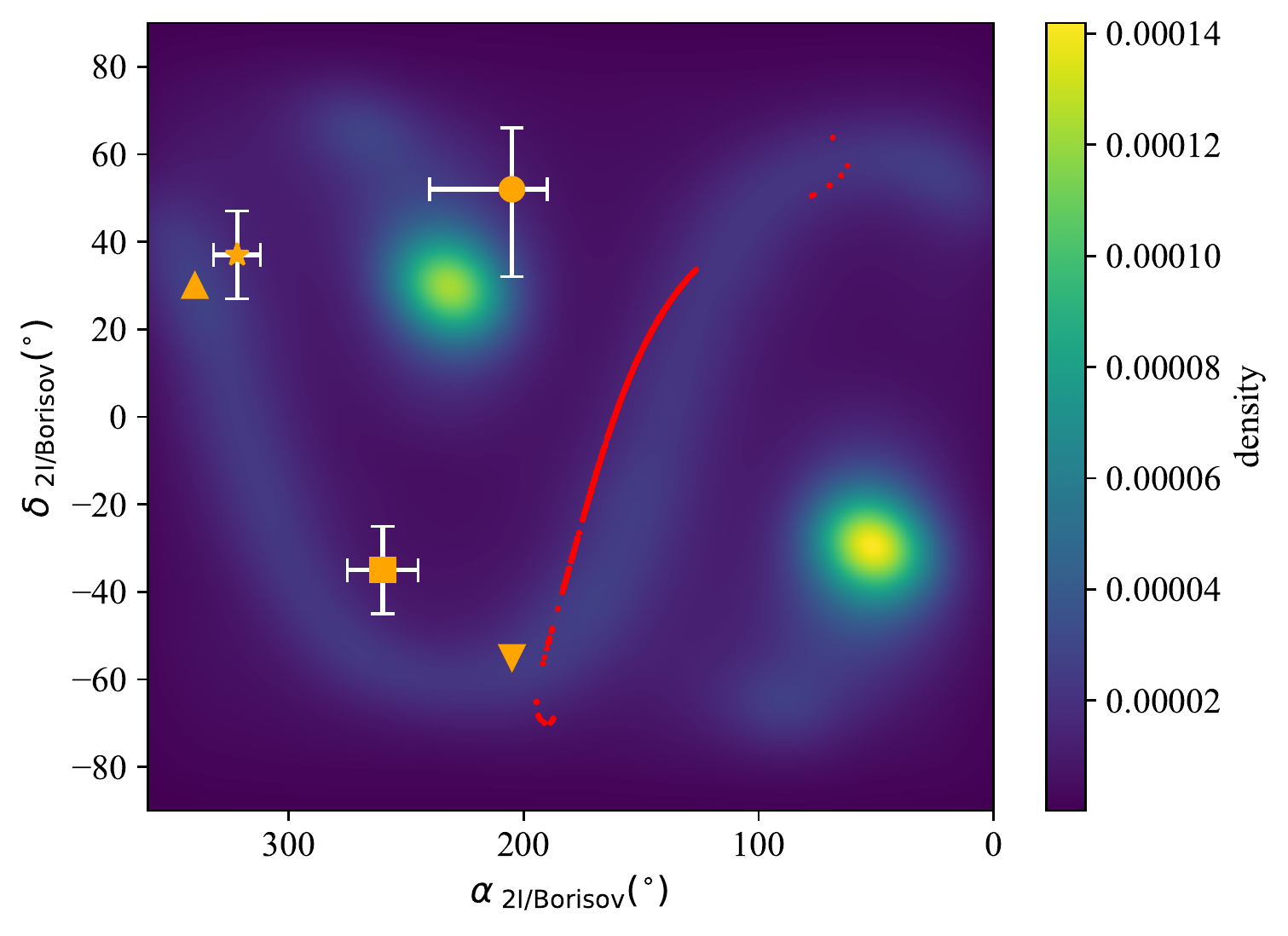}
         \caption{Same as Fig.~\ref{2Ipolecoor19Mar}, bottom panel, but based on data from Table~\ref{elements2Ic}.
                 }
         \label{2Ipolecoor21Aug}
      \end{figure}
%
%

   \section{Conclusions\label{Summary}} 
      In this paper, we have shown that the most probable values of the equatorial obliquity and the cometocentric longitude of the Sun at 
      perihelion as well as the value of the thermal lag angle and the oblateness of a comet can be computed from an orbit determination 
      that includes nongravitational parameters using a Monte Carlo random search approach within the framework of the forced precession 
      model of a nonspherical cometary nucleus. The algorithm receives input from a MCCM process that includes the uncertainties of the 
      orbit determination in a self-consistent way. The resulting combined algorithm is applied to investigate the spin-axis orientations of 
      extrasolar minor bodies 1I/2017~U1~(`Oumuamua) and 2I/Borisov. 

      For `Oumuamua and using data from Table~\ref{elements1I}, we found that its equatorial obliquity could be about 93{\degr}, if it has a 
      very prolate (fusiform) shape, or close to 16{\degr}, if it is very oblate (disk-like). Outgassing for `Oumuamua appears to have been 
      active only when the Sun was highest over the local horizon, as the thermal lag angle is small, if the body has a thin-disk appearance; 
      for a cigar-like shape, outgassing may have been evenly distributed in time. Our calculations suggest that the most probable spin-axis 
      direction of `Oumuamua in equatorial coordinates could be $(280{\degr},~+46{\degr})$ if very prolate or $(312{\degr},~-50{\degr})$ if 
      very oblate. Our analysis favors a prolate shape. No previous estimates of the direction of the polar axis of `Oumuamua are available.

      For 2I/Borisov, three different orbit determinations gave values of 59{\degr} (using data from Table~\ref{elements2Ia}) and 90{\degr} 
      (using data from Tables~\ref{elements2Ib} and ~\ref{elements2Ic}) for its obliquity, with most probable poles pointing near 
      $(275{\degr},~+65{\degr})$ and $(231{\degr},~+30{\degr})$, respectively. Although our analysis favors an oblate shape ($p$=0.34) for 
      2I/Borisov, a prolate one cannot be ruled out. Our estimates for the spin-axis direction of 2I/Borisov are inconsistent with published 
      determinations (except perhaps with the one in \citealt{2020ApJ...895L..34K}), which already were mutually exclusive. 

      No further data may be collected from `Oumuamua, but new data on 2I/Borisov will be made public and they may lead to better, more 
      reliable estimates of its rotational properties. The new data may improve what is already known and/or perhaps provide new snapshots 
      of the evolution of its rotation. 

   \begin{acknowledgements}
      We thank the referee for her/his constructive, detailed, and insightful report that included very helpful suggestions regarding the
      presentation of this paper and the interpretation of our results, J. de Le\'on, J. Licandro, and M. Serra-Ricart for comments on 
      2I/Borisov's ongoing cometary activity, and A.~I. G\'omez de Castro for providing access to computing facilities. Part of the 
      calculations and the data analysis were completed on the Brigit HPC server of the `Universidad Complutense de Madrid' (UCM), and we 
      thank S. Cano Als\'ua for his help during this stage. This research was partially supported by the Spanish `Ministerio de 
      Econom\'{\i}a y Competitividad' (MINECO) under grant ESP2017-87813-R. In preparation of this paper, we made use of the NASA 
      Astrophysics Data System, the ASTRO-PH e-print server, and the MPC data server. This research made use of 
      Astropy,\footnote{http://www.astropy.org} a community-developed core Python package for Astronomy 
      \citep{2013A&A...558A..33A,2018AJ....156..123A}.
   \end{acknowledgements}

   \bibliographystyle{aa}

   \begin{appendix}
      \section{Elements of the matrix \textbf{\textsf{A}}\label{Aelements}}
         If the elements of the covariance matrix \textbf{\textsf{C}} are written as $c_{ij}$ and those of \textbf{\textsf{A}} as $a_{ij}$ 
         (so \textbf{\textsf{C}} = \textbf{\textsf{A}} \textbf{\textsf{A}}$^{\textbf{\textsf{T}}}$), where those are the entries in the 
         $i$-th row and $j$-th column, they are related by the following expressions:
         \begin{equation}
            \nonumber
            \begin{aligned}
               a_{11} & = \sqrt{c_{11}} \\
               a_{21} & = c_{12} / a_{11} \\
               a_{31} & = c_{13} / a_{11} \\
               a_{41} & = c_{14} / a_{11} \\
               a_{51} & = c_{15} / a_{11} \\
               a_{61} & = c_{16} / a_{11} \\
               a_{71} & = c_{17} / a_{11} \\
               a_{81} & = c_{18} / a_{11} \\
               a_{91} & = c_{19} / a_{11} 
            \end{aligned}
         \end{equation}
         \begin{equation}
            \nonumber
            \begin{aligned}
               a_{22} & = \sqrt{c_{22} - a_{21}^{2}} \\
               a_{32} & = (c_{23} - a_{21}\,a_{31}) / a_{22} \\
               a_{42} & = (c_{24} - a_{21}\,a_{41}) / a_{22} \\
               a_{52} & = (c_{25} - a_{21}\,a_{51}) / a_{22} \\
               a_{62} & = (c_{26} - a_{21}\,a_{61}) / a_{22} \\
               a_{72} & = (c_{27} - a_{21}\,a_{71}) / a_{22} \\
               a_{82} & = (c_{28} - a_{21}\,a_{81}) / a_{22} \\
               a_{92} & = (c_{29} - a_{21}\,a_{91}) / a_{22} \\
            \end{aligned}
         \end{equation}
         \begin{equation}
            \nonumber
            \begin{aligned}
               a_{33} & = \sqrt{c_{33} - a_{31}^{2} - a_{32}^{2}} \\
               a_{43} & = (c_{34} - a_{31}\,a_{41} - a_{32}\,a_{42}) / a_{33} \\
               a_{53} & = (c_{35} - a_{31}\,a_{51} - a_{32}\,a_{52}) / a_{33} \\
               a_{63} & = (c_{36} - a_{31}\,a_{61} - a_{32}\,a_{62}) / a_{33} \\
               a_{73} & = (c_{37} - a_{31}\,a_{71} - a_{32}\,a_{72}) / a_{33} \\
               a_{83} & = (c_{38} - a_{31}\,a_{81} - a_{32}\,a_{82}) / a_{33} \\
               a_{93} & = (c_{39} - a_{31}\,a_{91} - a_{32}\,a_{92}) / a_{33} \\
            \end{aligned}
         \end{equation}
         \begin{equation}
            \nonumber
            \begin{aligned}
               a_{44} & = \sqrt{c_{44} - a_{41}^{2} - a_{42}^{2} - a_{43}^{2}} \\
               a_{54} & = (c_{45} - a_{41}\,a_{51} - a_{42}\,a_{52} - a_{43}\,a_{53}) / a_{44} \\
               a_{64} & = (c_{46} - a_{41}\,a_{61} - a_{42}\,a_{62} - a_{43}\,a_{63}) / a_{44} \\
               a_{74} & = (c_{47} - a_{41}\,a_{71} - a_{42}\,a_{72} - a_{43}\,a_{73}) / a_{44} \\
               a_{84} & = (c_{48} - a_{41}\,a_{81} - a_{42}\,a_{82} - a_{43}\,a_{83}) / a_{44} \\
               a_{94} & = (c_{49} - a_{41}\,a_{91} - a_{42}\,a_{92} - a_{43}\,a_{93}) / a_{44} \\
            \end{aligned}
         \end{equation}
         \begin{equation}
            \nonumber
            \begin{aligned}
               a_{55} & = \sqrt{c_{55} - a_{51}^{2} - a_{52}^{2} - a_{53}^{2} - a_{54}^{2}} \\
               a_{65} & = (c_{56} - a_{51}\,a_{61} - a_{52}\,a_{62} - a_{53}\,a_{63} - a_{54}\,a_{64}) / a_{55} \\
               a_{75} & = (c_{57} - a_{51}\,a_{71} - a_{52}\,a_{72} - a_{53}\,a_{73} - a_{54}\,a_{74}) / a_{55} \\
               a_{85} & = (c_{58} - a_{51}\,a_{81} - a_{52}\,a_{82} - a_{53}\,a_{83} - a_{54}\,a_{84}) / a_{55} \\
               a_{95} & = (c_{59} - a_{51}\,a_{91} - a_{52}\,a_{92} - a_{53}\,a_{93} - a_{54}\,a_{94}) / a_{55} \\
            \end{aligned}
         \end{equation}
         \begin{equation}
            \nonumber
            \begin{aligned}
               a_{66} & = \sqrt{c_{66} - a_{61}^{2} - a_{62}^{2} - a_{63}^{2} - a_{64}^{2} - a_{65}^{2}} \\
               a_{76} & = (c_{67} - a_{61}\,a_{71} - a_{62}\,a_{72} - a_{63}\,a_{73} - a_{64}\,a_{74} - a_{65}\,a_{75}) / a_{66} \\
               a_{86} & = (c_{68} - a_{61}\,a_{81} - a_{62}\,a_{82} - a_{63}\,a_{83} - a_{64}\,a_{84} - a_{65}\,a_{85}) / a_{66} \\
               a_{96} & = (c_{69} - a_{61}\,a_{91} - a_{62}\,a_{92} - a_{63}\,a_{93} - a_{64}\,a_{94} - a_{65}\,a_{95}) / a_{66} \\
            \end{aligned}
         \end{equation}
         \begin{equation}
            \nonumber
            \begin{aligned}
               a_{77} & = \sqrt{c_{77} - a_{71}^{2} - a_{72}^{2} - a_{73}^{2} - a_{74}^{2} - a_{75}^{2} - a_{76}^{2}} \\
               a_{87} & = (c_{78} - a_{71}\,a_{81} - a_{72}\,a_{82} - a_{73}\,a_{83} - a_{74}\,a_{84} - a_{75}\,a_{85} \\ 
                      & - a_{76}\,a_{86}) / a_{77} \\
               a_{97} & = (c_{79} - a_{71}\,a_{91} - a_{72}\,a_{92} - a_{73}\,a_{93} - a_{74}\,a_{94} - a_{75}\,a_{95} \\
                      & - a_{76}\,a_{96}) / a_{77} \\
            \end{aligned}
         \end{equation}
         \begin{equation}
            \nonumber
            \begin{aligned}
               a_{88} & = \sqrt{c_{88} - a_{81}^{2} - a_{82}^{2} - a_{83}^{2} - a_{84}^{2} - a_{85}^{2} - a_{86}^{2} - a_{87}^{2}} \\
               a_{98} & = (c_{89} - a_{81}\,a_{91} - a_{82}\,a_{92} - a_{83}\,a_{93} - a_{84}\,a_{94} - a_{85}\,a_{95} \\
                      & - a_{86}\,a_{96} - a_{87}\,a_{97}) / a_{88} \\
            \end{aligned}
         \end{equation}
         \begin{equation}
            \nonumber
            \begin{aligned}
               a_{99} & = \sqrt{c_{99} - a_{91}^{2} - a_{92}^{2} - a_{93}^{2} - a_{94}^{2} - a_{95}^{2} - a_{96}^{2} - a_{97}^{2} 
                                       - a_{98}^{2}} \,.
            \end{aligned}
         \end{equation}
   \end{appendix}

\end{document}